%% file: main.tex
\documentclass[acmtog,screen, nonacm]{acmart}
\acmArticle{116}
\newif\ifmarked

\author{Anka He Chen}
\orcid{0000-0002-5819-3453}
\email{ankachan92@gmail.com}
\affiliation{%
  \institution{University of Utah}
  \city{Salt Lake City}
  \state{UT}
  \country{USA}
}

\author{Ziheng Liu}
\orcid{0009-0008-5031-5074}
\email{NA}
\affiliation{%
  \institution{University of Utah}
  \city{Salt Lake City}
  \state{UT}
  \country{USA}
}

\author{Yin Yang}
\orcid{0000-0001-7645-5931}
\email{NA}
\affiliation{%
  \institution{University of Utah}
  \city{Salt Lake City}
  \state{UT}
  \country{USA}
}

\author{Cem Yuksel}
\orcid{0000-0002-0122-4159}
\email{cem@cemyuksel.com}
\affiliation{%
  \institution{University of Utah \& Roblox}
  \city{Salt Lake City}
  \state{UT}
  \country{USA}
}

\usepackage[font=normalsize]{caption}
\usepackage{booktabs} 
\usepackage{tabularx}
\usepackage{booktabs}
\usepackage[labelfont=bf,format=plain]{caption}
\usepackage{float}
\usepackage[ruled,linesnumbered]{algorithm2e}
\usepackage{amsmath}
\usepackage{stackengine}
\usepackage{comment}
\usepackage{bm}

\setcopyright{rightsretained}
\acmJournal{TOG}
\acmYear{2024} \acmVolume{43} \acmNumber{4} \acmArticle{} \acmMonth{7}\acmDOI{10.1145/3658179}

\DeclareGraphicsExtensions{.pdf,.jpg,.png}  

\usepackage{subcaption}
\newcommand{\commentText}[1]{}

\newcommand{\anka}[1]{\commentText{{\color{blue}[\textbf{\textsc{Anka}}: \textit{#1}]}}}
\newcommand{\ziheng}[1]{\commentText{{\color{brown}[\textbf{\textsc{ziheng}}: \textit{#1}]}}}
\newcommand{\cem}[1]{\commentText{{\color{orange}[\textbf{\textsc{Cem}}: \textit{#1}]}}}

\newcommand{\sref}[1]{\textbf{(#1)}}

\usepackage{soul,xcolor}
\soulregister\cite7
\soulregister\citet7
\soulregister\ref7
\soulregister\pageref7
\soulregister\shortcite7
\soulregister\autoref7
 
\newcommand{\add}[1]{#1}
\newcommand{\newrefbegin}[1]{}

\newcommand{\del}[1]{\ignorespaces}

\begin{document}
\title{Vertex Block Descent}



\begin{abstract}
We introduce vertex block descent, a block coordinate descent solution for the variational form of implicit Euler through vertex-level Gauss-Seidel iterations. It operates with local vertex position updates that achieve reductions in global variational energy with maximized parallelism. This forms a physics solver that can achieve numerical convergence with unconditional stability and exceptional computation performance. It can also fit in a given computation budget by simply limiting the iteration count while maintaining its stability and superior convergence rate.

We present and evaluate our method in the context of elastic body dynamics, providing details of all essential components and showing that it outperforms alternative techniques. In addition, we discuss and show examples of how our method can be used for other simulation systems, including particle-based simulations and rigid bodies.
\end{abstract}

%
%

\ccsdesc[500]{Computing methodologies~Physical simulation}
\ccsdesc[300]{Computing methodologies~Collision detection}

%
%

\keywords{physics-based simulation, elastic body, rigid body, time integration}

\newcommand{\TODO}[1]{\textcolor{red}{ToDo:#1}}
\renewcommand{\vec}[1]{\mathbf{#1}}
\newcommand{\rest}[1]{\overline{#1}}
\newcommand{\bound}[1]{\partial#1}
\newcommand{\inter}[1]{#1^\circ}

\newcommand{\restbf}[1]{\overline{\mathbf{#1}}}

\setlength{\fboxsep}{0pt}
\setlength{\fboxrule}{0.2pt}

\begin{teaserfigure}
\centering
\fbox{\includegraphics[height=0.313\linewidth]{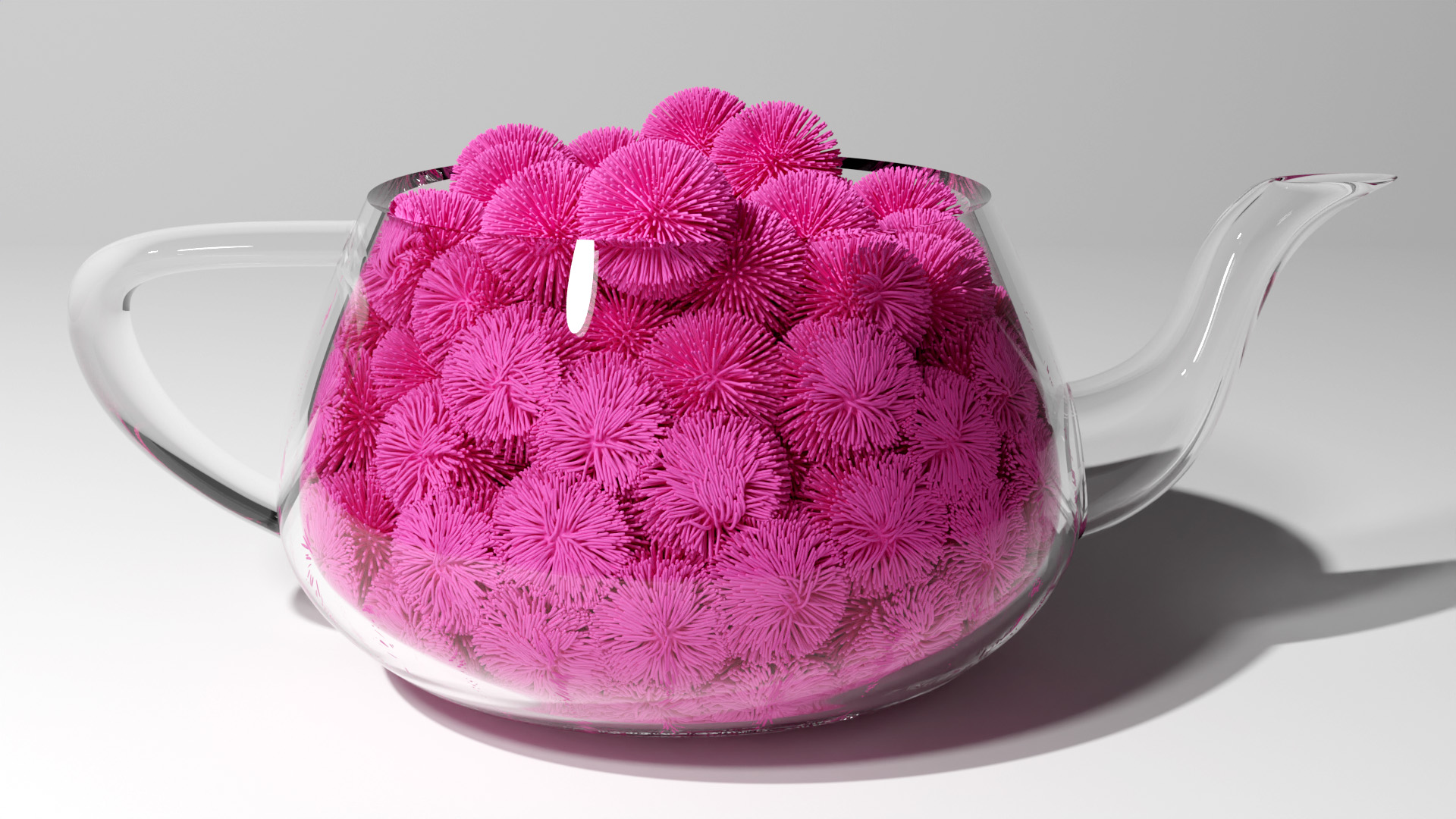}}\hfill%
\fbox{\includegraphics[height=0.313\linewidth,trim=280 0 300 120,clip]{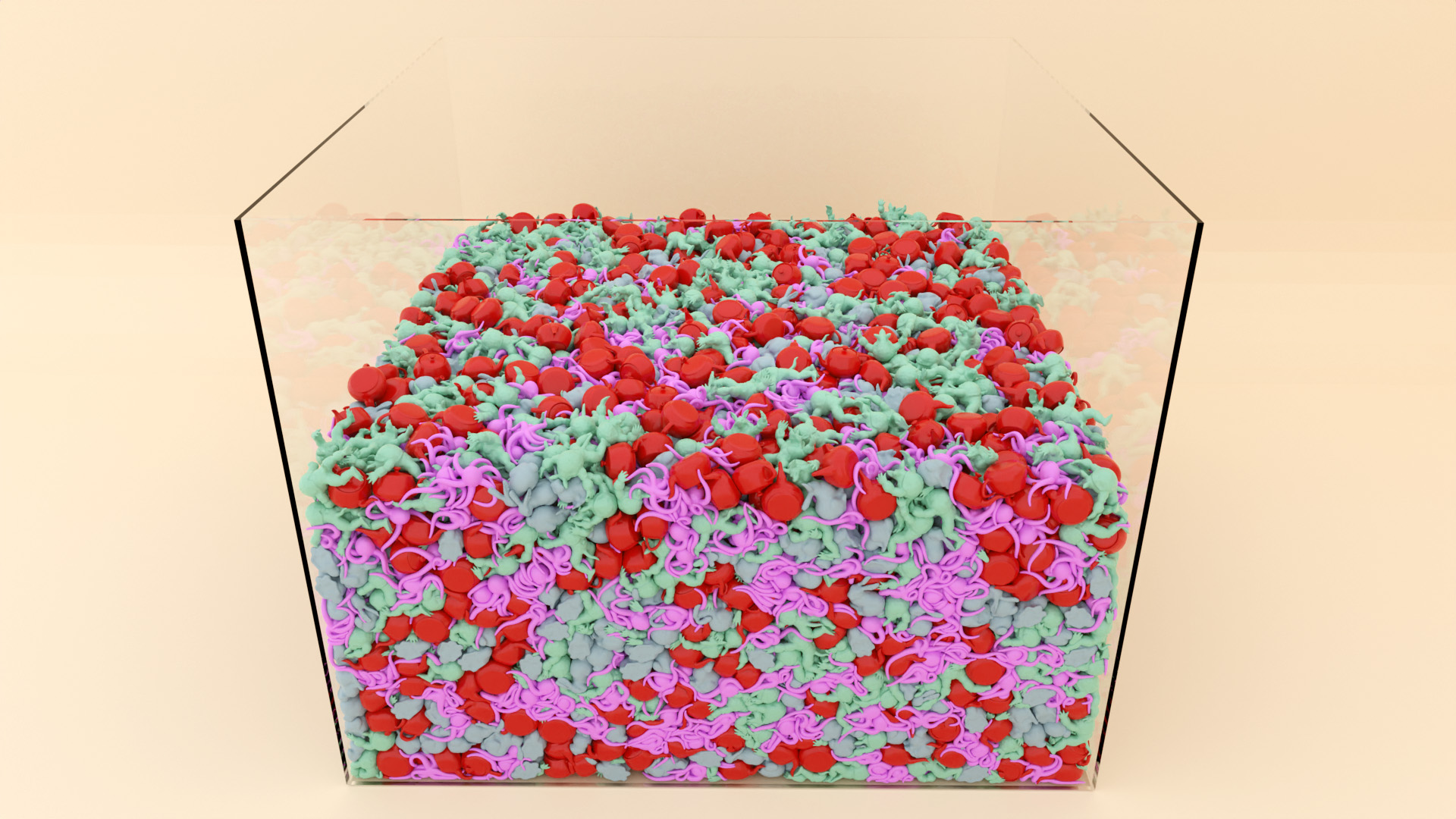}}%
\captionof{figure}{Example simulation results using our solver, both of those methods involve more than 100 million DoFs and 1 million active collisions. }
\label{fig:teaser}
\end{teaserfigure}
\maketitle
\input{01_Introduction}

\input{02_RelatedWorks}

\input{03_VertexBlockDescent}

\input{04_Results}

\input{05_Discussion}

\input{06_Conclusion}

\begin{acks}
We thank Miles Macklin and Theodore Kim for an inspiring discussion on primal solvers.
This work was supported in part by NSF grant 1764071.
\end{acks}

\bibliographystyle{ACM-Reference-Format}
\bibliography{references, References_custom}

\end{document}

%% file: 01_Introduction.tex
\section{Introduction}

Physics-based simulation is the cornerstone of most graphics applications and the demands from simulation systems to deliver improved stability, accelerated computational performance, and enhanced visual realism are ever-growing. Particularly in real-time graphics applications, the stability and performance requirements are so strict that realism can sometimes be begrudgingly considered of secondary importance.

Notwithstanding the substantial amount of research and groundbreaking discoveries made on physics solvers over the past decades, existing methods still leave some things to be desired. They either deliver high-quality results, but fail to meet the computational demands of many applications or fit in a given computation time by sacrificing quality. Stability, on the other hand, is always a challenge, particularly with strict computation budgets.

In this paper, we introduce \emph{vertex block descent (VBD)}, a physics solver that offers unconditional stability, superior computational performance than prior methods, and the ability to achieve numerical convergence \add{to an implicit Euler integration}. 
Though our method is a general solution that can be used for a variety of simulation problems, 
we present and evaluate it in the context of elastic body dynamics. Then, we briefly discuss how our method can be applied to some other simulation systems, including particle-based simulations and rigid bodies.

Our VBD method is based on block coordinate descent that performs vertex-based Gauss-Seidel iterations to solve the variational form of implicit Euler.
For elastic body dynamics, each iteration runs a loop over the mesh vertices, adjusting the position of a single vertex at a time, temporarily fixing all others. 
This offers maximized parallelism when coupled with vertex-based mesh coloring, which can achieve an order of magnitude fewer colors (i.e. serialized workloads) as compared to element-based parallelization.
Our local position-based updates can ensure that the variational energy is always reduced. 
Therefore, our method maintains its stability even with a single iteration per time step and large time steps, operating with unconverged solutions containing a large amount of residual. With more iterations, it converges faster than its alternatives. Thus, it can more easily fit in a given computation budget, while maintaining stability with improved convergence.

\begin{figure}
\newcommand{\img}[1]{\includegraphics[width=0.5\linewidth,trim=100 230 150 150, clip]{Figures/BeamRotation/#1-cc}}
\centering
\img{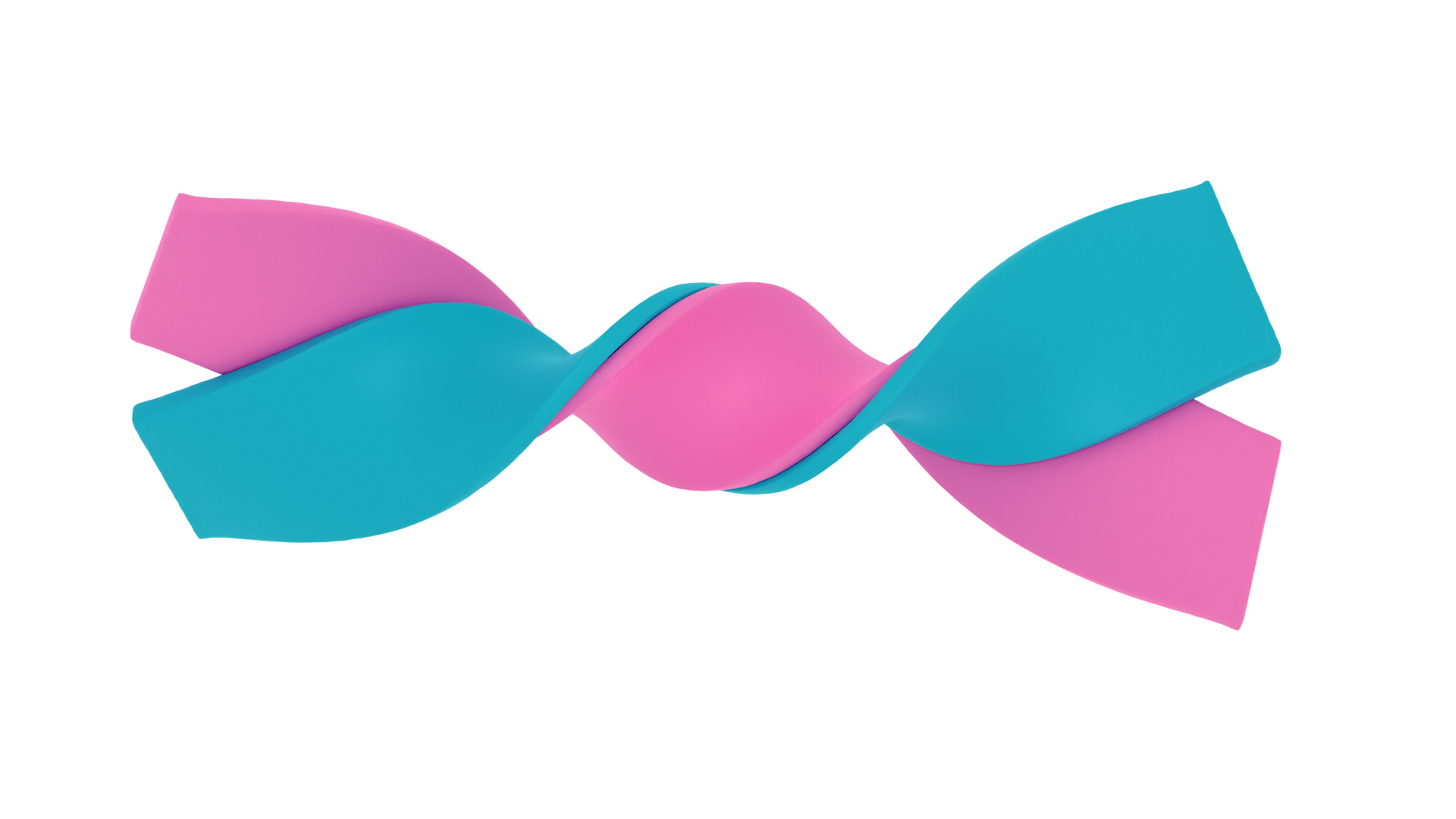}\hfill\img{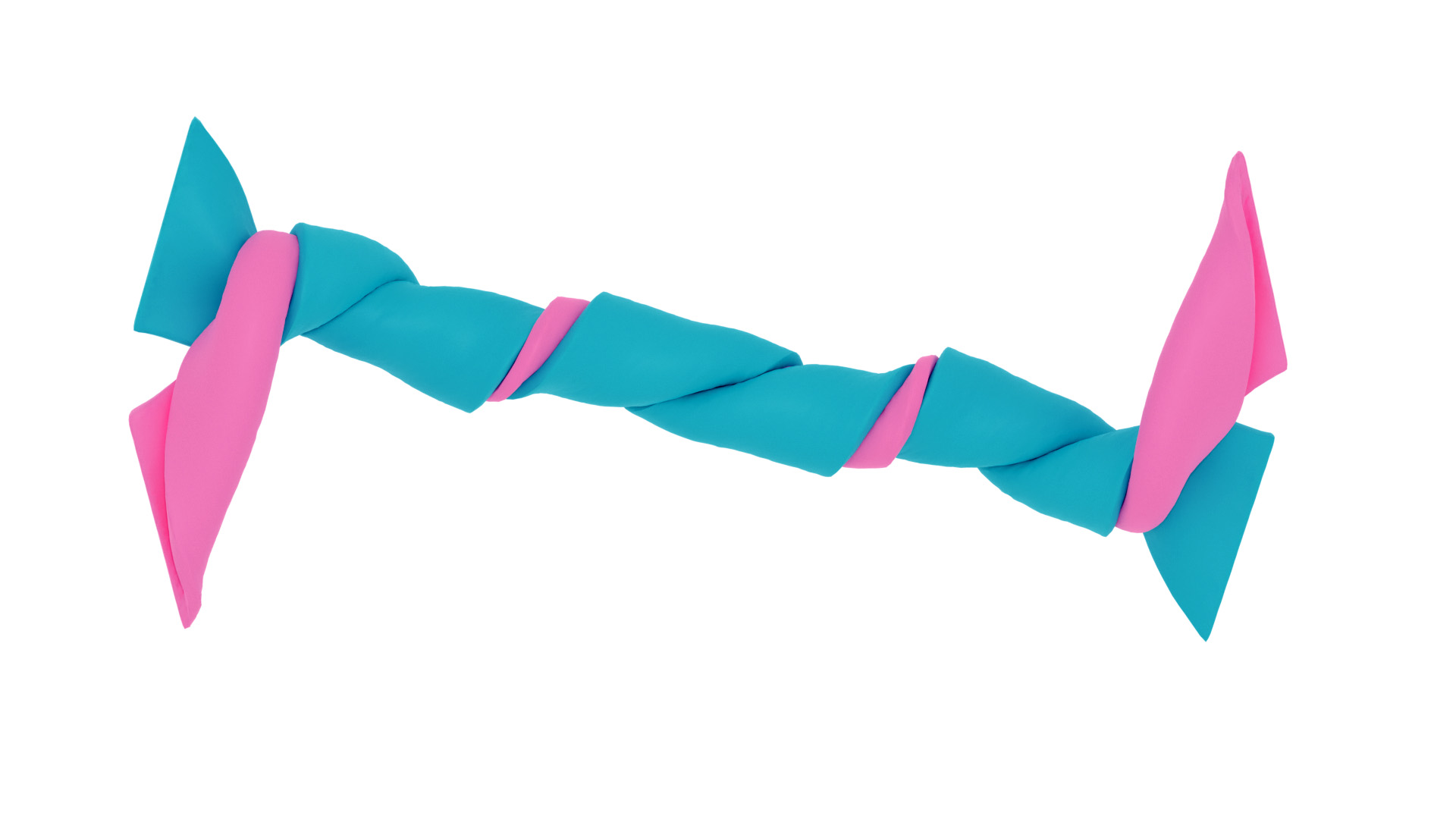}
\caption{Twisting two beams together, totaling 97 thousand vertices and 266 thousand tetrahedra, demonstrating complex frictional contact and buckling.}
\label{fig:ExtremeTwist}
\end{figure}

We present all essential components of using VBD for elastic body dynamics, including formulations for damping, constraints, collisions, and friction. We also introduce a simple initialization scheme to warm-start the optimization and improve convergence. In addition, we discuss momentum-based acceleration techniques and parallelization in the presence of dynamic collisions.
Our evaluation includes large simulations (\autoref{fig:teaser}) and stress tests (e.g. \autoref{fig:ExtremeTwist} and \ref{fig:ExtremeInitialization}) that demonstrate VBD's performance, scalability, and stability.



%% file: 02_RelatedWorks.tex
\section{Related Work}

There is a large body of work on physics-based simulation in computer graphics. Here we only discuss the recent and the most relevant work to our method.

Implicit time integrators are widely accepted as the primary methods for simulating elastic bodies in computer graphics due to their exceptional stability, especially when addressing stiff problems.
Among these options, backward Euler \cite{baraff_large_1998, hirota_implicit_2001, volino_comparing_2001, martin2011example} is the most commonly utilized method, though other approaches like implicit Newmark \cite{kane2000variational, bridson2002robust, bridson2005simulation}, BDF2 \cite{choi_stable_2005, hauth_high_2001}, and implicit-explicit \cite{hansmann_implicit-explicit_2000, stern_implicit-explicit_2009} have also been explored.
Backward Euler is often approximated as a single Newton step, solving a linear system of equations \cite{baraff_large_1998}. Line search can be applied to improve robustness \cite{hirota_implicit_2001}.
Preconditioning \cite{ascher2003modified} or positive-definite projection \cite{teran2005robust} can be used to improve convergence.
To circumvent a full linear solve for every Newton step, Cholesky factorization \cite{hecht_updated_2012} emerges as a viable strategy. 
Techniques like multi-resolution \cite{grinspun2002charms, capell_multiresolution_2002} or multigrid \cite{bolz2003sparse, tamstorf_smoothed_2015, xian_scalable_2019, wang_hierarchical_2020, zhu_efficient_2010} solvers 
project
finer details onto a coarser grid with fewer degrees of freedom, effectively reducing the computational cost of the linear system solver.
Our solution linearizes the forces locally, avoiding the global linear system, and it converges to the same result as backward Euler with multiple Newton steps.

Additionally, stiffness warping \cite{muller2002stable} repurposes the stiffness matrix from the rest shape to handle significant rotational deformations. Using a quasi-Newton method \cite{liu_quasi-newton_2017} with an approximate Hessian can be far more cost-effective for computing its inverse with prefactoring than the actual Hessian. Example-based dynamics has also been explored \cite{chao_simple_2010, muller2005meshless, martin2011example}, where the deformation energy is defined based on the nearest point in the example space.
Projective Dynamics \cite{bouaziz2023projective} represents deformation energy via a series of constraints that can be solved independently, then synchronized either through a prefactorized global step to accelerate convergence.

The relation between dynamics, energy, and minimization has been leveraged in variational integrators \cite{kane2000variational, kane_symplectic-energy-momentum_1999, simo_exact_1992,lew_variational_2004, 10.2312:SCA:SCA06:043-051}.  Reformulating backward Euler to an optimization problem combined with optimization techniques to allows the usage of large steps \cite{martin2011example, gast_optimization_2015}. Domain-decomposed optimization for solving the nonlinear problems of implicit numerical time integration can accelerate convergence \citet{li_decomposed_2019}. Yet, the optimization formulation has its drawbacks, notably the varying problem formulation and initial positions in each step. Consequently, achieving consistent convergence within a fixed time budget becomes challenging. Real-time simulators compromise by accepting partially converged results, which visually resemble the final solution. Unfortunately, this compromise can lead to significant visual artifacts mainly due to retained residuals from earlier steps, which can accumulate across frames and can jeopardize the solver's stability.
Our method, as illustrated in \autoref{fig:ExtremeStretch2}, demonstrates exceptional stability even when retaining a substantial amount of residual with a limited number of iterations.

\begin{figure}
\centering
\newcommand{\fig}[2][700 100 640 400]{\includegraphics[width=0.248\linewidth,trim=#1, clip]{Figures/RandomizeAndFlattening/#2-cc}}
\fig{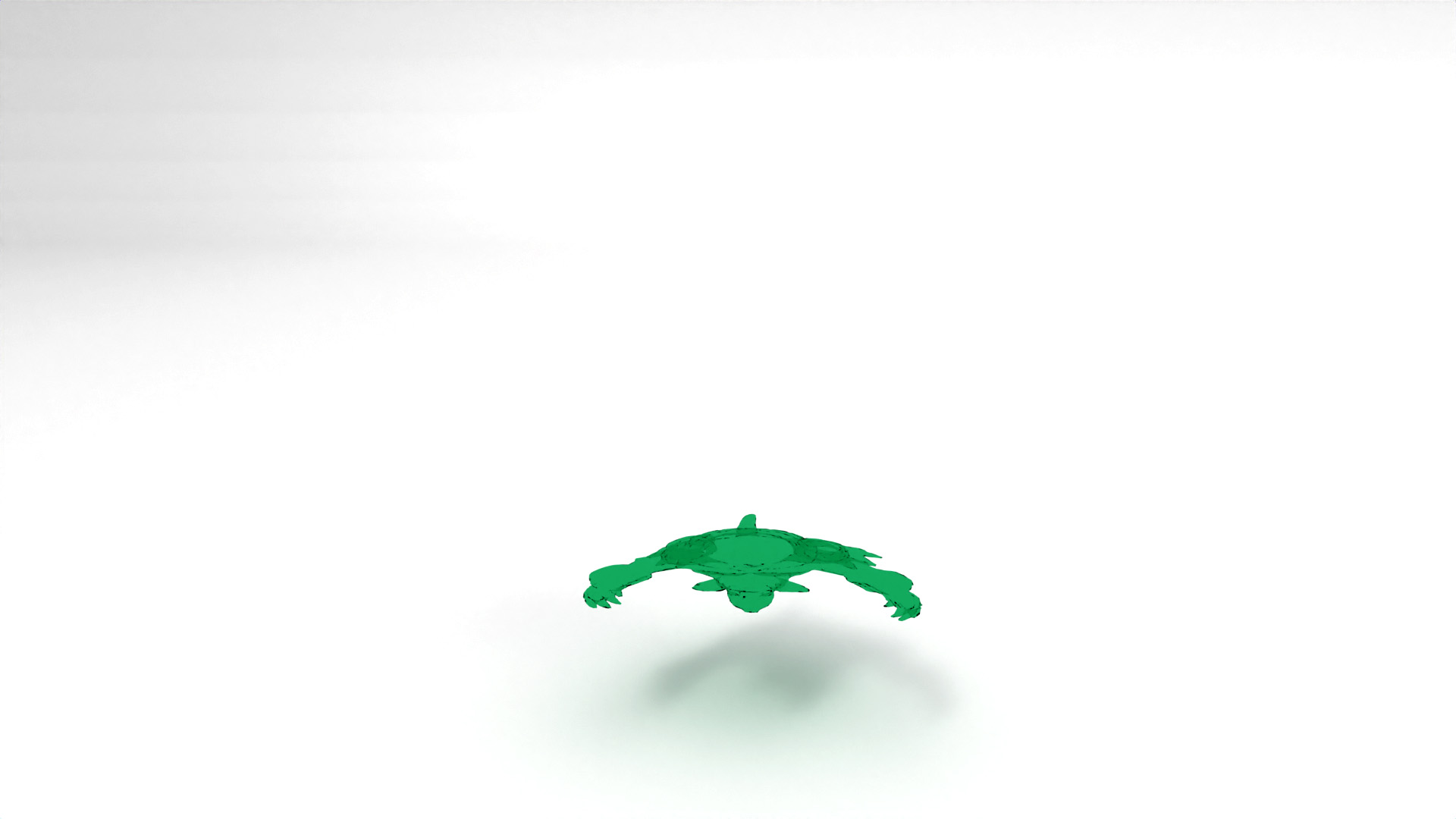}\hfill%
\fig{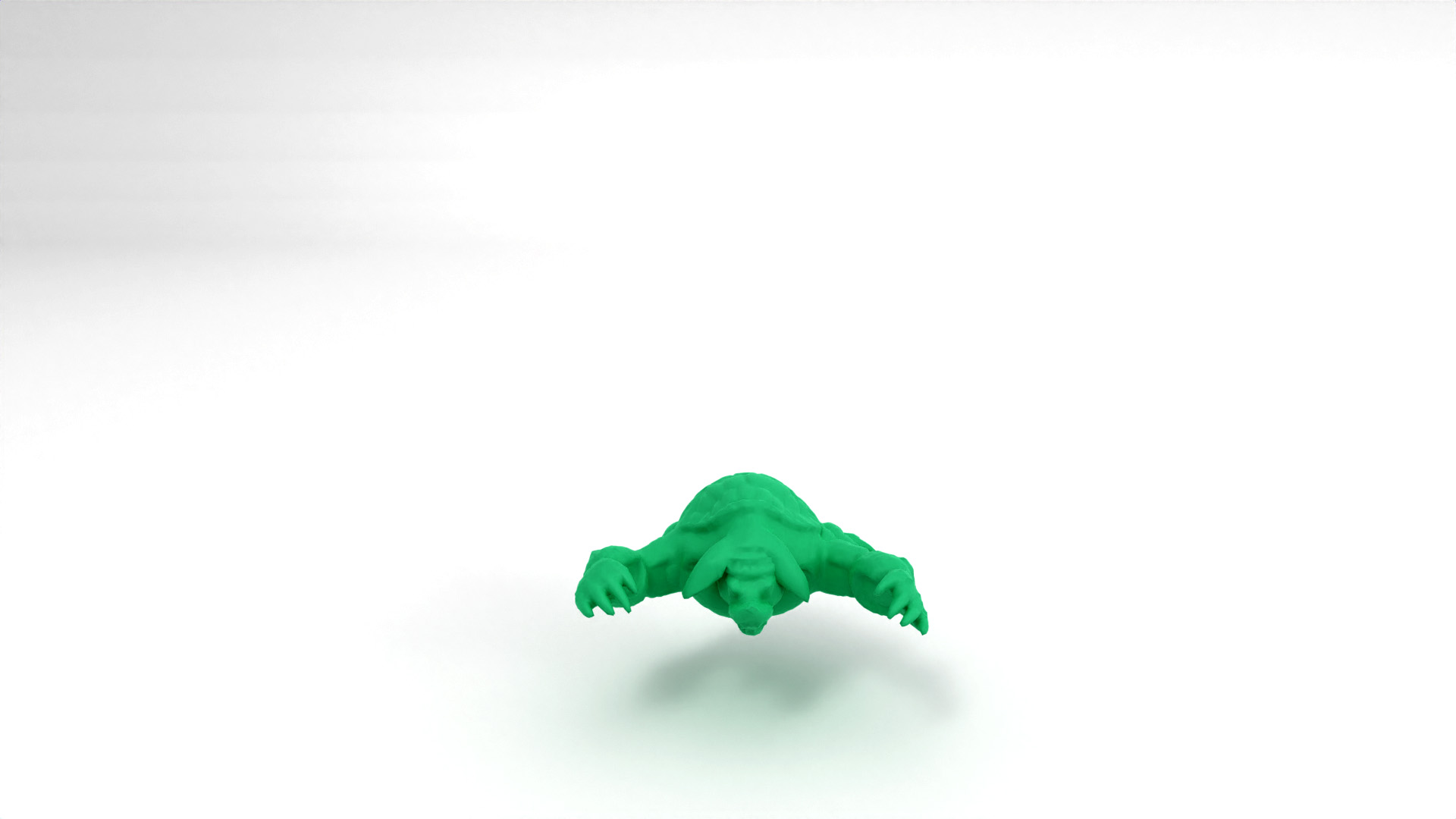}\hfill%
\fig{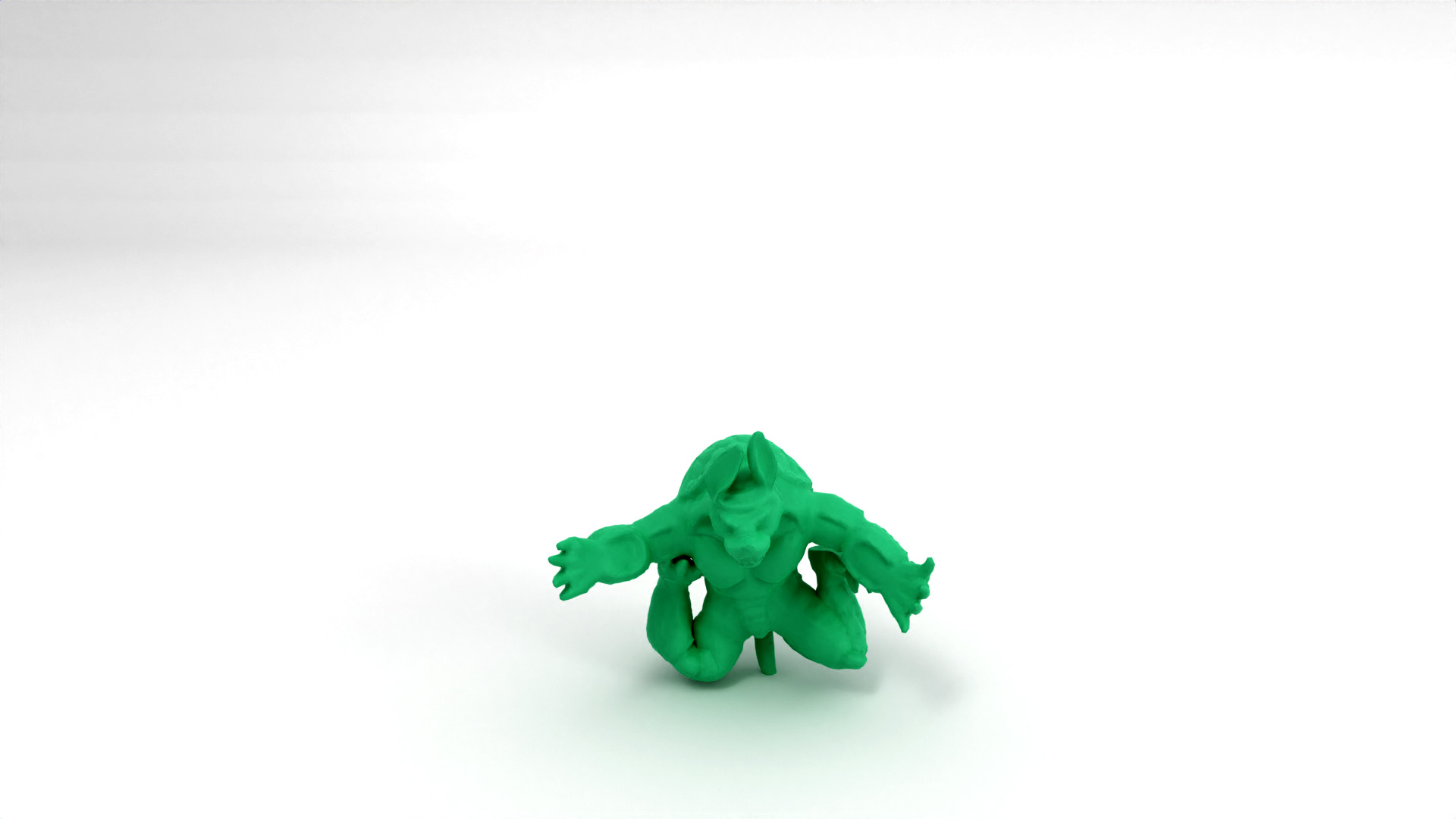}\hfill%
\fig[700 300 640 200]{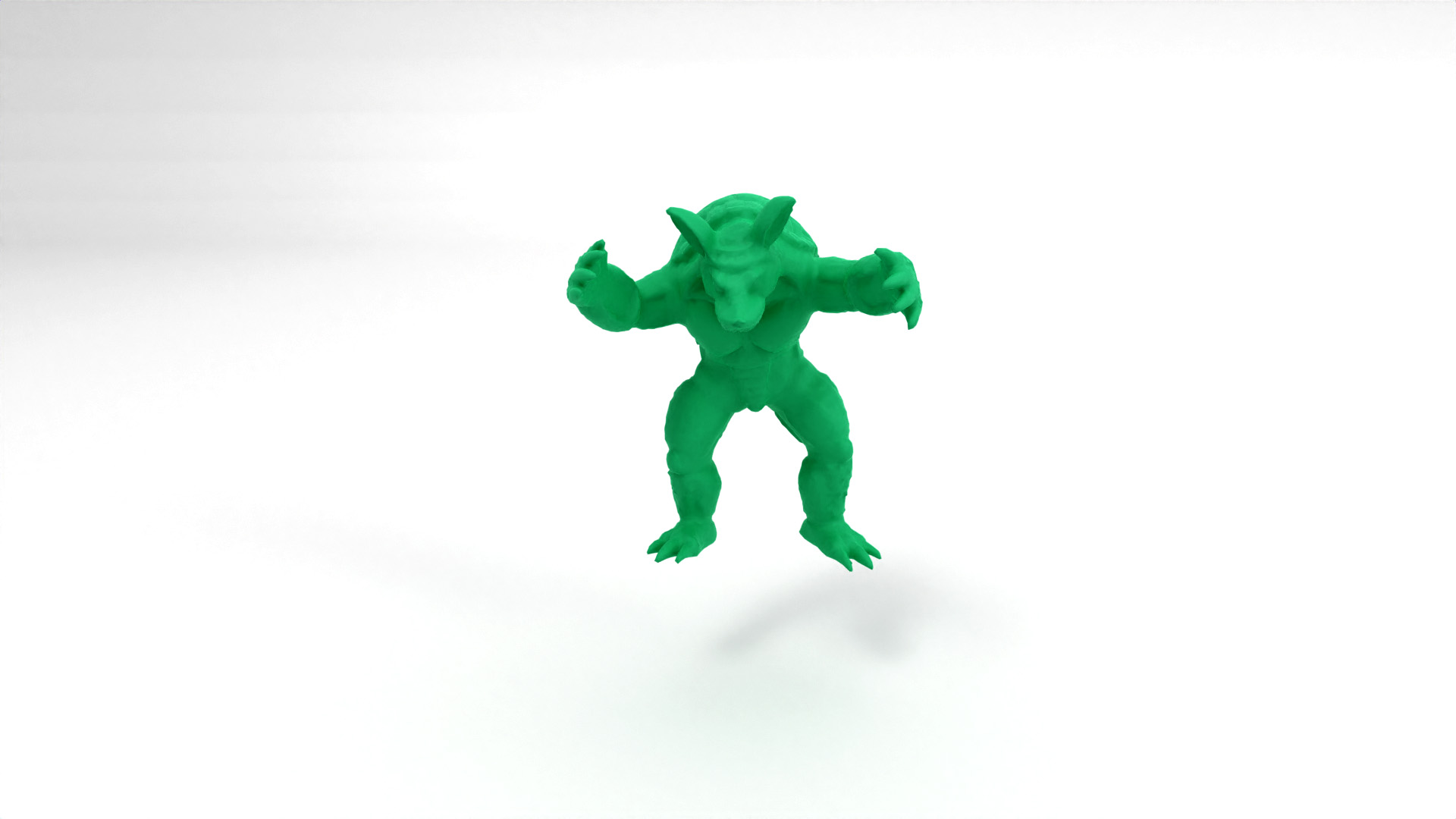}
\newcommand{\figT}[1]{\fig[400 200 400 80]{#1}}
\figT{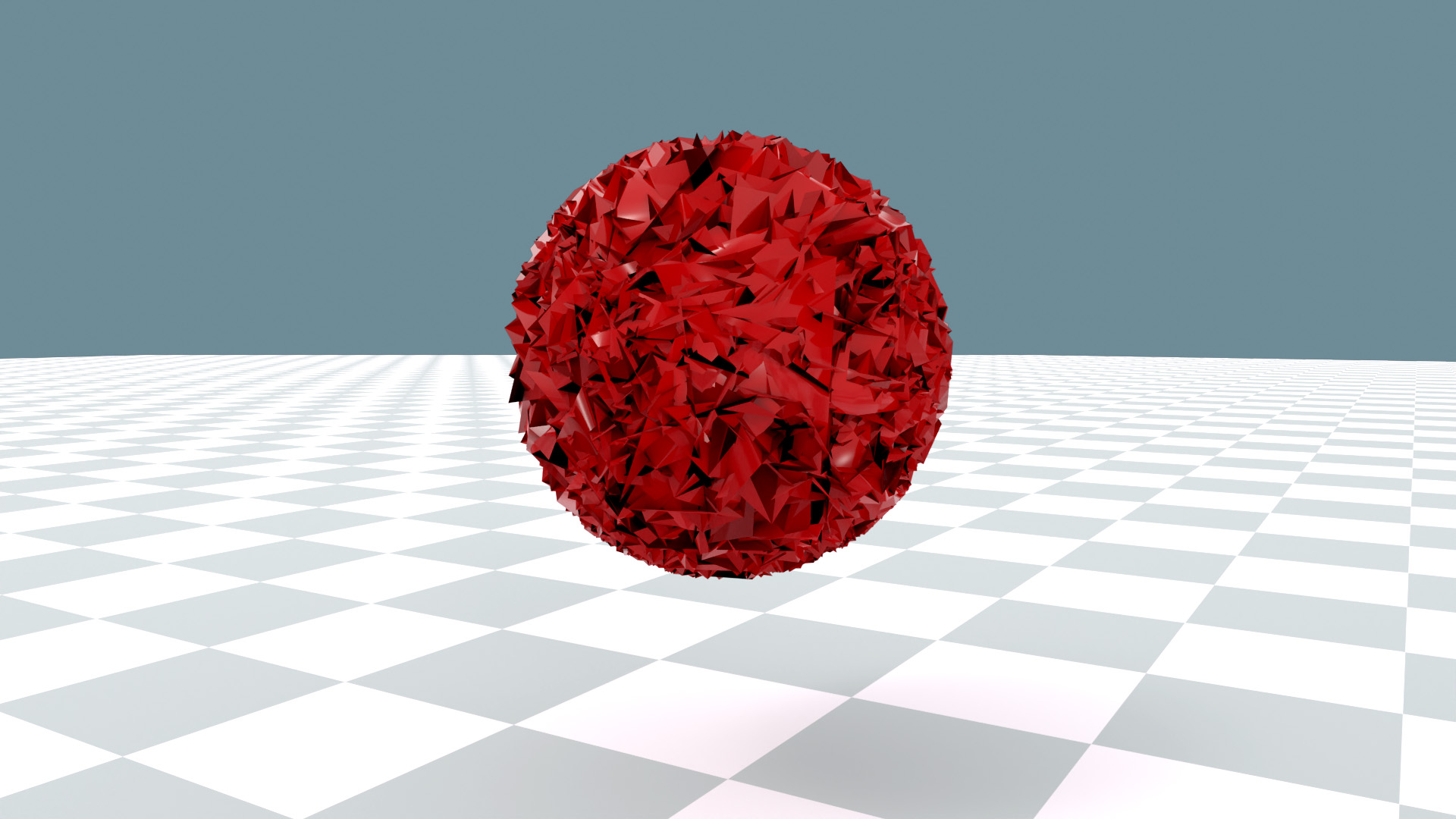}\hfill%
\figT{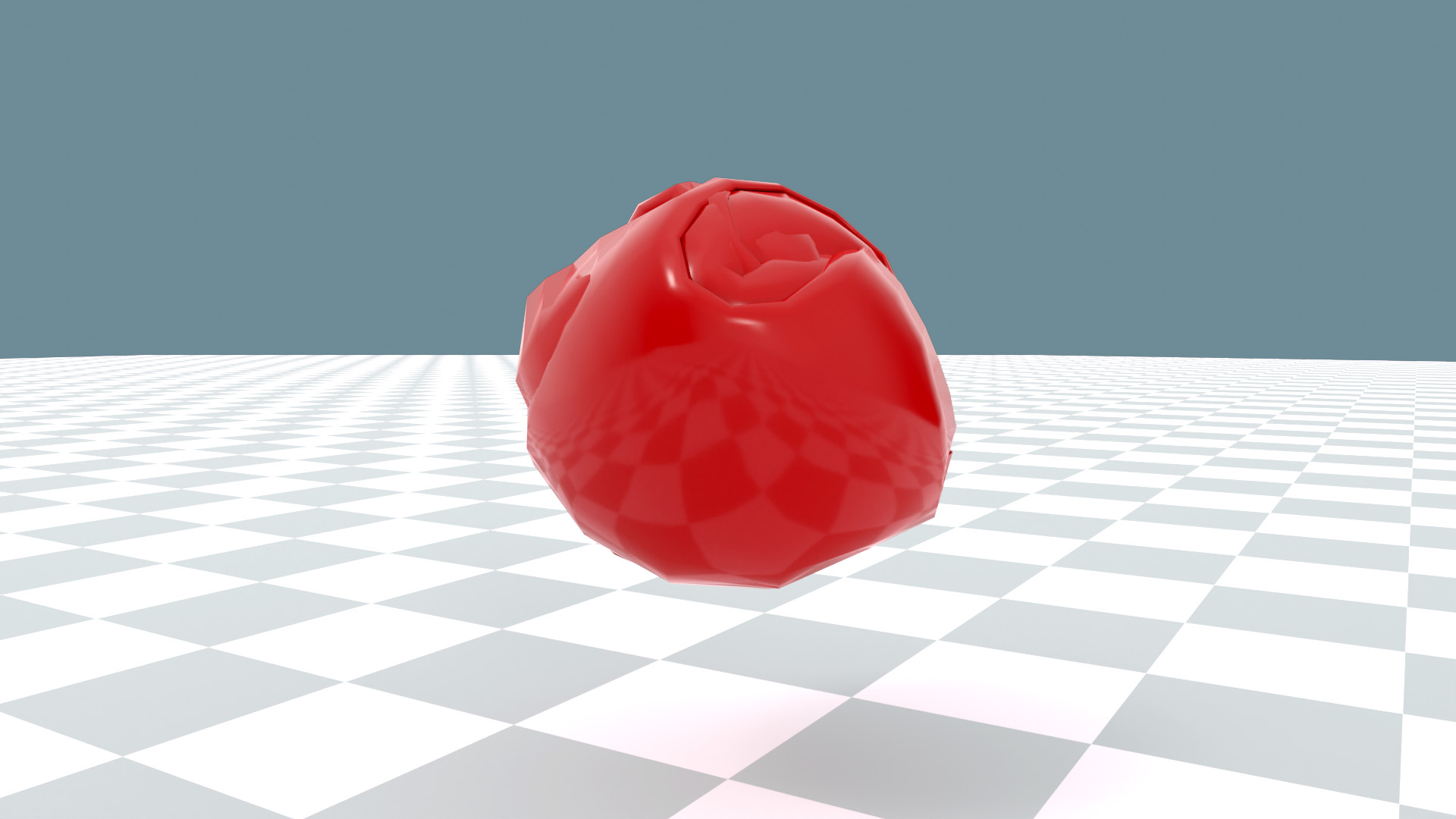}\hfill%
\figT{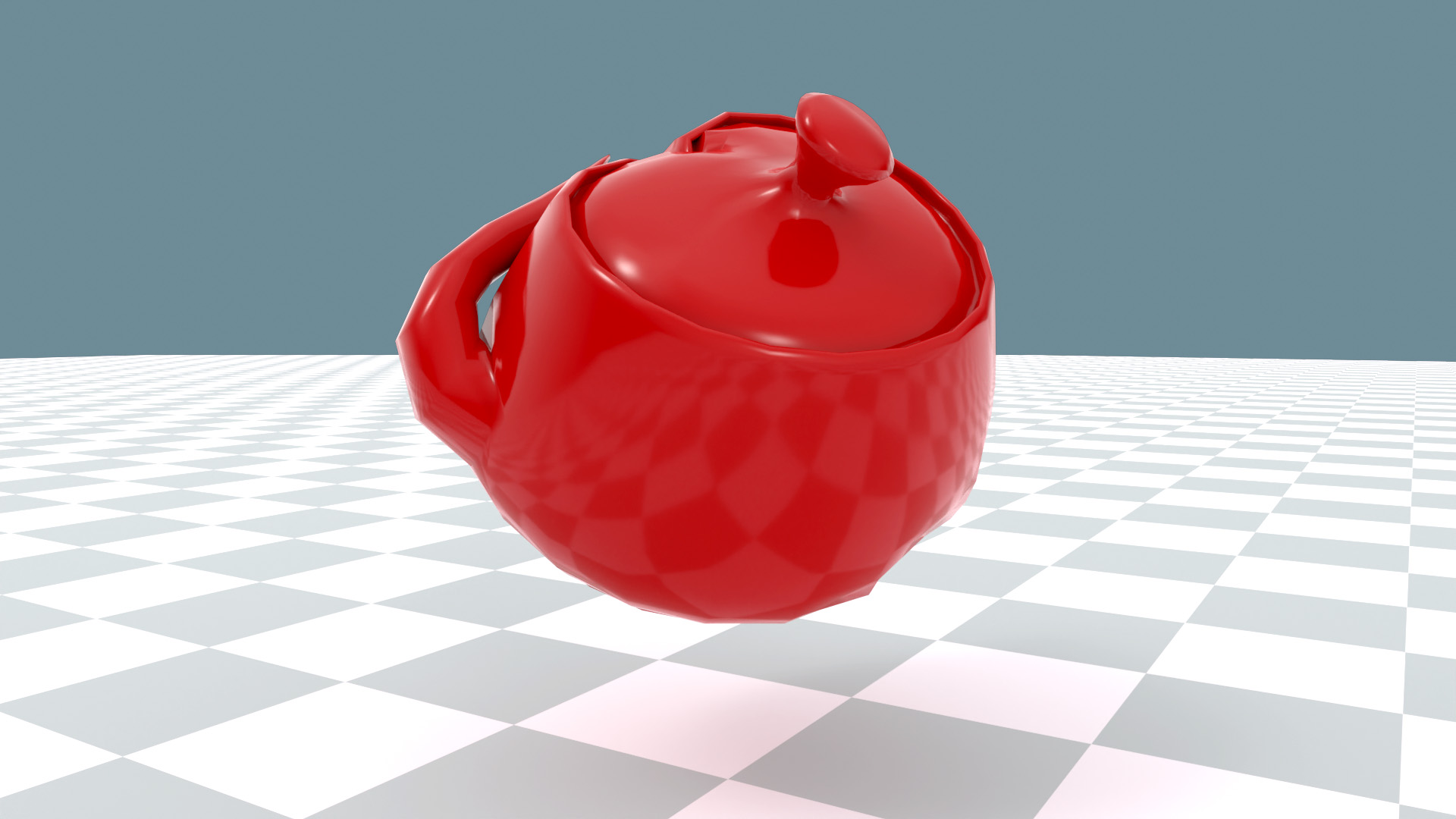}\hfill%
\figT{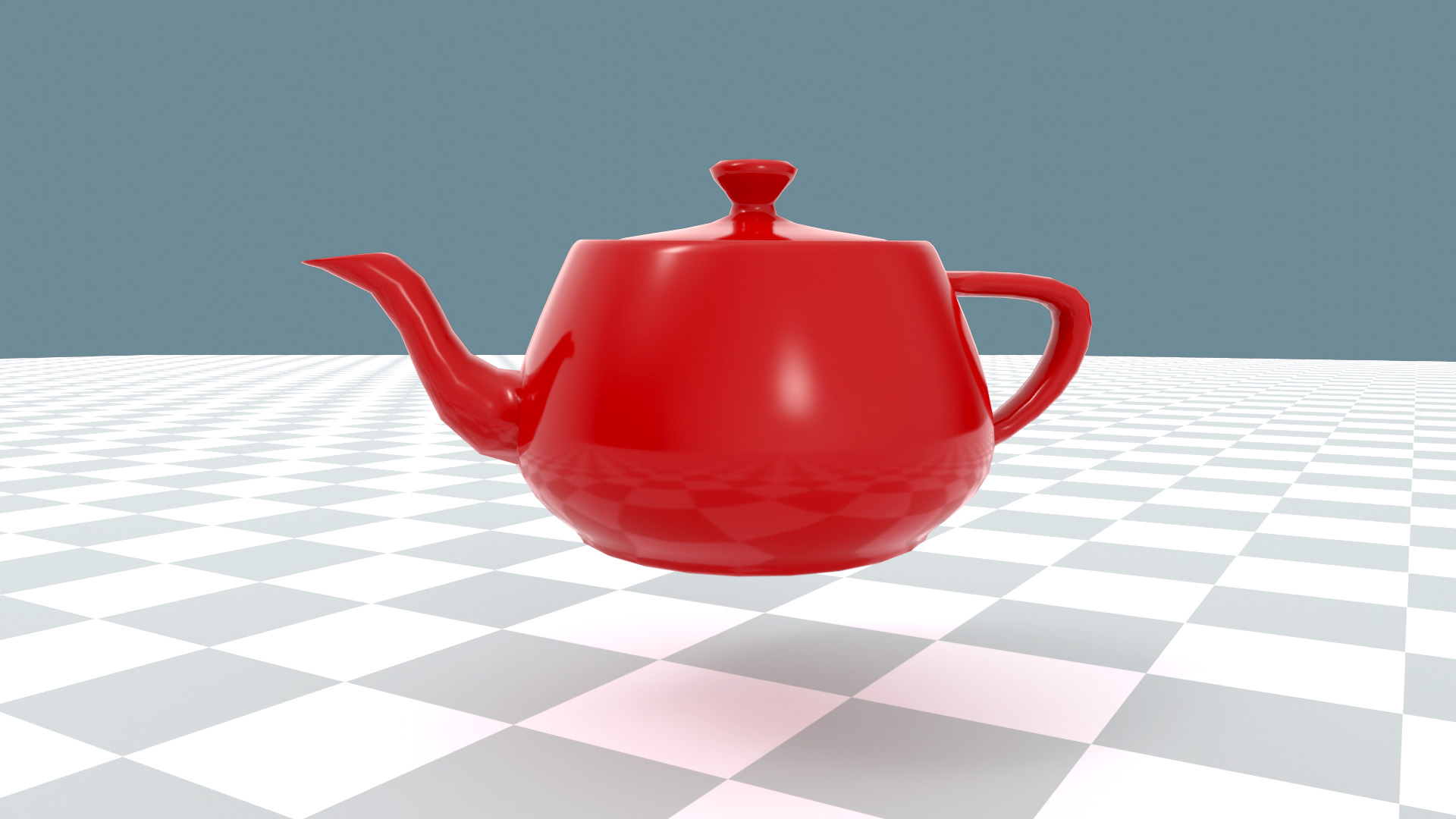}
\caption{Stress tests that begin simulations under extreme deformations: \sref{top}~a perfectly flattened armadillo model with 15 thousand vertices and 50 thousand tetrahedra, and \sref{bottom}~a Utah teapot model with 2 thousand vertices and 8.5 thousand tetrahedra, deformed by randomly placing its vertices onto the surface of a sphere. Both models quickly recover to their original shape shortly after the simulation starts. Both simulations use accelerated iterations with $\rho=0.95$.} 
\label{fig:ExtremeInitialization}
\vspace{1em}
\end{figure}

Position-based dynamics methods (PBD \cite{muller2007position} and XPBD \cite{macklin_xpbd_2016}) convert forces into (soft) constraints and directly update the positions with Gauss-Seidel iterations operating on one constraint at a time. These position-based updates result in exceptional stability, which is often exploited by limiting the number of iterations to fix the computation cost, an effective strategy for real-time simulations. Akin to PBD, our method also works with position updates, but operates directly using the force formulations without converting them to constraints. Parallelization with XPBD is achieved by graph coloring the constraints (i.e. the dual graph) \cite{fratarcangeli_scalable_2015, fratarcangeli_vivace_2016, ton-that_parallel_2023}. However, this dual graph contains multiple times more connections (depending on the types of constraints) than the original graph of vertices, severely limiting the level of parallelism. In comparison, our method is parallelized by coloring the original graph, which leads to much fewer colors (i.e. computation groups that must be processed sequentially) and thereby better parallelism.
More importantly, the approximations in XPBD's formulation introduce errors that make it diverge from the solution of implicit Euler and can degrade realism, particularly with large time steps and limited iteration count, which are common in practice. In addition, XPBD particularly struggles with high mass ratios. Our method, on the other hand, has none of these problems.

In recent years, a growing effort has been placed on accelerating simulations using GPUs  \cite{wang_chebyshev_2015, huthwaite2014accelerated, wang_descent_2016, lan2022penetration, macklin_primaldual_2020, li2020p, li2023subspace}. Among them, the first-order descent methods \cite{wang_descent_2016, macklin_primaldual_2020} have gained popularity due to their excellent parallelism. These methods employ a Jacobi-style preconditioned first-order descent on the backward Euler minimization formulation, enabling full vertex-level parallelization.
However, Jacobi-style iterations typically converge substantially slower than Gauss-Seidel iterations.
Also, such methods necessitate a line search to avoid overshooting and ensure stability.

Our method can be categorized as a coordinate descent method for optimization \cite{wright2015coordinate}. In graphics, this technique has been used for geometric processing \cite{naitsat2020adaptive} and simulation with a barrier function \cite{lan_second-order_2023}.  
Recently, \citet{lan_second-order_2023} proposed a hybrid scheme where Gauss-Sediel and Jacobi iterations are combined at each parallel call.
In comparison, our method uses blocks of coordinates based on vertices instead of blocks based on elements, which results in much better parallelism and smaller local linear systems to solve, leading to faster convergence.

%% file: 03_VertexBlockDescent.tex
\section{Vertex Block Descent for Elastic Bodies}
\label{sec:vbd}

Our vertex block descent method is essentially a solver for optimization problems. Therefore, it can be applied to various simulation problems, particularly if they can formulated as an optimization problem.
In this section, we explain our method in the context of elastic body dynamics for objects represented by a set of vertices that carry mass and a set of force elements and constraints that act on them. Generalization of our method to other example simulation problems is discussed later in \autoref{sec:othersims}.


We begin with deriving our global optimization method that splits the simulated system into vertex-level local systems (\autoref{sec:global_opt}). Then, we discuss the methods we use for solving the local systems (\autoref{sec:local_solver}) and describe how we incorporate damping (\autoref{sec:damping}) and constraints (\autoref{sec:constraints}). We present our collision formulation (\autoref{sec:collisions}) and friction formulation (\autoref{sec:friction}). 
After explaining our methods for warm-starting our optimization to improve convergence 
(\autoref{Sec:Initialization}), we describe how to incorporate momentum-based acceleration to improve convergence (\autoref{sec:acceleration}). Finally, we discuss the improved parallelization that our method provides
along with
methods for efficiently parallelizing dynamically-introduced force elements due to varying collision events (\autoref{sec:parallelism}).

\subsection{Global Optimization}
\label{sec:global_opt}

Our vertex block descent method is formulated based on the optimization form of implicit Euler time integration \cite{martin2011example,gast_optimization_2015}.
Given a system with $N$ vertices,
we represent the state of our simulation at step $t$ as ($\Vec{x}^t, \Vec{v}^t)$, where $\Vec{x}^t \in  \mathbb{R}^{3N}$ and $\Vec{v}^t\in \mathbb{R}^{3N}$ are the concatenated position and velocity vectors, respectively.
The resulting optimization problem can be written as
\begin{align}
\Vec{x}^{t+1} &= \underset{\Vec{x}}{\operatorname{argmin}} \; G(\Vec{x}) \;,
\end{align}
for evaluating the positions at the end of the time step $\Vec{x}^{t+1}$ that minimizes the \emph{variational energy} $G(\Vec{x})$ in the form of
\begin{align}
G(\Vec{x}) &= \frac{1}{2h^2} \lVert \Vec{x}-\Vec{y} \rVert _M^2 + E(\Vec{x}) \;.
\label{Eq:VariantionalFormOfImplicitEuler}
\end{align}
The first term here is the \emph{inertia potential}, which contains the time step size $h$,
the mass-weighted norm $\lVert \cdot \rVert_M$, and
%
\begin{align}
\Vec{y} &= \Vec{x}^t + h \Vec{v}^t + h^2 \Vec{a}_{\text{ext}}
\label{Eq:inertia}
\end{align}
with a fixed external acceleration $\Vec{a}_{\text{ext}}$, such as gravity. The second term $E(\Vec{x})$ is the \emph{total potential energy} evaluated at $\Vec{x}$.

We propose an optimization technique that falls under the category of coordinate descent methods to efficiently minimize this energy $G$.
If we only modify a single vertex at a time, fixing all other vertices, 
the part of the energy term $E(\vec{x})$ that is affected only includes the set of force elements $\mathcal{F}_i$ that are acting on (or using the position of) vertex $i$. Thus, we define the \emph{local variational energy} $G_i$ around vertex $i$ as
\begin{align}
    G_i(\Vec{x}) &= \frac{m_i}{2h^2} \lVert \Vec{x}_i-\Vec{y}_i \rVert ^2 + \sum_{j\in \mathcal{F}_i} E_j(\Vec{x}) \;,
     \label{Eq:VBDEnergy}
\end{align}
where $m_i$ is the mass of the vertex and $E_j$ is the energy of force element $j$ in $\mathcal{F}_i$.

Note that $G$ is not equal to the sum of local variational energies, i.e. $G(\Vec{x})\neq\sum_i G_i(\Vec{x})$, simply because the force elements appear multiple times in this sum (once for each of its vertices). However, when we modify the position of a single vertex only, the reduction in $G_i$ is equal to the resulting reduction in $G$.

Based on this observation, our method operates on a single vertex at a time and updates its position by minimizing the local energy
\begin{align}
\Vec{x}_i &\gets  \underset{\Vec{x}_i}{\operatorname{argmin}} \; G_i(\Vec{x})
\label{eq:position_update}
\end{align}
and solves the global system using Gauss-Seidel iterations. Each local minimization for a vertex effectively finds a descent step for $G$ using the degrees of freedom (DoF) for the vertex as a block of coordinates, hence the name \emph{vertex block descent} (VBD).
After each iteration, the total reduction in $G$ is equal to the sum of all reductions in $G_i$.
In other words, the energy change of each vertex position adjustment is accumulated to the system energy.
Consequently, if we can make sure that each local energy drops $G_i$ when we are adjusting vertex $i$, we can guarantee that we are descending the system energy $G$.

Thus, our system directly operates on vertex positions 
and the resulting velocities are calculated following the implicit Euler formulation
\begin{align}
    \Vec{v}^{t+1} = \frac{1}{h}\big(\Vec{x}^{t+1}-\Vec{x}^t\big) \;.
    \label{eq:velocity_update}
\end{align}

%

\subsection{Local System Solver}
\label{sec:local_solver}

The position updates per vertex (in \autoref{eq:position_update}) involve solving a local system that only depends on the position change of a single vertex, represented by $G_i$.
Note that \autoref{Eq:VBDEnergy} only has 3 DoF, 
so the cost of evaluating and inverting its Hessian is much cheaper compared to the global problem in \autoref{Eq:VariantionalFormOfImplicitEuler}. Therefore, we can fully utilize the second-order information and use Newton's method to minimize the localized energy $G_i$, which involves solving the 3D linear system
\begin{align}
   \Vec{H}_i \, \Delta\Vec{x}_i &= \vec{f}_i \;,
   \label{Eq:VBDLocalSystem}
\end{align}
where $\Delta\vec{x}_i$ is the change in position,
$\vec{f}_i$ is the total force acting on the vertex, calculated using
\begin{align}
    \vec{f}_i &= - \frac{\partial G_i(\Vec{x})}{\partial \Vec{x}_i} 
    = - \frac{m_i}{h^2} \left(\vec{x}_i-\vec{y}_i\right) \;- \sum_{j\in\mathcal{F}_i} \frac{\partial E_j(\Vec{x})}{ \partial \Vec{x}_i} \;,
    \label{Eq:VBDForce}
\end{align}
and $\Vec{H}_i \in \mathbb{R}^{3 \times 3}$ is the Hessian of $G_i$ with respect to the DoF of vertex $i$, such that
\begin{align}
\Vec{H}_i &= \frac{\partial^2 G_i(\Vec{x})}{\partial \Vec{x}_i \partial \Vec{x}_i} 
    = \frac{m_i}{h^2} \vec{I} \;+ \sum_{j\in\mathcal{F}_i} \frac{\partial^2 E_j(\Vec{x})}{\partial \Vec{x}_i \partial \Vec{x}_i} \;.
    \label{Eq:VBDHessian}
\end{align}
%
Here, the first term is
a diagonal matrix and the second one is the sum of Hessians of the
force elements with respect to
vertex $i$. Intuitively, the solution of this linear system is 
the extreme point of the quadratic approximation for the localized energy $G_i$.
By reducing $G_i$ with each iteration, we can guarantee a reduction in $G$, thereby iteratively solving the global system in \autoref{Eq:VariantionalFormOfImplicitEuler}.



We can analytically solve this system using $\Delta\vec{x}_i=\vec{H}_i^{-1}\vec{f}_i$.
For such a small system, the analytical solver is very efficient and stable. We found it to be faster than solvers based on Conjugate Gradient or LU/QR decomposition. 
Another advantage of the analytical solver is that it does not require the Hessian to be positive-definite. Of course, when the Hessian is not positive-definite, the direction given by \autoref{Eq:VBDLocalSystem} may not be a descent direction. Nevertheless, we opt for this direction regardless, recognizing that even when $\Vec{H}_i$ is not positive-definite, solving \autoref{Eq:VBDLocalSystem} still brings us towards the extremum of the quadratic approximation for the localized energy $G_i$.  This solution is close to where the gradient of the inertia and the potential terms balance out and it is usually a stable state of the system. Also, the motivation of the variational form of implicit Euler (\autoref{Eq:VariantionalFormOfImplicitEuler}), is to find a point where ${d G(\Vec{x})/d \Vec{x}=0}$. Therefore, any extreme point is a valid solution of implicit Euler, and it does not have to be a local minimum.
In all our experiments, including those specifically designed stress tests (see \autoref{fig:ExtremeTwist} and \autoref{fig:ExtremeInitialization}), we have consistently observed that this scheme does not pose any issues concerning system stability or convergence.

An alternative solution to this is the PSD Hessian projection \cite{teran2005robust}. However, it is exceptionally rare for the Hessian to not be positive-definite, and the PSD projection process is notably costly due to multiple SVD decompositions. Engaging in this costly operation to prevent such rare events seems unjustified, especially considering that these occurrences do not jeopardize system stability or convergence significantly.

Another challenge with the analytical solver arises when encountering a (nearly) \del{low-rank}\add{rank-deficient} Hessian. 
To address this, we propose a simple solution: if $|\det(\Vec{H}_i)| \leq \epsilon$ for some small threshold $\epsilon$, we opt to bypass adjusting this particular vertex for that iteration. Given that its neighboring vertices are likely to undergo adjustments before the next iteration, it is improbable that its Hessian will remain \del{low-rank}\add{rank-deficient} in subsequent iterations.
With this simple solution, in the extreme scenario where all vertices possess a \del{low-rank}\add{rank-deficient} Hessian, the system could potentially become frozen. Yet, it is crucial to note that such a case is highly improbable, since the Hessian of the inertia potential is always full-rank.
One potential remedy for this would be switching to the modified Conjugate Gradient solver \cite{lan_second-order_2023} when such a case happens. However, doing this will add additional branching to the code and can slow down the solver. Thus, we have not included it in our implementation, but, depending on specific use cases, there is always the flexibility to opt for the Conjugate Gradient solver as a backup solution.


The linear system in \autoref{Eq:VBDLocalSystem} corresponds to a single Newton step, so it does not necessarily provide the optimal solution for \autoref{eq:position_update}. In fact, since it is just a second-order approximation of $G_i$, it does not even guarantee a reduction in $G_i$.
To ensure the descent of energy with this single step,
we can incorporate a backtracking line search along the descent direction.
Note that, unlike global line searches in descent-based simulation methods (e.g. \citet{wang_descent_2016}), this line search operates locally. It specifically verifies the descent of $G_i$,
which in turn guarantees a descent in $G$ without the need for evaluating the global system.

Line search avoids over-shooting and, thereby, ensures stability. 
In practice, 
however, the additional computation cost of line search may not be justified. In our experiments,
we found that line search costs an extra 40\% computation time, while not providing any measurable benefits.
This is because VBD can maintain 
stability even without line search. 
Therefore, the results we present in this paper do not include line search, though it is an option available.

\subsection{Damping}
\label{sec:damping}

Damping plays a crucial role in simulations. It prevents excessive oscillations and also enhances system stability. Despite the inherent numerical damping introduced by the implicit Euler method, providing users with manual control over damping is highly desirable. To address this, we have integrated a simplified Rayleigh damping model into our solver~\cite{sifakis_fem_2012}. This process is both straightforward and efficient, as it also operates locally within the $3 \times 3$ system and utilizes the precomputed stiffness matrix.

Since we are relying on implicit Euler, we can represent the velocity as position change, using $\Vec{v}_i=(\Vec{x}_i-\Vec{x}_i^t)/h$. 
Then, we can add the damping term to the Hessian in \autoref{Eq:VBDHessian}, resulting
\begin{align}
\Vec{H}_i &= \frac{m_i}{h^2} \vec{I} \;+ \sum_{j\in\mathcal{F}_i} \frac{\partial^2 E_j(\Vec{x})}{\partial \Vec{x}_i \partial \Vec{x}_i} 
    + \left(\sum_{j\in\mathcal{F}_i} \frac{k_d}{h} \frac{\partial^2 E_j(\Vec{x})}{\partial \Vec{x}_i \partial \Vec{x}_i}\right) \;,
    \label{Eq:VBDHessianWithDamping}
\end{align}
where $k_d$ is the damping coefficient. Finally, we add the damping 
force to \autoref{Eq:VBDForce} using the same damping term, such that
\begin{align}
    \vec{f}_i &= - \frac{m_i}{h^2} \left(\vec{x}_i-\vec{y}_i\right) -\!\! \sum_{j\in\mathcal{F}_i} \frac{\partial E_j(\Vec{x})}{ \partial \Vec{x}_i}
    - \left(\sum_{j\in\mathcal{F}_i} \frac{k_d}{h} \frac{\partial^2 E_j(\Vec{x})}{\partial \Vec{x}_i \partial \Vec{x}_i}\right)
    \left(\Vec{x}_i-\Vec{x}_i^t\right)
    .
    \label{Eq:VBDForceWithDamping}
\end{align}

\subsection{Constraints}
\label{sec:constraints}

Since our method directly manipulates the position of each vertex, managing a constraint on a vertex becomes straightforward. Constraints generally fall into two categories: unilateral ($C(\vec{x}) \leq 0$) or bilateral ($C(\vec{x}) = 0$). With \del{biliteral}\add{bilateral} constraints, if a vertex position is directly set to a specific value, we simply skip updating its position. Otherwise, it is constrained to a (usually linear) subspace. Our approach involves representing the constrained vertex position using the subspace basis. This transforms both the vertex position and gradient into an $L$-dimensional vector, where $L$ is the subspace dimension. Consequently, handling local steps for constrained vertices involves solving an $L\times L$ system.
Regarding unilateral constraints, we allow compromises and define potential energy to be solved alongside other potentials. This method handles world box constraints in our simulations.



\begin{figure}\vspace{-0.8em}
\centering
\newcommand{\img}[2]{\begin{subfigure}{0.35\linewidth}\centering
\includegraphics[width=\linewidth]{Figures/CollisionPotential/CollisionPotential#1}\vspace{-1.2em}
\caption{#2}\end{subfigure}}
\img{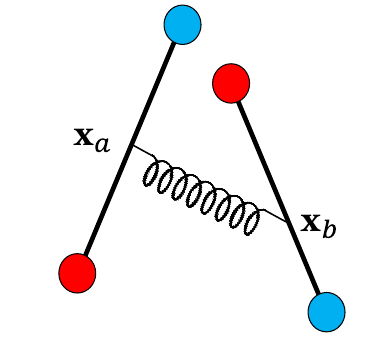}{Edge-edge}\hspace{0.1\linewidth}%
\img{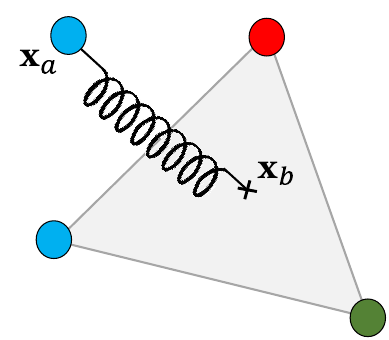}{Vertex-triangle}%
\caption{Two collisions types: \sref{a}~edge-edge can have at most two pairs and \sref{b}~vertex-face can have at most one pair with the same color, since vertices on the same side of a collision must have different colors.}
\label{fig:CollisionPotential}
\end{figure}

\subsection{Collisions}
\label{sec:collisions}

Collisions can be handled by simply introducing a quadratic collision energy per vertex, based on the penetration depth $d$, such that
\begin{align}
    E_c(\Vec{x}) &= \frac{1}{2} \, k_c \, d^2
    &&\text{with}&
    d &= \max\Big( 0, (\Vec{x}_b - \Vec{x}_a) \cdot \Vec{\hat{n}}\Big) \;,
    \label{eq:collisionEnergy}
\end{align}
where $k_c$ is the collision stiffness parameter, $\vec{x}_a$ and $\vec{x}_b$ are the two \emph{contact points} on either side of the collision, and $\vec{\hat{n}}$ is the contact normal.

There are two collision types for triangle meshes (\autoref{fig:CollisionPotential}):
\begin{itemize}
    \item Edge-edge collisions use continuous collision detection (CCD). $\vec{x}_a$ and $\vec{x}_b$ correspond to the intersection points on either edge and the contact normal is the direction between them, i.e. ${\vec{\hat{n}}=\vec{n}/\|\vec{n}\|}$, where ${\vec{n}=\vec{x}_b-\vec{x}_a}$.
    \item Vertex-triangle collisions are detected either by CCD or discrete collision detection (DCD). In this case, $\vec{x}_a$ is the colliding vertex and $\vec{x}_b$ is the corresponding point on either the collision point (for CCD) or the closest point (for DCD) on the triangle \cite{10.1145/3592136}. $\vec{\hat{n}}$ is the surface normal at $\vec{x}_b$.
\end{itemize}


In our implementation, we perform a DCD at the beginning of the time step using $\vec{x}^t$ to identify vertices that have already penetrated, and the rest of the collisions use CCD.
%
We simplify the computation of the gradient and the Hessian of the collision energy by not differentiating through $\vec{\hat{n}}$, i.e. assuming that $\vec{\hat{n}}$ is constant. 

Performing collision detection at every iteration using CCD can be expensive and can easily become the bottleneck. Therefore, in our implementation, we perform CCD once every $n_{\text{col}}$ iterations.
While this has the risk of missing some collision events, they are likely to be captured via DCD in the next time step.
All detected collisions remain as force elements until the next collision detection.
For vertex-triangle collisions detected with DCD, it is important to recompute the closest point ($\vec{x}_b$) before computing the gradient and Hessian of $E_c$.



\subsection{Friction}
\label{sec:friction}



To compute friction for collision $c$, we must consider the relative motion of the contact points defined as
\begin{equation}
    \delta\vec{x}_c  = \big(\vec{x}_a - \vec{x}_a^t\big) - \big(\vec{x}_b - \vec{x}_b^t\big) \;,
\end{equation}
where $\vec{x}_a^t$ and $\vec{x}_b^t$ are the positions of $\vec{x}_a$ and $\vec{x}_b$ at the beginning of the time step.

With this $\delta\vec{x}_c$, we can use the friction model of \emph{incremental potential contact} (IPC)~\cite{li2020incremental}.
First, we project $\delta\vec{x}_c$ to the 2D contact tangential space,
using a transformation matrix $\vec{T}_c \in \mathbb{R}^{3 \times 2}$, to evaluate the tangential relative translation $\vec{u}_c=\Vec{T}_c^T\delta\Vec{x}_c$, where $\vec{T}_c=[\vec{\hat{t}}~\vec{\hat{b}}]$ is formed by a tangent $\vec{\hat{t}}$ and binormal $\vec{\hat{b}}$ vectors orthogonal to $\vec{\hat{n}}$.
The signed magnitude of the collision force applied on vertex $i$ is $\lambda_{c,i}=\frac{\partial E_c}{\partial \Vec{x}_i}\cdot \vec{\hat{n}}$. Note that the sign of $\lambda_{c,i}$ is different for vertices on different sides of the collision.
Let $\mu_c$ be the friction coefficient. We can then calculate the friction force using
\begin{align}
   \vec{f}_{c,i}
   &= -\mu_c\,\lambda_{c,i}\,\frac{\partial \delta \Vec{x}_c}{\partial \Vec{x}_i}\,\Vec{T}_c\,f_1(\|\Vec{u}_c\|)\frac{\Vec{u}_c}{\|\Vec{u}_c\|}\;\text{,  where} \\
   f_1(u) &=
    \begin{cases}
      2\left(\frac{u}{\epsilon_vh}\right) - 
       \left(\frac{u}{\epsilon_vh}\right)^2, &u \in (0,h\epsilon_v)\\
      1, &u \geq h\epsilon_v
    \end{cases}.
   \label{Eq:IPC_friction}
\end{align}
Here, $f_1$ serves as a smooth transition function between static and dynamic friction. 
When the relative velocity exceeds a small threshold $\epsilon_v$, dynamic friction is applied. Conversely, if the relative velocity is below this threshold, static friction is applied, scaling between the range of $[0,1]$. 

In our formulation, we need the Hessian of the friction term, which is the derivative of this function. We approximate the derivative by not differentiating through $\|\Vec{u}_c\|$ for a more stable friction force formulation \cite{macklin_primaldual_2020}, such that
\begin{align}
\frac{\partial \vec{f}_{c,i}}{\partial \vec{x}_i} &\approx -\mu_c\,\lambda_{c,i}\,\frac{\partial \delta\vec{x}_c}{\partial \vec{x}_i}\,\Vec{T}_c\,\frac{f_1(\|\Vec{u}_c\|)}{\|\Vec{u}_c\|} \Vec{T}_c^T\,  \left(\frac{\partial \delta\vec{x}_c}{\partial \vec{x}_i}\right)^T\;.
\end{align}
Without this approximation, PSD projection and line search might be needed to ensure stability.
Finally, all friction forces $\vec{f}_{c,i}$ and their derivatives $\partial \vec{f}_{c,i}/\partial\vec{x}_i$ are added to $\vec{f}_i$ and $\vec{H}_i$, respectively.

\subsection{Initialization}
\label{Sec:Initialization}


Since our method is an iterative solver, we begin with an \emph{initial guess} for $\Vec{x}$. The closer this initial guess is to the resulting $\Vec{x}^{t+1}$, the fewer number of iterations we would need to converge. 
\del{Regardless of the initialization strategy for determining this initial guess, because we are solving an optimization problem, the system is expected to converge to  result, though it may require a different number of iterations.}\add{Typically, if the simulation converges, different initializations should not significantly affect the final results, although they may influence the number of iterations required to achieve convergence. }


Providing a good initial guess (one that is close to $\Vec{x}^{t+1}$) is particularly important for applications with a limited computation budget, e.g. using a fixed number of iterations. In such applications, the initial guess can strongly impact the remaining residual at the end of the final iteration.

\begin{figure}
\centering
\includegraphics[width=\linewidth,trim=0 0 0 0, clip]{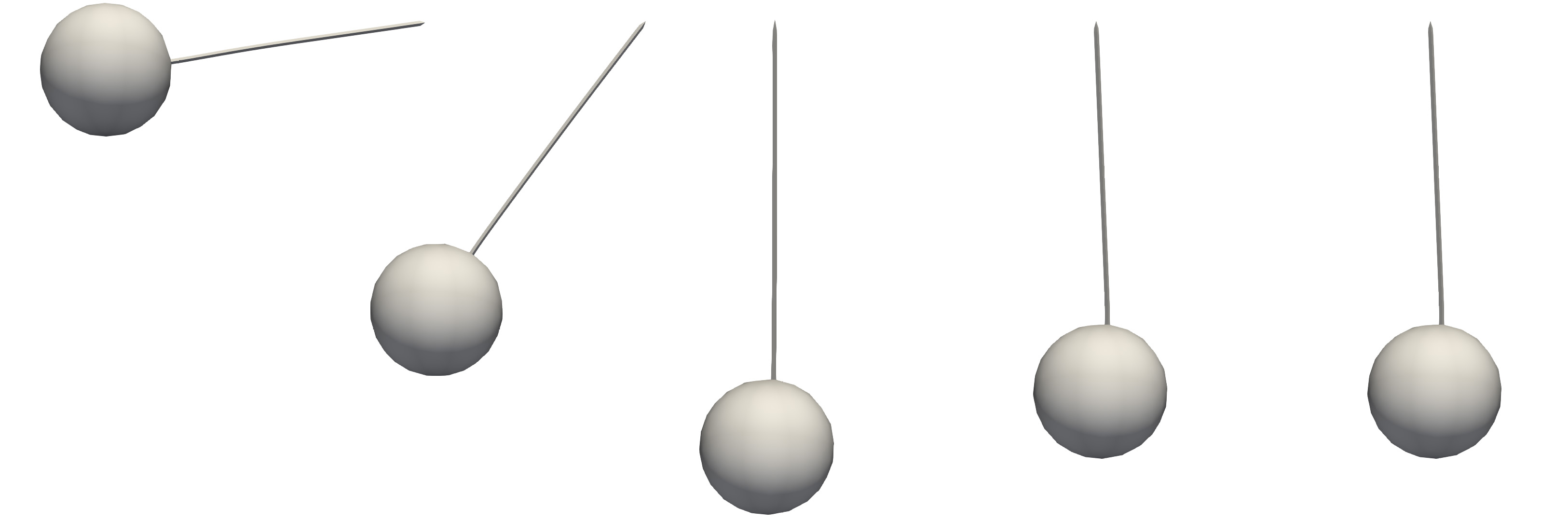}
\newcommand{\scap}[2]{\begin{subfigure}{0.2\linewidth}\caption{#2}\label{fig:Initialization:#1}\end{subfigure}}
\scap{a}{Previous\\\hspace*{1em}position}\hfill%
\scap{b}{Inertia\\~}\hfill%
\scap{c}{Inertia \& acceleration}\hfill%
\scap{d}{\textbf{Adaptive}\\~}\hfill%
\scap{e}{Reference\\~}
\caption{Different initialization options for a swinging elastic pendulum dropped from the same height (blue line) simulated with our method using only 20 iterations per frame, showing the same frame of the simulation. Notice that initializing using \sref{a}~previous position and \sref{b}~inertia fail to properly move under gravity, while \sref{c}~inertia and acceleration leads to accessive stretching (red line) when VBD does not run to convergence. \sref{d}~Our adaptive solution closely matches \sref{e}~the reference generated by fully converged Newton's method (green line).}
\label{fig:Initialization}
\end{figure}

We begin with considering three simple options for initialization: 
\begin{align*}
    \text{(a)~}&\text{\textbf{Previous position:}}&
    \Vec{x} &= \Vec{x}^t \\
    \text{(b)~}&\text{\textbf{Inertia:}}&
    \Vec{x} &= \Vec{x}^t + h\vec{v}^t \\
    \text{(c)~}&\text{\textbf{Inertia and acceleration:}}&
    \Vec{x} &= \vec{x}^t + h\vec{v}^t + h^2\vec{a}_{\text{ext}} = \Vec{y}
\end{align*}

Option (a) struggles with substantially stiff materials. As we adjust each vertex separately, assuming the others remain fixed, our method must lean on the inertia potential's gradient (i.e. ${m_i(\vec{x}_i-\vec{y}_i)/2h^2}$) to march toward the local minimum. With stiff materials, this method results in notably slower convergence rates due to the inertia potential being considerably less stiff. Consequently, it encounters challenges in simulating scenarios that resemble free fall, like a swinging elastic pendulum at its maximum height, as shown in \autoref{fig:Initialization:a}.

Option (b) allows the system to start with the inertia of the previous step, which helps, but terminating the iterations prior to convergence can again result in local material stiffness overpowering external acceleration, as shown in \autoref{fig:Initialization:b}.

Option (c) is similar to the initialization of position-based dynamics, and performs notably better as it effectively preserves inertia and properly reacts to external acceleration.
However, with a limited number of iterations, materials behave softer than they should, often resulting in overstretching or collapsing under gravity.
An example of this can be seen in \autoref{fig:Initialization:c}, where the pendulum extends more than it should.
Most notably, it struggles with steady objects at rest in contact, consistently initializing them into a penetrating state, as if they are in free fall. In such cases, the contact forces must entirely undo the position change of initialization during the iterations. This not only creates unnecessary computation, but also places considerable strain on the accuracy of the collision detection and handling methods (including friction). Therefore, properly simulating objects that are stacked on top of each other becomes a major challenge with this initialization option.


\begin{figure}
\centering
\newcommand{\spc}{~\vspace{0.2em}}
\includegraphics[width=\linewidth,trim=0 0 0 0, clip]{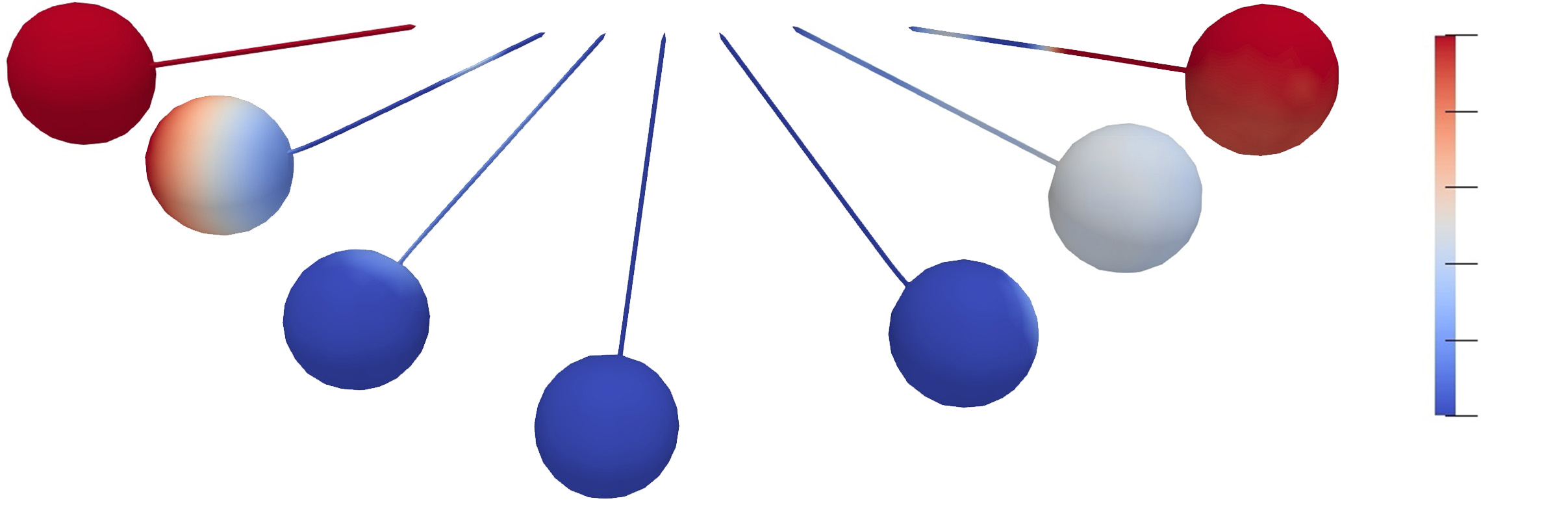}\vspace{-0.34\linewidth}\\
\begin{flushright}\small 
1.0\spc\\
0.8\spc\\
0.6\spc\\
0.4\spc\\
0.2\spc\\
0.0\spc\\
gravity initialization factor ($\tilde{a}$)
\end{flushright}
\caption{The ratio of gravity $\tilde{a}$ used with adaptive initialization during the swinging of an elastic pendulum. The model is a single-piece tetrahedral mesh. It automatically distinguishes vertices in approximate free-fall (red) and those where elasticity counteracts gravity (blue).}
\label{fig:AdaptiveInitialization}
\end{figure}

We propose an adaptive initialization scheme that combines options (b) and (c), taking advantage of the freedom that VBD provides in the choice of initialization.
This scheme uses
\begin{align*}
    \text{(d)~}&\text{\textbf{Adaptive:}}&
    \Vec{x} &= \vec{x}^t + h\vec{v}^t + h^2\vec{\tilde{a}}
    &\hspace{0.2\linewidth}
\end{align*}
replacing the external acceleration $\vec{a}_{\text{ext}}$ in (c) with an estimated acceleration term $\vec{\tilde{a}}$, determined by exploiting the typical similarity between two consecutive time steps.
We begin with the acceleration of the previous frame
${\Vec{a}^t=(\vec{v}^t-\vec{v}^{t-1})/h}$ and compute its component $a^t_{\text{ext}}$ along the external acceleration direction ${\vec{\hat{a}}_{\text{ext}}=\vec{a}_{\text{ext}}/\|\vec{a}_{\text{ext}}\|}$, such that ${a^t_{\text{ext}}=\vec{a}^t\cdot\vec{\hat{a}}_{\text{ext}}}$.
Then, we simply make sure that the estimated acceleration does
not exceed the external acceleration, using
\begin{align}
    \vec{\tilde{a}} &= \tilde{a}\,\vec{a}_{\text{ext}} \;, &&\text{where}&
    \tilde{a} &=
    \begin{cases}
      1\;, & \text{if $a^t_{\text{ext}}>\|\vec{a}_{\text{ext}}\|$}\\
      0\;, & \text{if $a^t_{\text{ext}}<0$}\\
      a^t_{\text{ext}}/\|\vec{a}_{\text{ext}}\|\;, & \text{otherwise.}
    \end{cases}       
\end{align}
This adaptive approach includes external acceleration in the initialization when the motion of a vertex resembles free fall. When an object is stationary, however, as in rest-in-contact, it maintains the previous position in the initialization, preventing undesired penetrations before the first iteration. It also successfully avoids excessive stretching, as can be seen in \autoref{fig:Initialization:d}. Visualizations of different $\tilde{a}$ values in this simulation are shown in \autoref{fig:AdaptiveInitialization}. In short, our adaptive initialization is a simple but effective strategy and it is possible because VBD does not dictate a particular initialization (unlike XPBD, for example).

\subsection{Accelerated Iterations}
\label{sec:acceleration}

We use the Chebyshev semi-iterative approach \cite{wang_chebyshev_2015} to improve the convergence of our method, though other momentum-based acceleration techniques, such as the Nesterov’s method \cite{golub2013matrix} can be applied as well.
The Chebyshev method
iteratively computes an acceleration ratio based on the approximation of the system's spectral radius. Instead of directly using the output vertex positions of Gauss-Seidel ${\vec{\bar{x}}}^{(n)}$ after iteration $n$, it recomputes the positions at the end of the iteration
using
\begin{align}
    \vec{x}^{(n)} &= \omega_{n} ({\vec{\bar{x}}}^{(n)} - \vec{x}^{(n-2)} ) + \vec{x}^{(n-2)} \;,
    \label{eq:accelerator}
\end{align}
where 
$\omega_{n}$ is the acceleration ratio that changes at each iteration as
\begin{align}
    \omega_{n}&=\frac{4}{4-\rho^2\omega_{n-1}}
    &\text{with}&&
    \omega_1 &= 1
    &\text{and}&&
    \omega_2 &= \frac{2}{2-\rho^2} \;,
\end{align}
where $\rho \in(0,1)$ is the estimated spectral radius, which can be set manually, or automatically tuned using the technique introduced in \cite{wang_descent_2016}.
Note that this position recomputation procedure is performed globally after each Gauss-Seidel iteration is completed, 
not after each local solver step.

\begin{figure}
\centering
\newcommand{\figA}[3]{\begin{subfigure}{0.315\linewidth}\centering
\includegraphics[width=\linewidth,trim=450 300 550 50,clip]{Figures/AccelerationComparison/#1}
\caption{#2}\label{fig:AccelerationComparison:#3}
\end{subfigure}}
\begin{subfigure}{0.36\linewidth}\centering
\includegraphics[width=\linewidth,trim=450 200 450 150,clip]{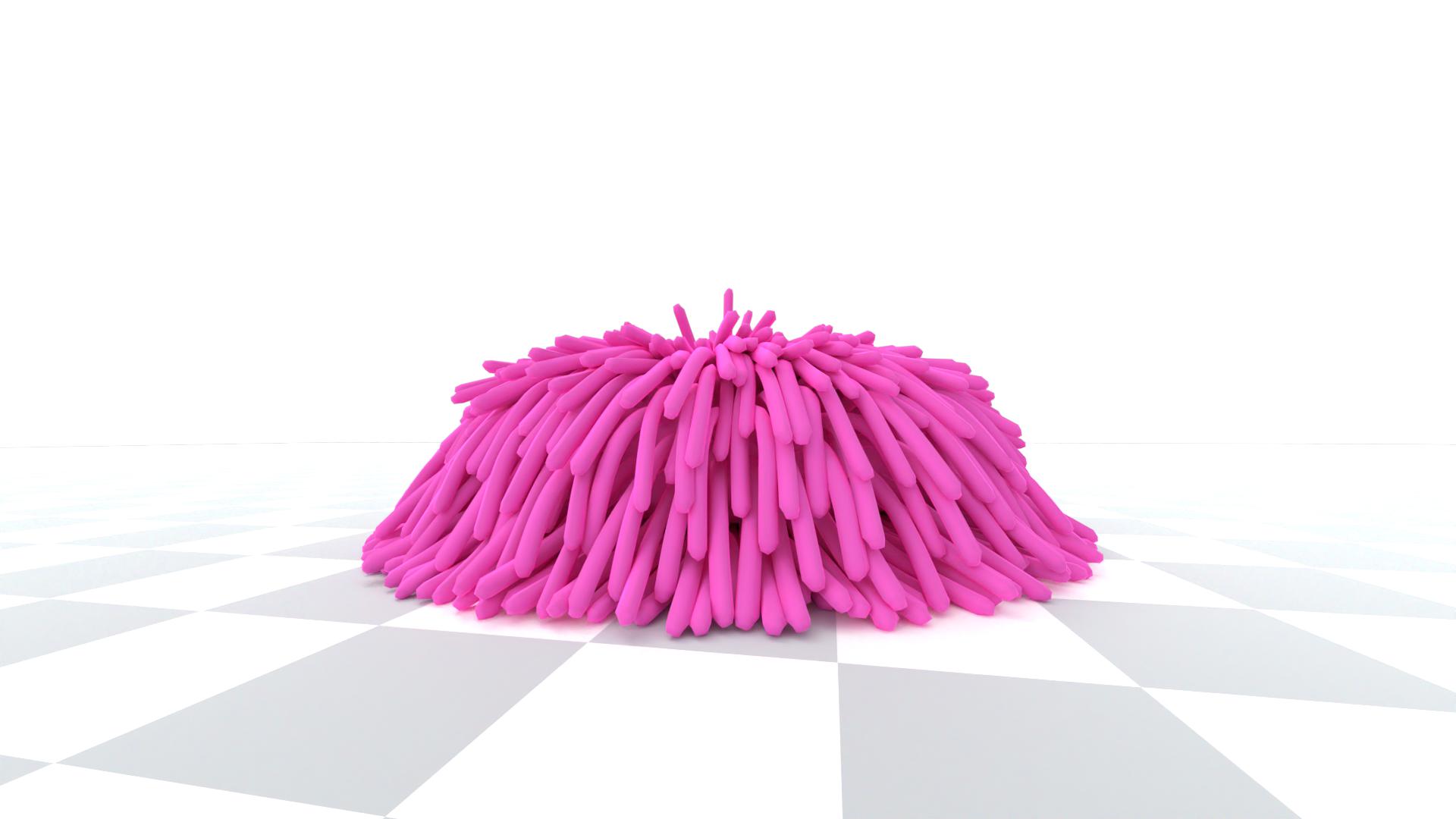}
\caption{Not accelerated}\label{fig:AccelerationComparison:a} 
\end{subfigure}\hfill%
\figA{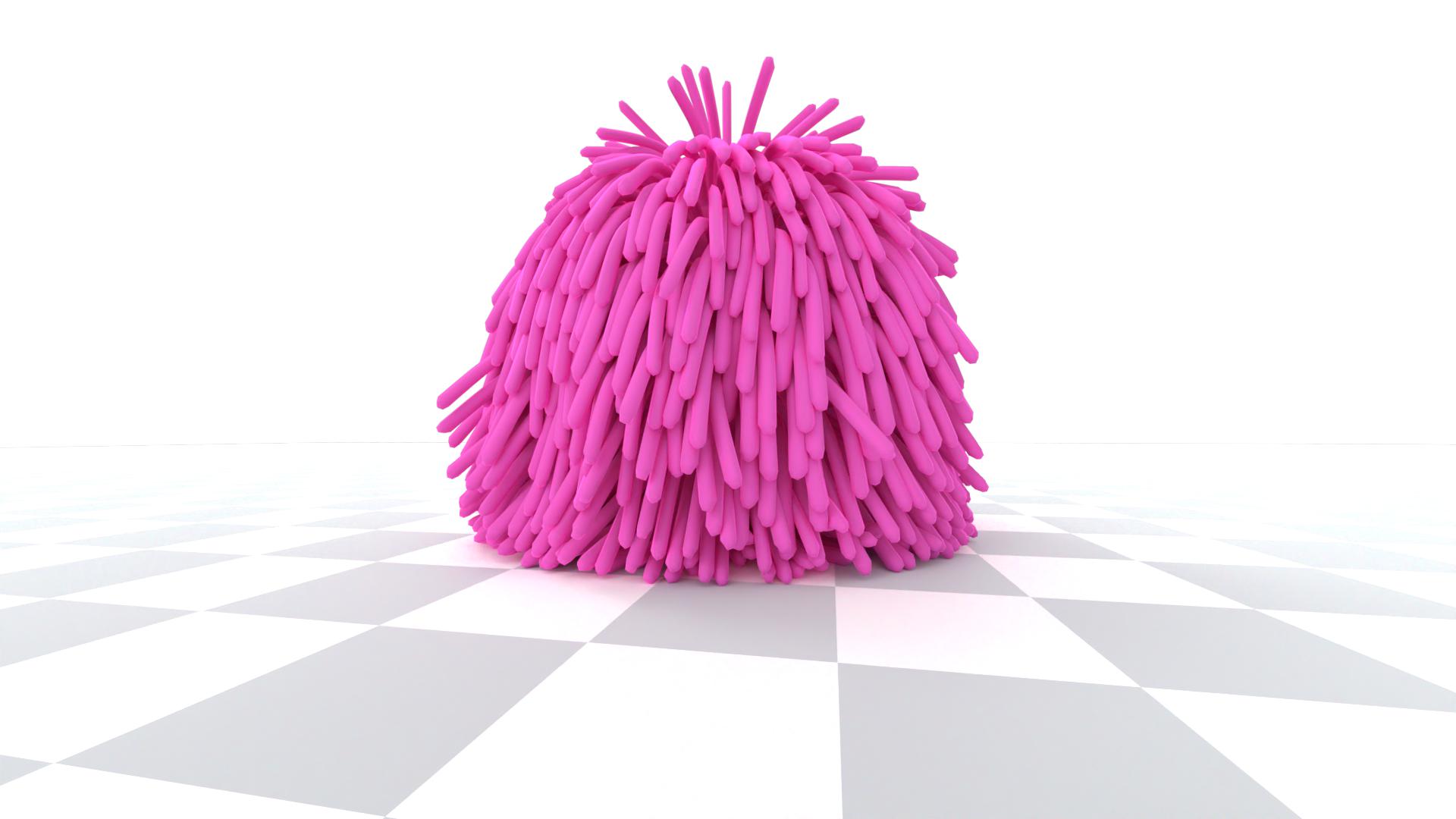}{Accelerated}{b} \hfill%
\figA{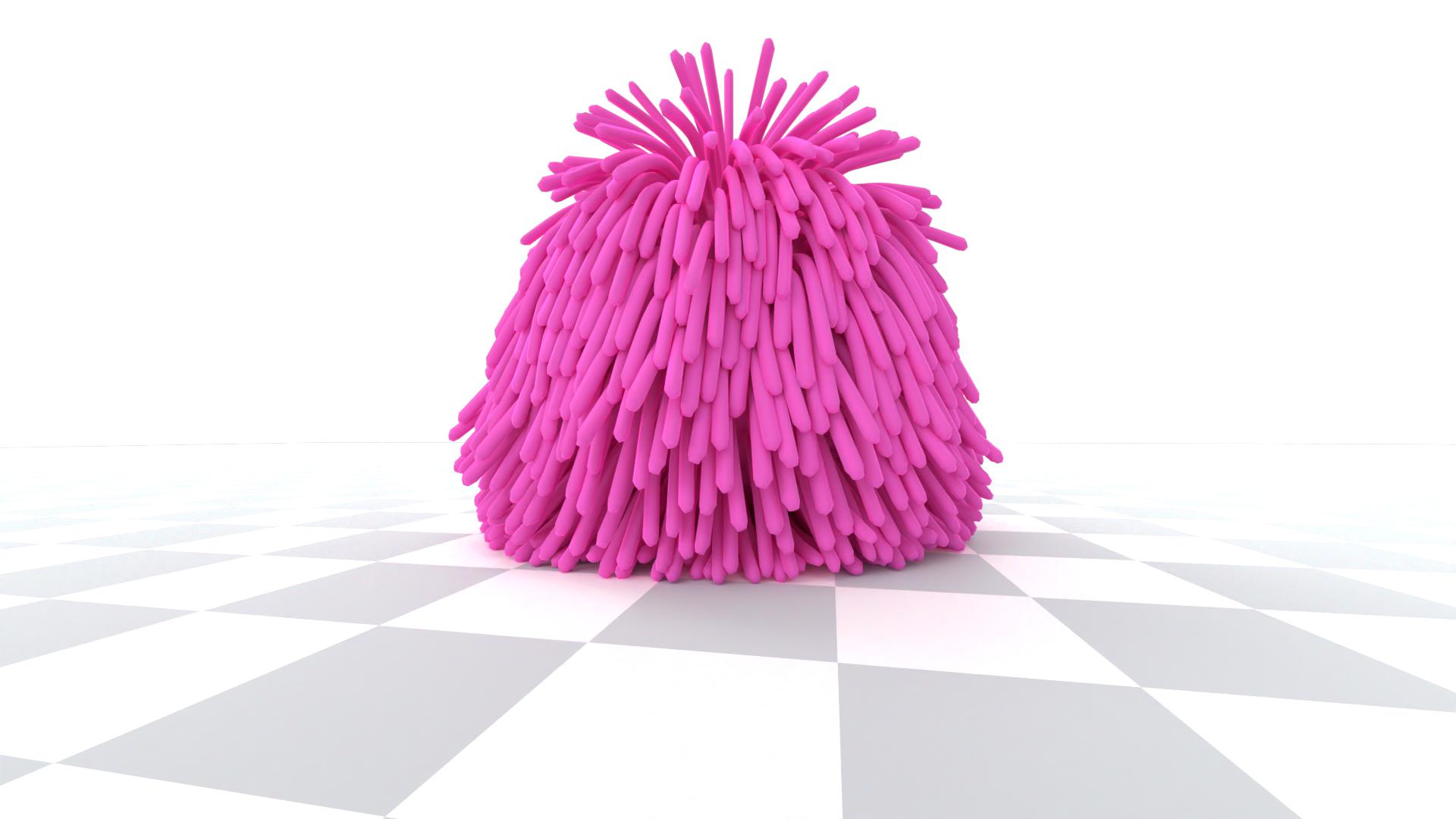}{Reference}{c} 
\caption{
Demonstrating the accelerator's impact in a collision-intensive scene: a squishy ball (230K vertices, 700K tetrahedra) dropping and bouncing. Both \sref{a} and \sref{b} use $h=1/120\text{ seconds}$ with a constant number of $120$ iterations per time step, taking $0.11\text{ seconds}$ of average computation time per frame. \sref{a}~Without acceleration 120 iterations appear to be insufficient. \sref{b}~Our acceleration scheme ($\rho=0.95$), skipping colliding vertices, manages to resolve complex collisions, notably enhancing elasticity convergence for much stiffer outcomes, 
closely matching \sref{c}~the reference computed using 2000 iterations.}
\label{fig:AccelerationComparison}
\end{figure}

This accelerator was originally developed for solving linear systems, assuming the energy to be smooth and (nearly) quadratic. Elastic energy generally fulfills these criteria. However, collision energy tends to be discontinuous and highly stiff, making the use of an accelerator in collision-intensive scenes prone to overshot and compromise the system's stability. To address this, we propose a simple yet highly effective solution for accelerating scenes with collisions: skipping the accelerations for actively colliding vertices. 
Note that the acceleration must be a continuous process. If a vertex is detected colliding at a certain iteration, we will skip the acceleration for it in all the following iterations in the same step, regardless of whether the collision has been resolved. 
This approach has minimal impact on the convergence of elasticity, since 
typically only a small fraction of vertices are in collision.
Also, for those colliding vertices, the elasticity is usually overpowered by the collision forces.
Thus, this solution maintains the stability of the system, while effectively accelerating the convergence of elasticity, as shown in \autoref{fig:AccelerationComparison}.

\subsection{Parallelization}
\label{sec:parallelism}

Gauss-Seidel-type iterative methods are often parallelized using graph coloring by determining groups (i.e. colors) that can be handled in parallel without impacting the sequential nature of the Gauss-Seidel loop. Obviously, the same can be applied to VBD by simply coloring vertices such that no force element uses multiple vertices of the same color.

\begin{figure}
\centering
\newcommand{\figWithWidth}[3]{\begin{subfigure}{0.495\linewidth}\centering
\includegraphics[width=0.8\linewidth]{Figures/GraphAndColoring/GraphAndColoring#3}\vspace{-0.5em}
\caption{#2}\label{fig:GraphAndColoring:#1}
\end{subfigure}}
\newcommand{\figA}[3]{\begin{subfigure}{0.495\linewidth}\centering
\includegraphics[width=\linewidth,trim=0 240 0 0,clip]{Figures/GraphAndColoring/#3}\vspace{-0.5em}
\caption{#2}\label{fig:GraphAndColoring:#1}
\end{subfigure}}
\figWithWidth{a}{Vertex colors: 3}{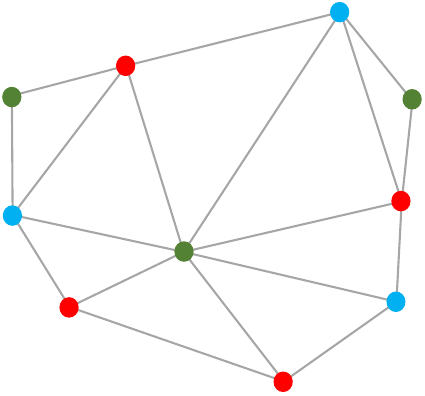}\hfill%
\figWithWidth{b}{Dual graph \& element colors: 7}{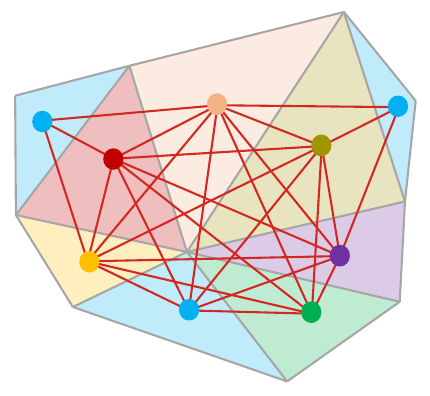}
\figA{c}{Vertex colors: 8}{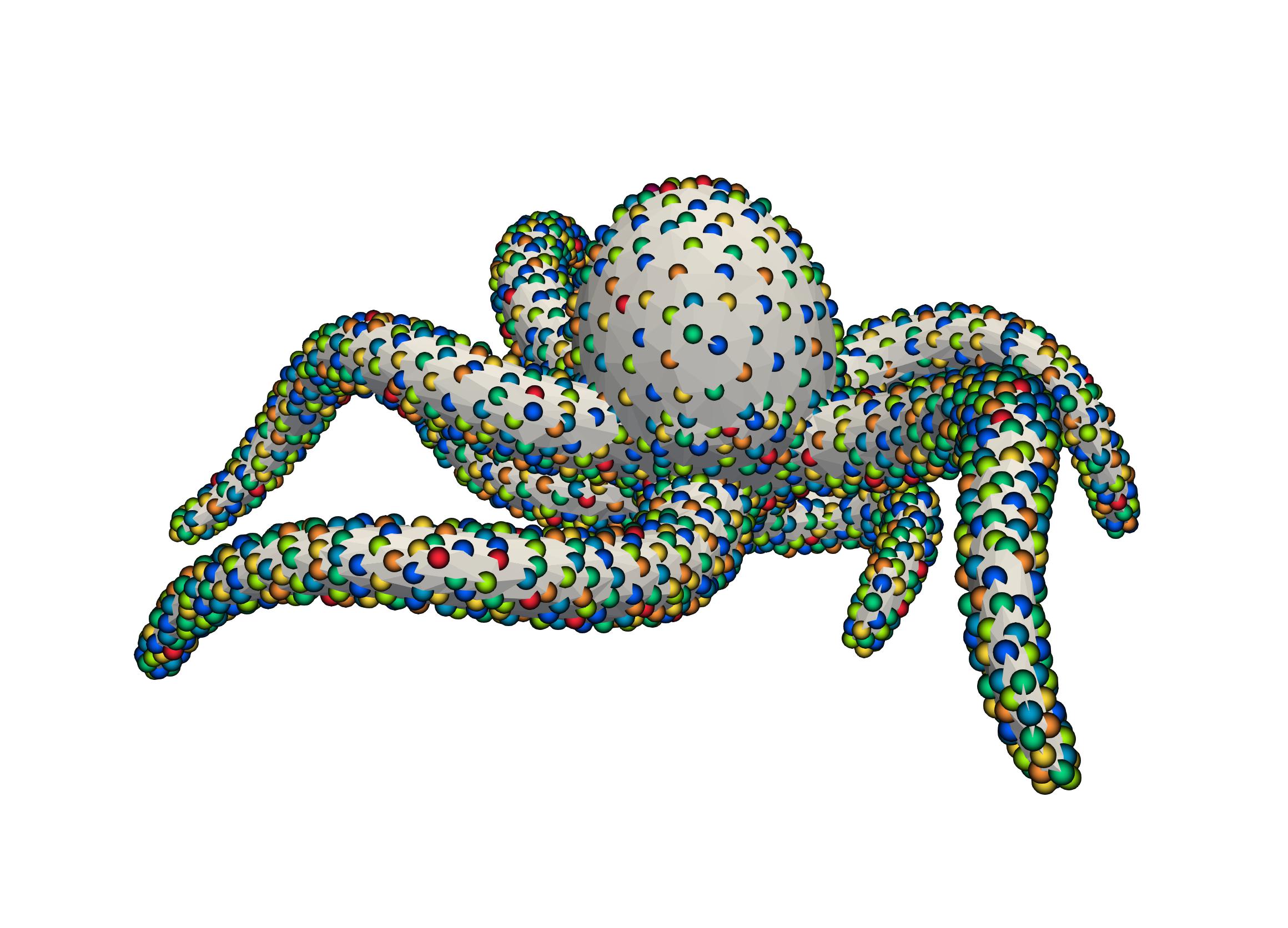}\hfill%
\figA{d}{Element colors: 76}{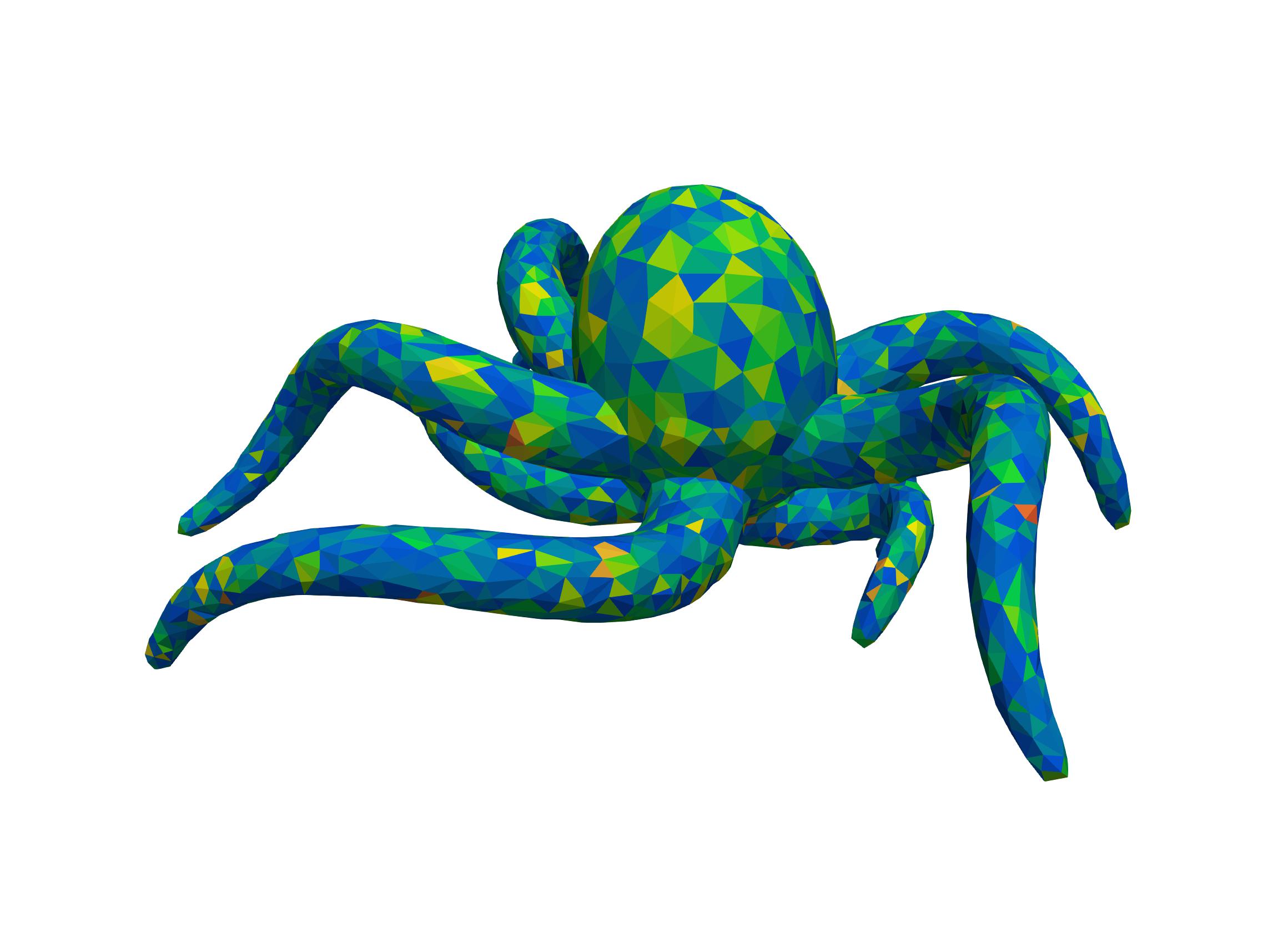}
\caption{Coloring vertices vs. elements: \sref{a}~vertex coloring needs 3 colors for 10 vertices while \sref{b}~element coloring (i.e. coloring the vertices of the dual graph) needs 7 colors for 9 triangles in this example. The difference is more pronounced for tet-meshes: \sref{c}~our vertex coloring uses only 8 colors for 3,891 vertices while \sref{d}~our element coloring implementation needs 76 colors for 14,802 tetrahedra in this example.
}
\label{fig:GraphAndColoring}
\end{figure}


The advantage of VBD here is that, because it colors vertices, it typically results in much fewer colors as compared to techniques that color constraints/force elements, such as PBD. This is because coloring these elements is equivalent to coloring the vertices of the dual graph, which not only has more vertices, but, more importantly, also has a lot more connections per vertex in general. Examples of this are shown in \autoref{fig:GraphAndColoring}. Since different colors must be handled sequentially, fewer colors means better parallelism.

When all force elements are known ahead of the simulation, graph coloring can be performed as a preprocess. However, dynamically generated constraints/force elements, such as ones due to collisions, cannot be known ahead of time, requiring dynamic recoloring.

In our implementation for elastic body dynamics, we avoid the cost of recoloring by precomputing coloring only for material forces, ignoring dynamically generated force elements due to collisions.
This means that these collision forces may use multiple vertices of the same color. Therefore, we cannot simply run a parallel loop over all vertices of the same color and update them, because handling a vertex with a dynamically generated force element may run into race conditions (with partially updated vertex positions) when accessing other vertices of the same color.

\del{We resolve this by splitting the processing of a color into two phases, as shown in Algorithm 1. The first phase computes the gradients and Hessians of all force elements acting on vertices of the color without updating any vertex positions. The second phase uses the computed data in the first phase to solve the local system and update the vertex positions. Thus, dynamically generated force elements using multiple vertices of the same color result in (partially) Jacobi-style iterations for those vertices, as opposed to Gauss-Seidel. Everything else follows the Gauss-Seidel order.}

\add{We resolve this by having an auxiliary vertex position buffer ($\vec{x}^{\text{new}}$) that stores the updated vertex position. When executing each local VBD position update, we write the updated vertex positions to the auxiliary buffer, instead of directly overwriting the original vertex position buffer. Then, we copy the updated positions from the auxiliary buffer to the original vertex position buffer after each color pass. This prevents the race conditions that arise from simultaneous read and write operations on vertex positions.

With this process, dynamically generated force elements using multiple vertices of the same color result in (partially) Jacobi-style iterations for those vertices, because they rely on the positions from the previous iteration of those vertices. 
For vertices with different colors, it is equivalent to Gauss-Seidel iterations, considering the updated positions of the vertices with different colors.
Note that our algorithm does not explicitly switch between Jacobi and Gauss-Seidel iterations, but the resulting iteration we describe above corresponds to either (partially) Jacobi or Gauss-seidel, depending on the colors of the colliding vertices.
}

\begin{figure}
\centering
\newcommand{\img}[3]{\begin{subfigure}{0.49\linewidth}\centering
\includegraphics[width=\linewidth,trim=#2 100, clip]
{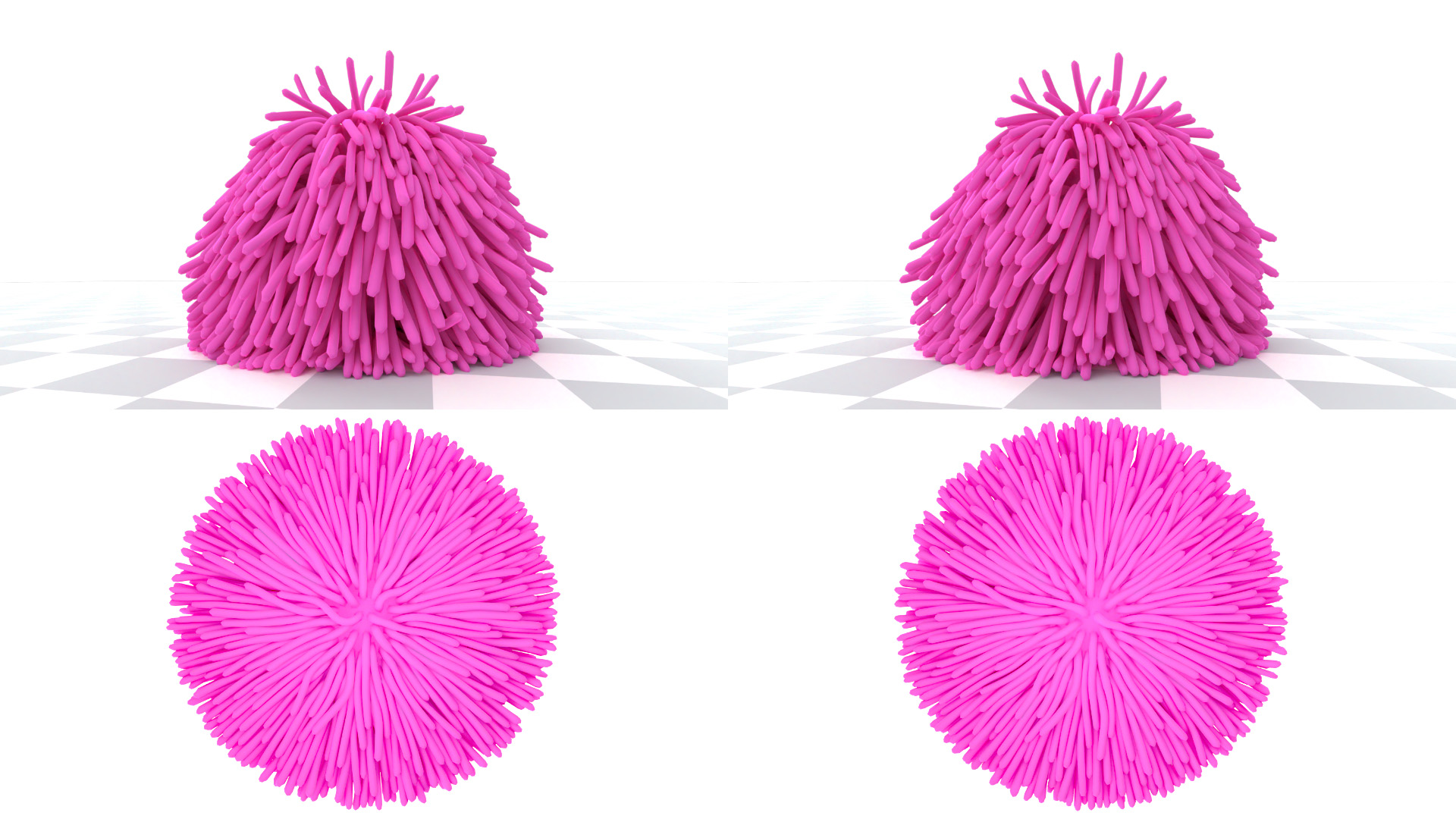}\vspace{-0.5em}
\caption{#3} \label{CollisionResolutionScheme:#1}
\end{subfigure}}
\img{a}{0 0 960}{No recoloring \textbf{(ours)}}\hfill%
\img{b}{960 0 0}{Recoloring}
\caption{Handling collisions using \sref{a}~our scheme without recoloring and \sref{b}~recoloring to achieve perfect Gauss-Seidel iterations, 
both simulated using friction forces and accelerated iterations with $\rho=0.95$. Notice that the results are highly similar, though not identical.}
\label{fig:CollisionResolutionScheme}
\end{figure}

One might expect this solution to negatively impact the convergence of VBD. Fortunately, however, such Jacobi-style information exchanges are relatively rare. This is because, as shown in \autoref{fig:CollisionPotential}, with vertex-face collisions at most one pair and with edge-edge collisions at most two pairs of vertices can have the same color. Also, vertices with the same colors must be located on different sides of the collision; therefore, their elastic energies are usually decoupled. This ensures that the majority of the information exchange follows the Gauss-Seidel order and thereby the impact of our solution on convergence is minimal.
\autoref{fig:CollisionResolutionScheme} presents an example with a large number of collisions, comparing our solution of skipping recoloring to proper Gauss-Seidel iterations with recoloring, showing that the differences are minor.

\del{Another advantage of our two-phase computation is that it results in a workload that is better suited for the GPU. If a single parallel loop is used, for each vertex we must loop over all of its force elements to compute their gradient and Hessian. Since vertices can have different numbers of force elements, this can result in poor occupation with SIMD computation. Our two-phase solution minimizes this and only requires storing 12 numbers (3 for the gradient and 9 for the Hessian of $E_j$) per force element.}

\begin{figure}
\centering
\includegraphics[width=0.28\linewidth,trim=250 200 250 150,clip]{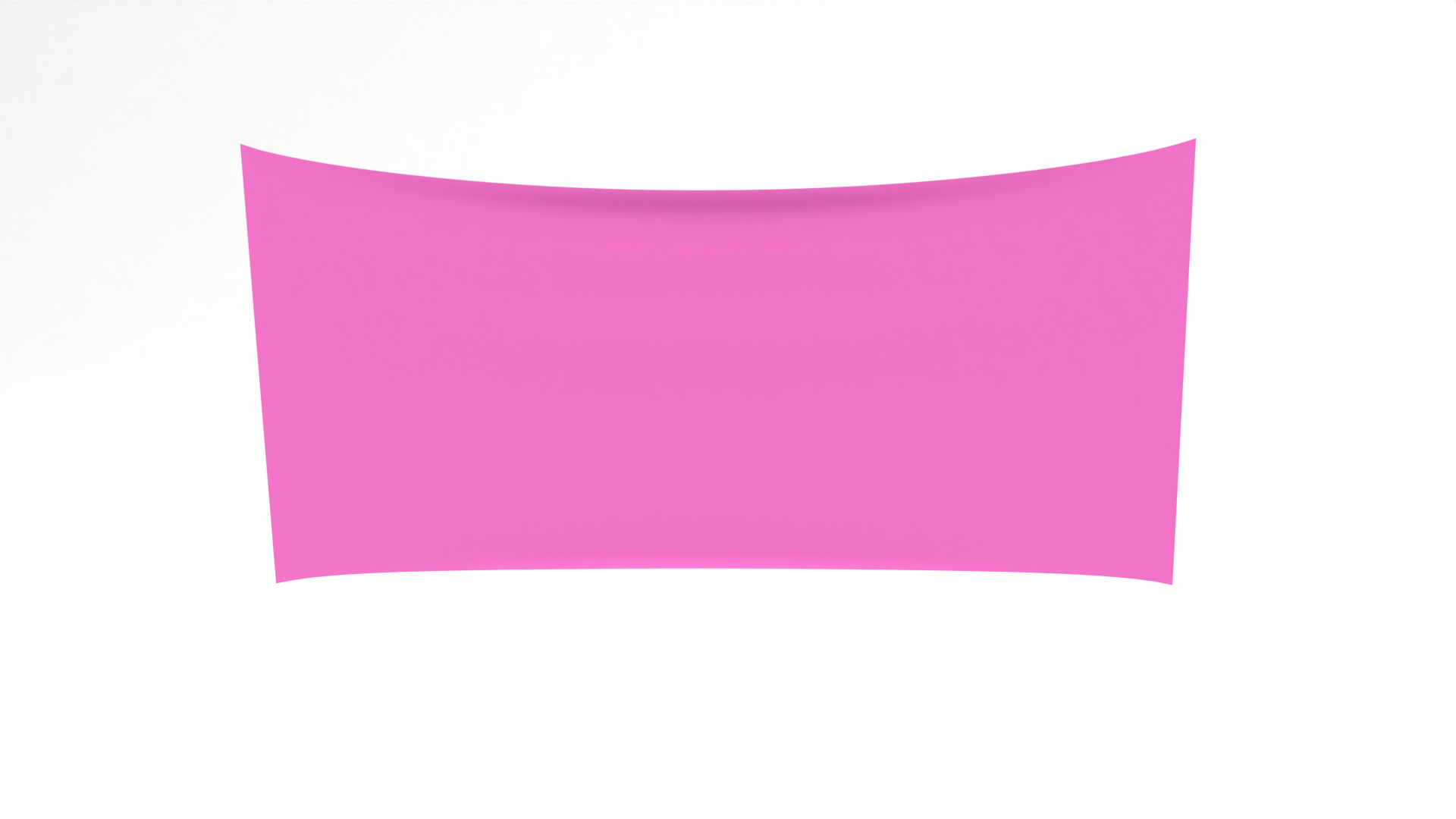} \hfill%
\includegraphics[width=0.33\linewidth,trim=138 200 148 150,clip]{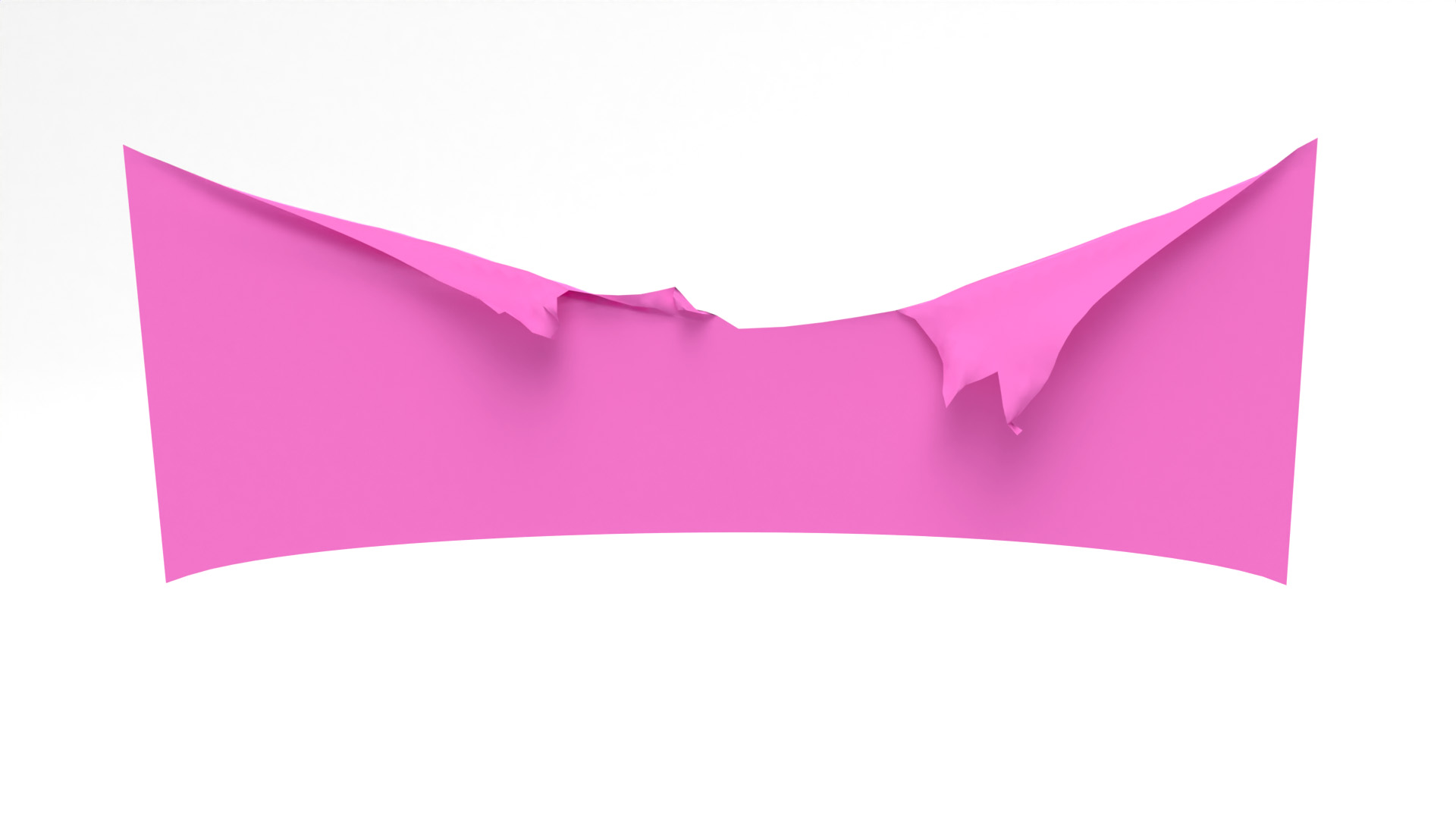} \hfill%
\includegraphics[width=0.35\linewidth,trim=128 200 128 150,clip]{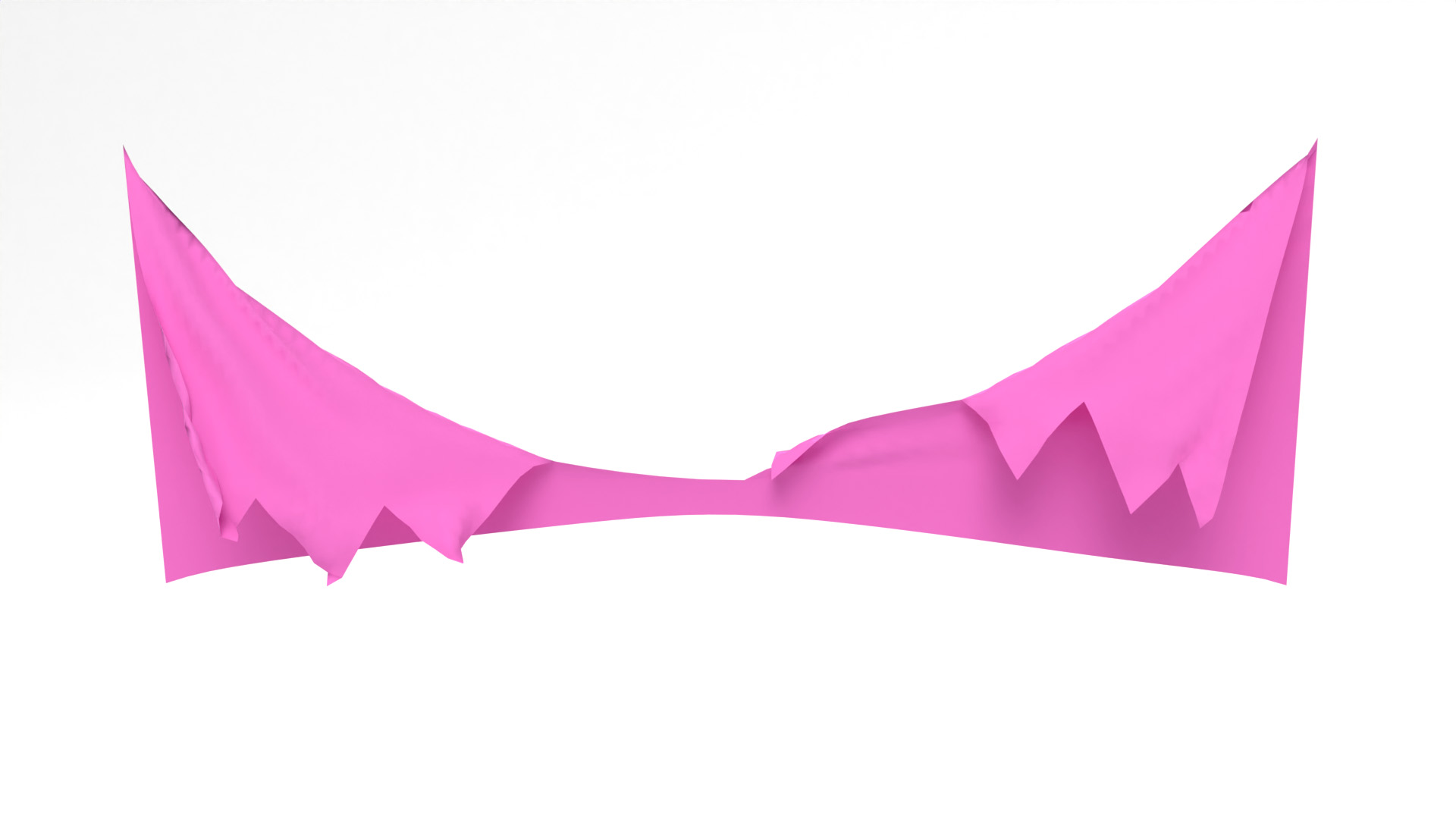}
\caption{Tearing a piece of cloth with 2500 vertices and 4800 triangles.}
\label{fig:TopologicalChange}
\end{figure}

Our solution also works with other types of topological changes, such as tearing and fracturing. Deleting force elements does not require any changes to vertex coloring. When an object is split by duplicating vertices, as in the case of tearing a piece of cloth along some edges (see \autoref{fig:TopologicalChange}), duplicated vertices can safely inherit the colors of their original vertices.

\add{
\section{GPU Implementation}


In this section, we describe a GPU implementation specifically designed to leverage the inherent parallelization mechanism of modern GPUs, which consists of two hierarchical levels: block-level and thread-level parallelism. Block-level parallelism facilitates large-scale parallel operations, assuming that each block operates independently. On the other hand, thread-level parallelism provides finer, single-instruction-multiple-thread (SIMT) style parallelism, allowing for inter-thread communication and synchronization within the same block.

Reflecting on VBD, we observed that it naturally aligns with this hierarchical architecture. We have thousands of vertices within a single color category that operate independently, and each vertex is associated with multiple force elements, which can be processed concurrently.
\autoref{alg:VBDStep} shows the pseudocode of our implementation. It uses block-level parallelism for processing each vertex. The threads within each block are used for computing the forces and Hessians, storing them in local shared memory, and computing the sums via reduction. We use a fixed number of threads for each block.
When the total number of adjacent force elements exceeds the number of threads for each block, individual threads will loop over their assigned elements. During this process, they calculate the forces and Hessians for these force elements and then sum them to their assigned shared memory. 


In our experiments, we observed nearly \textit{an order of magnitude performance improvement}, as compared to processing each vertex with a single thread.
The primary advantage lies in the optimization of the memory access pattern, a common bottleneck in GPU programs. This implementation reduces memory divergence within blocks.  Because the neighboring force elements of a vertex often share multiple vertices, the threads within the same block can share a  global memory access to those shared vertices. 
Furthermore, this strategy improved the parallelism of the algorithm by allowing parallel evaluation of the force and Hessian of adjacent force elements, which are then written to the significantly faster shared memory. This bypasses the slower global memory and enables the parallel aggregation of force and Hessian values, enhancing the overall efficiency of the process.


}

\begin{algorithm}[t]
\SetKwFor{Foreach}{for each}{do}{end}
\SetKwFor{DoParallel}{parallel for each}{do}{end}
\newcommand\mycommfont[1]{\footnotesize\textcolor[RGB]{0 128 0}{#1}}
\SetCommentSty{mycommfont}
\LinesNumbered
\DontPrintSemicolon
\SetAlgoNoLine
\KwIn{
$\Vec{x}^t$:  the positions of the previous step;
$\Vec{v}^t$:  the velocities of the previous step;
$\Vec{a}_{\text{ext}}$: the external acceleration
}
\KwOut{This step's position $\Vec{x}^{t+1}$ and velocity $\Vec{v}^{t+1}$.}
$\Vec{y}\gets\Vec{x}^t + h \Vec{v}^t + h^2 \Vec{a}_{\text{ext}}$\;
Initial DCD using $\Vec{x}^t$\;
$\Vec{x}\gets$ initial guess with adaptive initialization\;
\Foreach{\upshape iteration $n\leq n_{\max}$}{
   \lIf{\upshape $n\!\!\mod{n_{\text{col}}}=1$}{CCD using $\Vec{x}$}
   
   \Foreach{\upshape color $c$}{
        

        \add{
        \tcp{Block-level parallelization} 
        \DoParallel{\upshape vertex $i$ in color $c$}{
            \tcp{Thread-level parallelization }
            \DoParallel{$j\in \mathcal{F}_i$} {
                \tcp{Variables in shared memory}
                $\vec{f}_{i,j} = -\frac{\partial E_j}{\partial  \vec{x}_i}$\; 
                $\vec{H}_{i,j} = \frac{\partial^2 E_j}{\partial \vec{x}_i \partial \vec{x}_i}$\;
            }
            \tcp{Local reduction sums}
            $\vec{f}_i = \sum_{j\in\mathcal{F}_i}\vec{f}_{i,j}$\;
            $\vec{H}_i = \sum_{j\in\mathcal{F}_i}\vec{H}_{i,j}$\;
            
            $\Delta \Vec{x}_i \gets \vec{H}_i^{-1} \vec{f}_i$\;
                $\Delta \Vec{x}_i \gets$  optional line search from $\Vec{x}_i + \Delta \Vec{x}_i$ to $\vec{x}_i$\;
            $\Vec{x}_i^\text{new} \gets \Vec{x}_i + \Delta \Vec{x}_i$\;
        }

        \tcp{Copy updated positions back to the vertex buffer}
        \DoParallel{\upshape vertex $i$ in color $c$}{
            $\vec{x}_i =  \Vec{x}_i^\text{new}$
        }
        }
    }
    
    \tcp{Optional: accelerated iteration}
    \DoParallel{\upshape vertex $i$}{
        Update $\vec{x}_i$ using \autoref{eq:accelerator}.
    }
}        
$\Vec{v} = (\Vec{x} - \Vec{x}^t)/h$\;
return $\Vec{x}$, $\Vec{v}$\;
\caption{VBD simulation for one time step.}
\label{alg:VBDStep}
\end{algorithm}

%% file: 04_Results.tex
\begin{figure*}
\centering
\newcommand{\fig}[1]{\fbox{\includegraphics[width=0.245\linewidth,trim=350 0 450 0,clip]{Figures/SquishyBallsToTeapot/#1-cc}}}
\fig{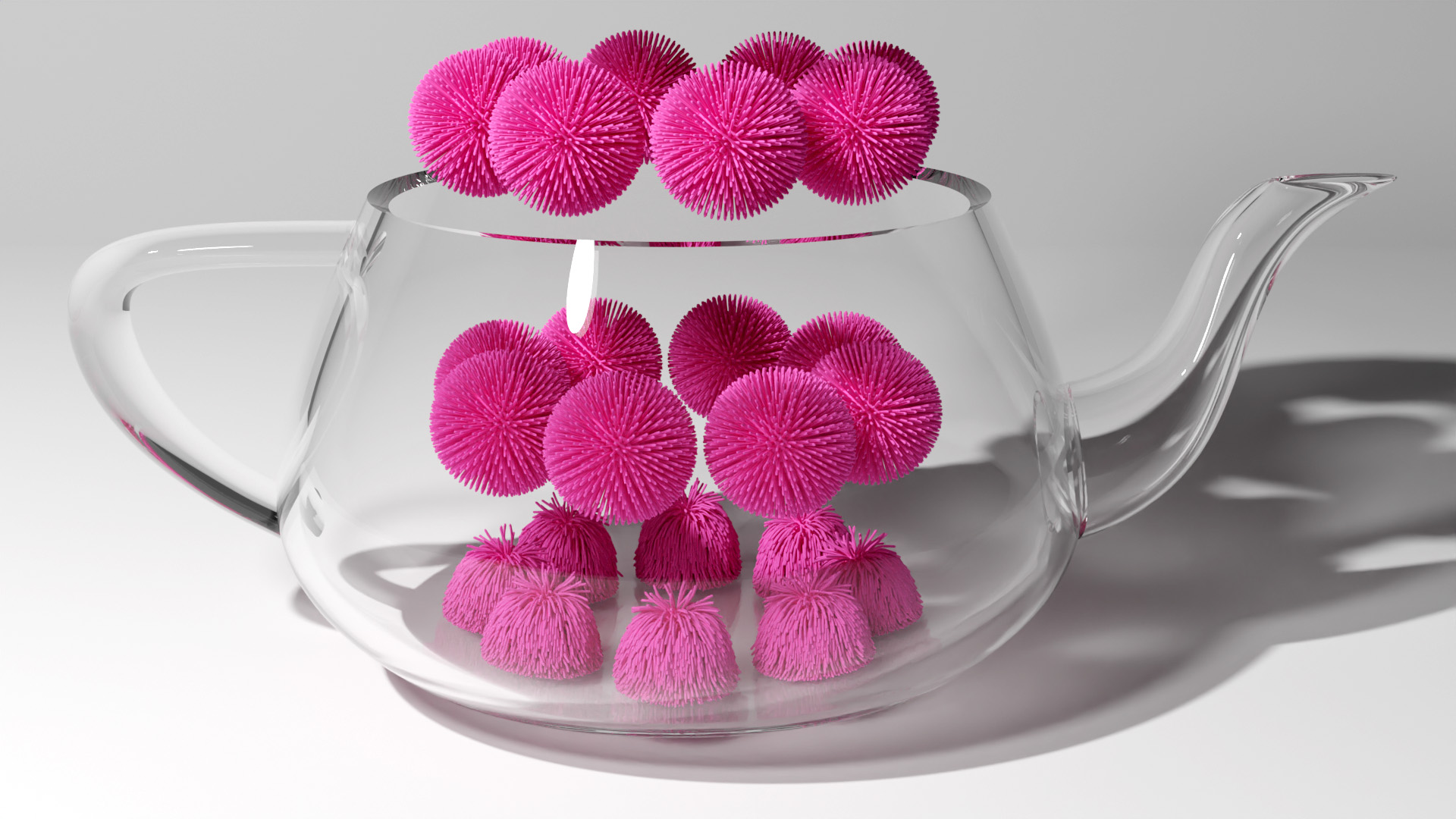}\hfill%
\fig{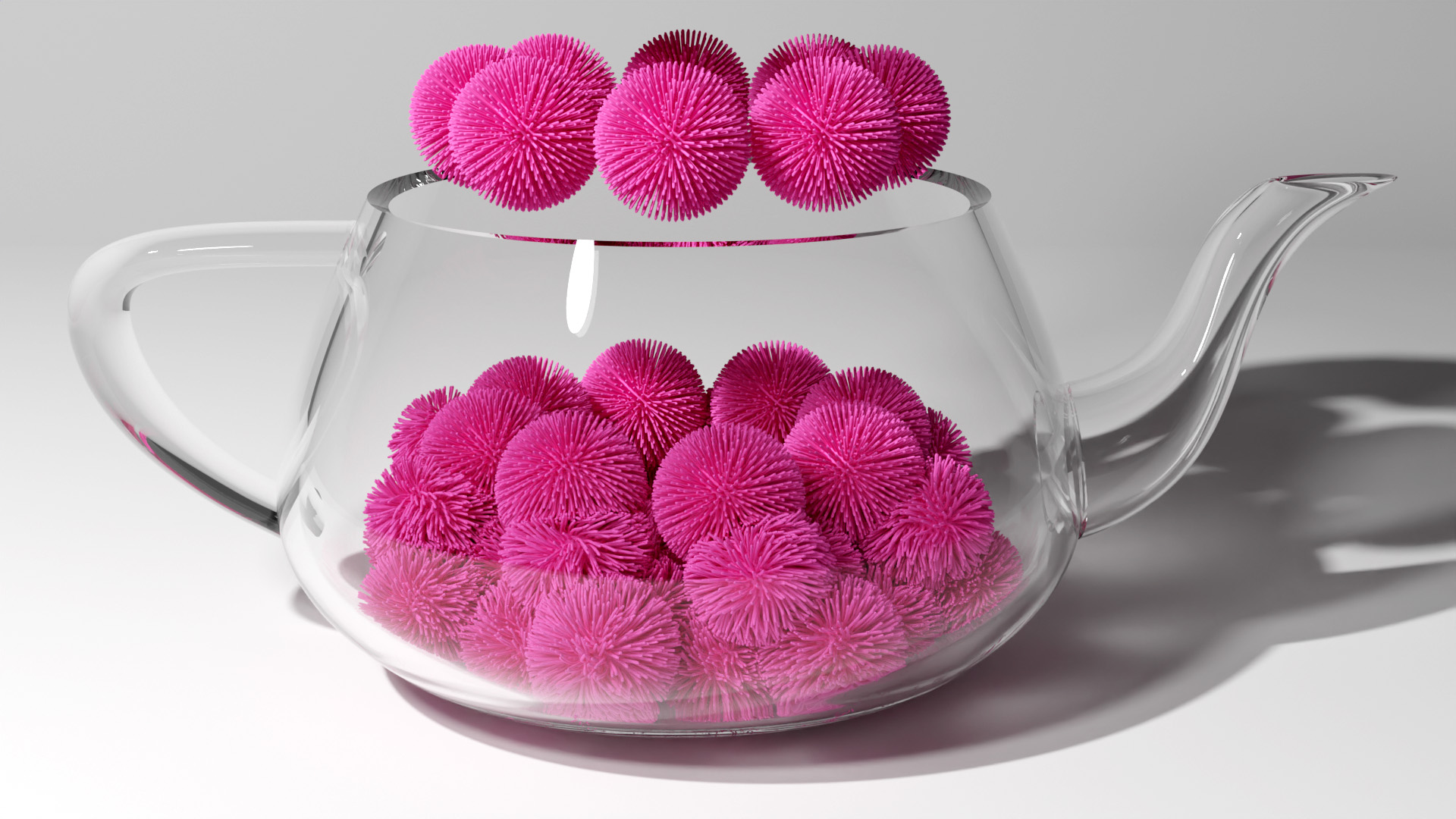}\hfill%
\fig{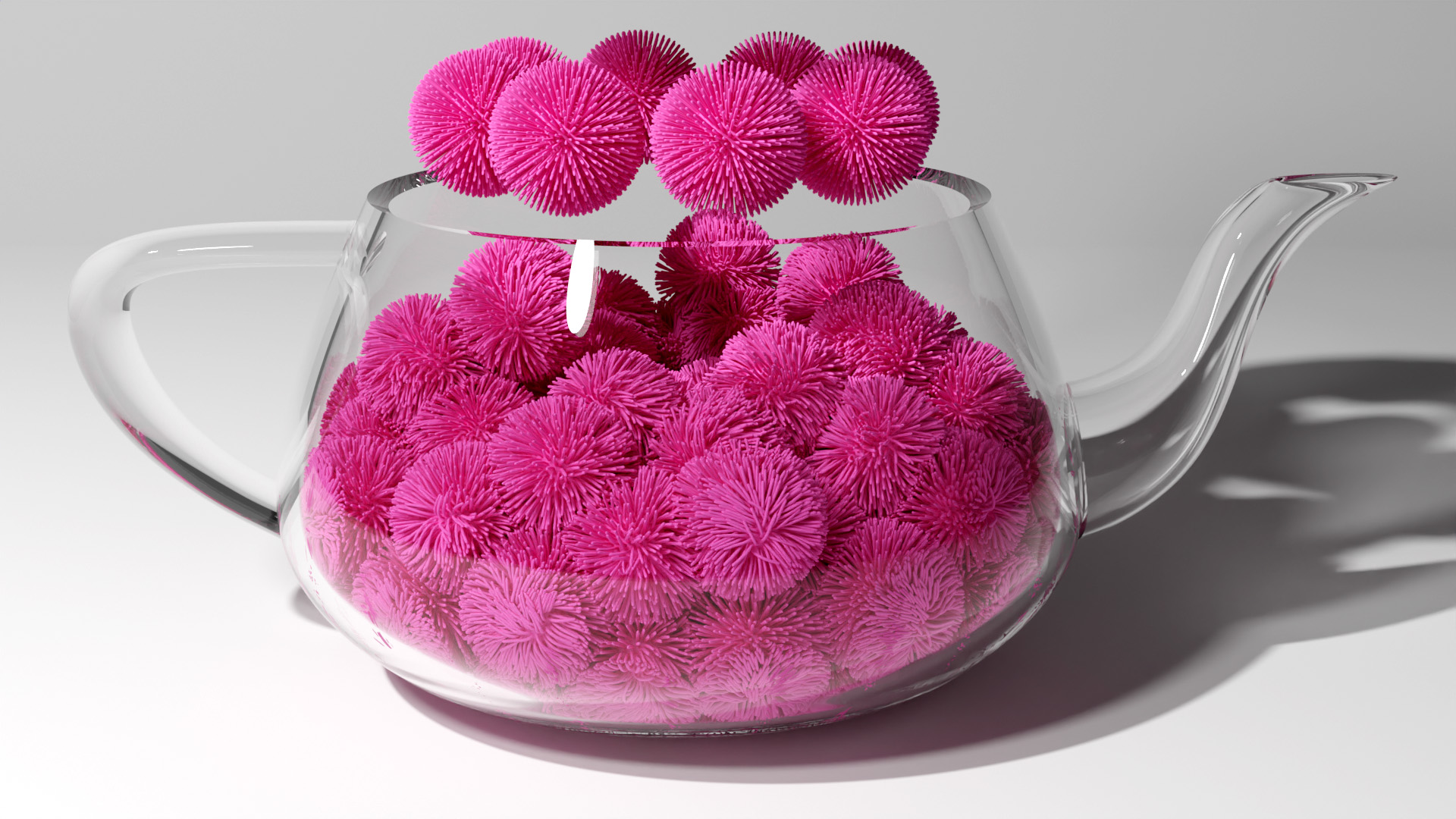}\hfill%
\fig{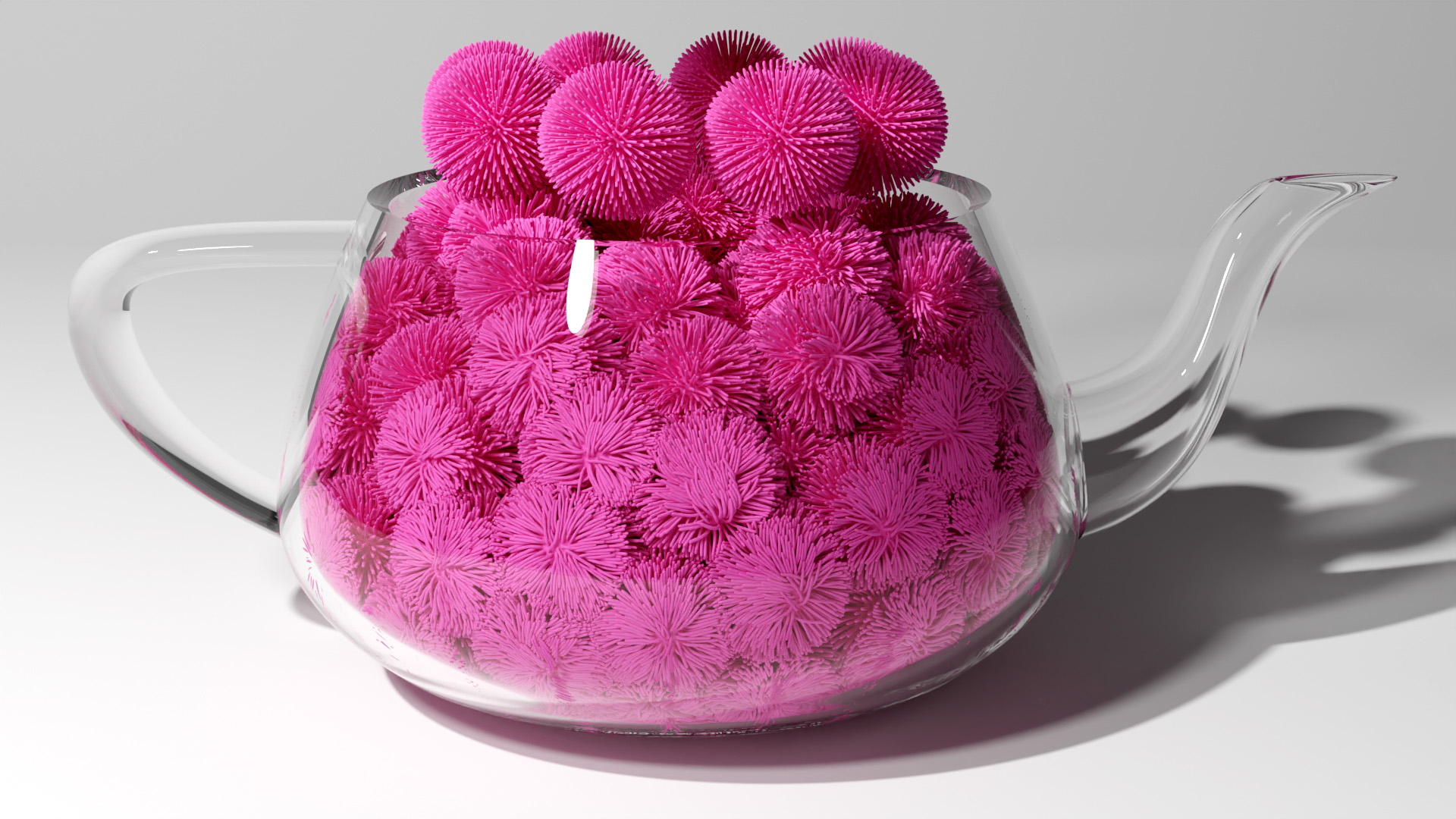}
\caption{Simulation of 216 squishy balls with tentacles, a total of 48 million vertices and 151 million tetrahedra, dropped into a Utah teapot, forming a stable pile with active frictional contacts. The average and maximum computation time per time step is \del{10 and 16} \add{3.6 and 3.9} seconds, respectively, using $S=4$ substeps per frame and $n_{\max}=40$ iterations per step. The final frame of this simulation is shown in \autoref{fig:teaser}.}
\label{fig:SquishyBallsToTeapot}
\end{figure*}

\begin{figure*}
\centering
\newcommand{\fig}[1]{\fbox{\includegraphics[width=0.245\linewidth,trim=250 0 250 0,clip]{Figures/HybrridModelsDropping/#1-cc}}}
\fig{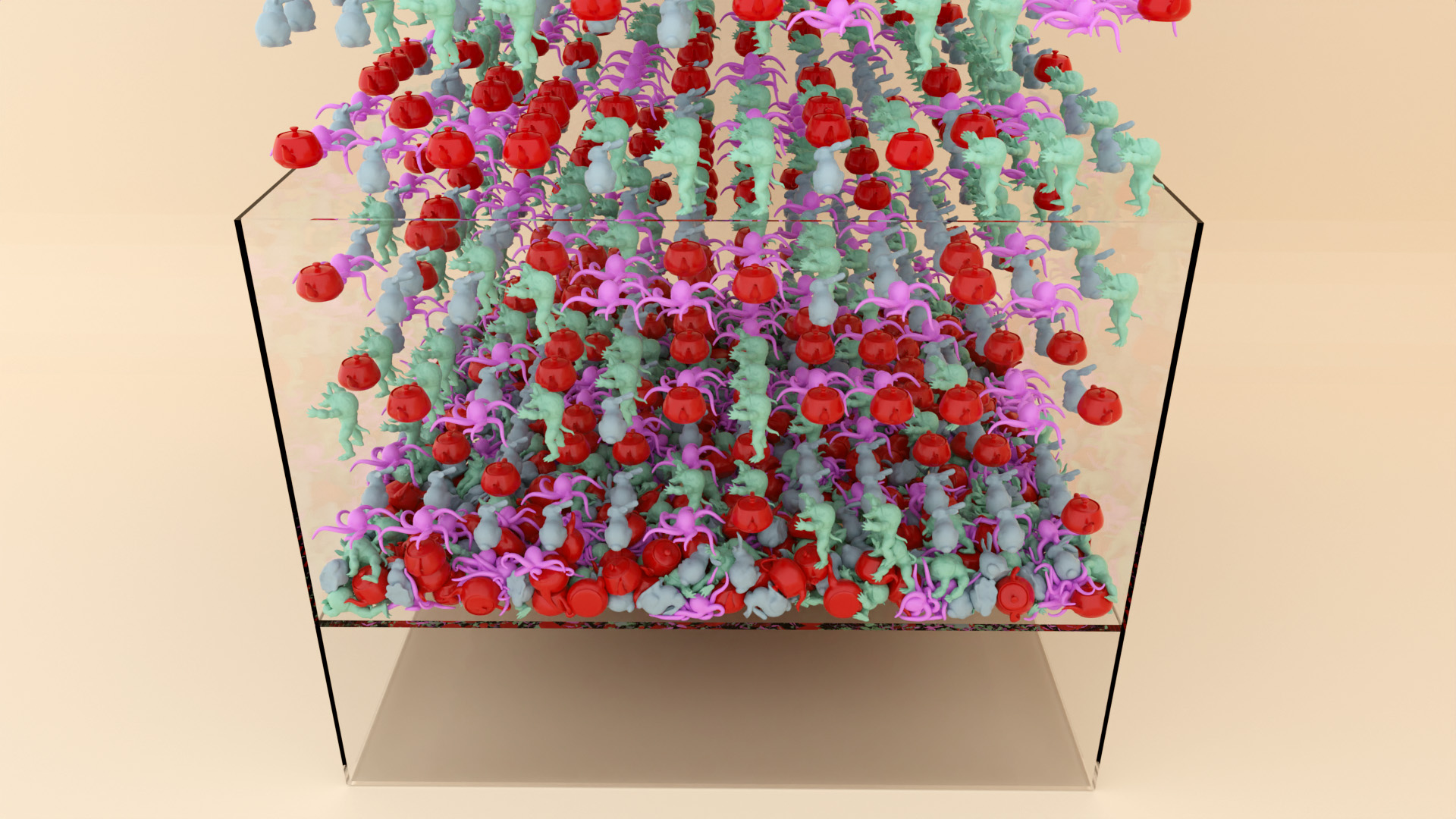}\hfill%
\fig{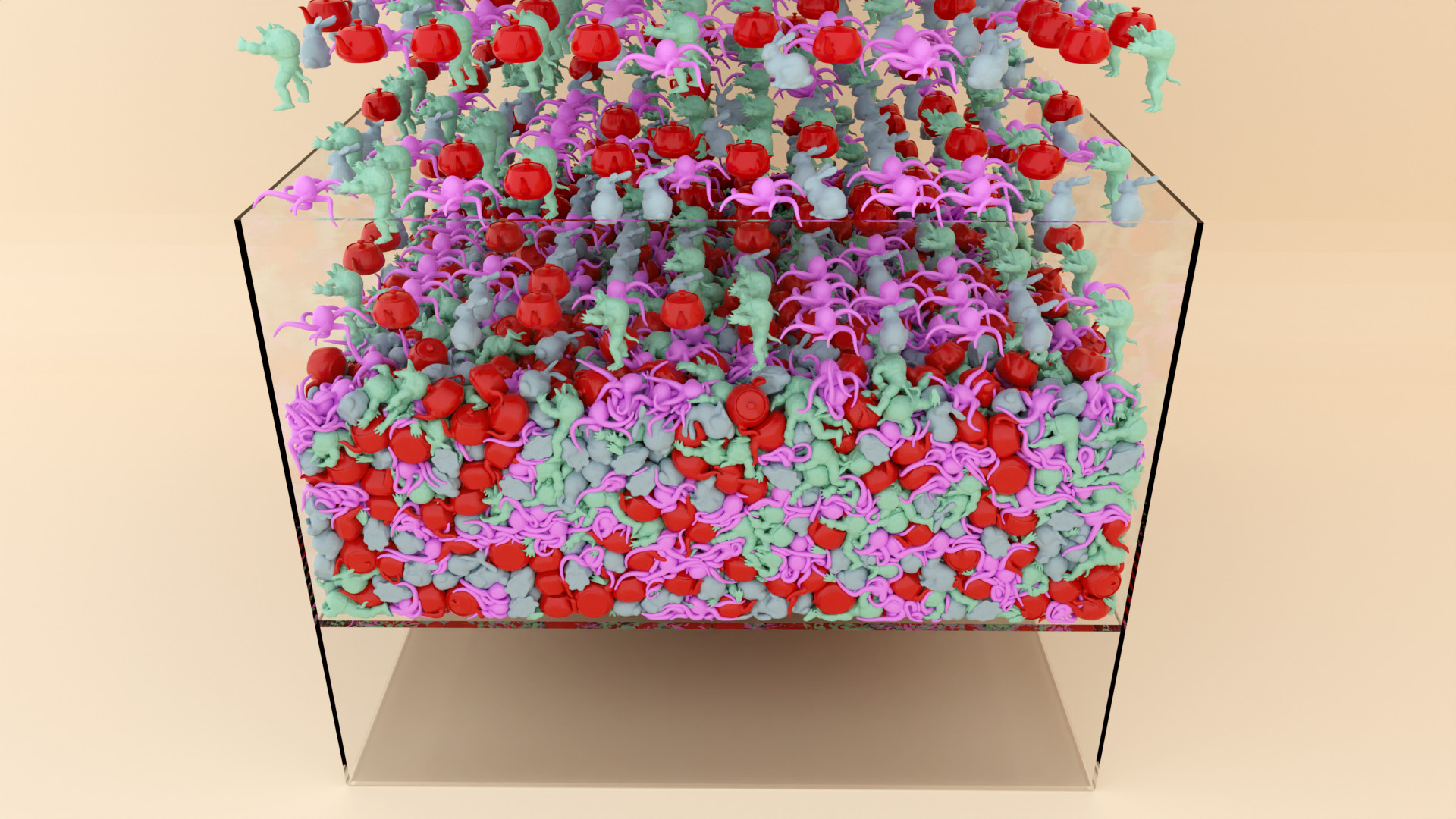}\hfill%
\fig{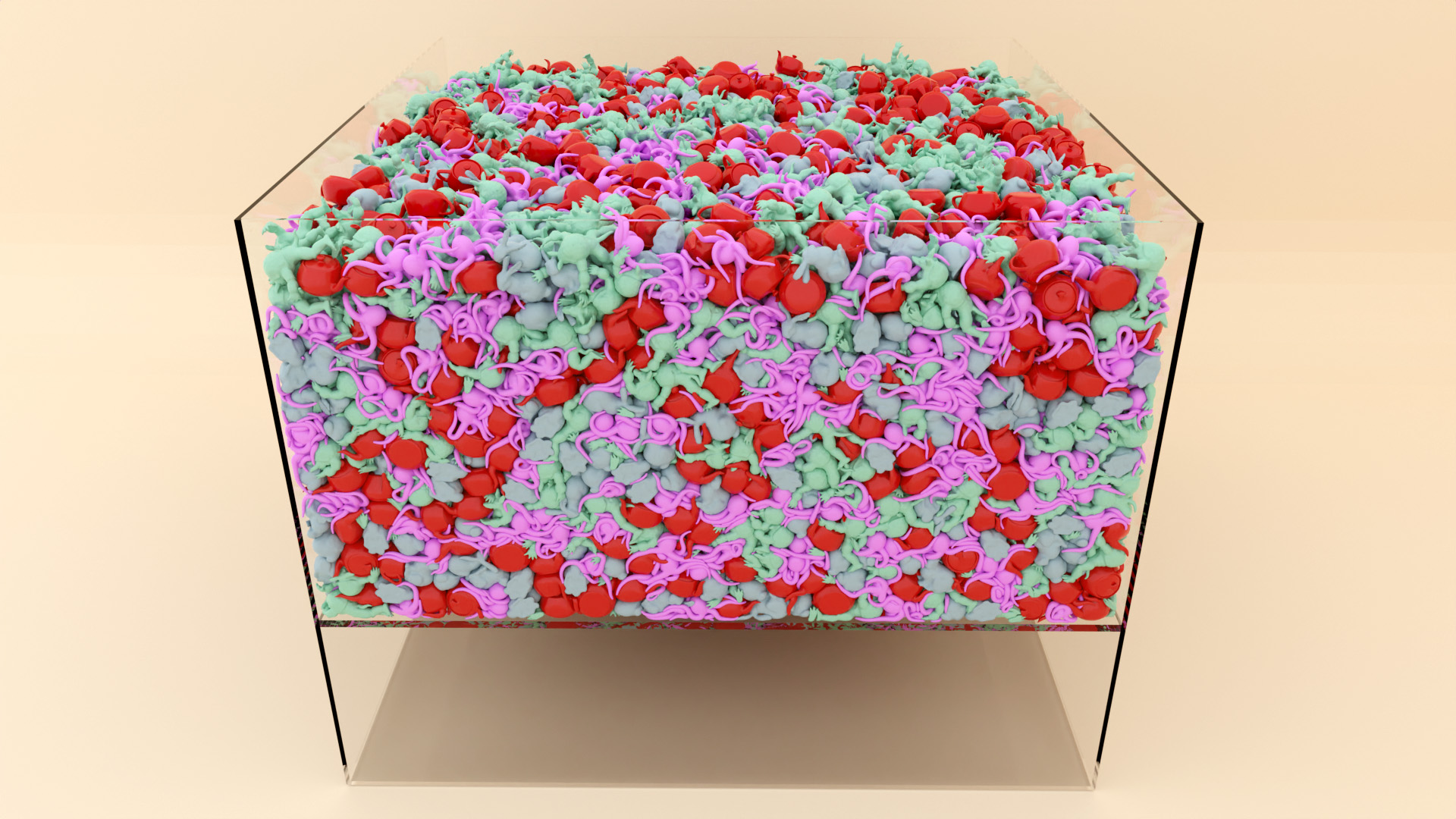}\hfill%
\fig{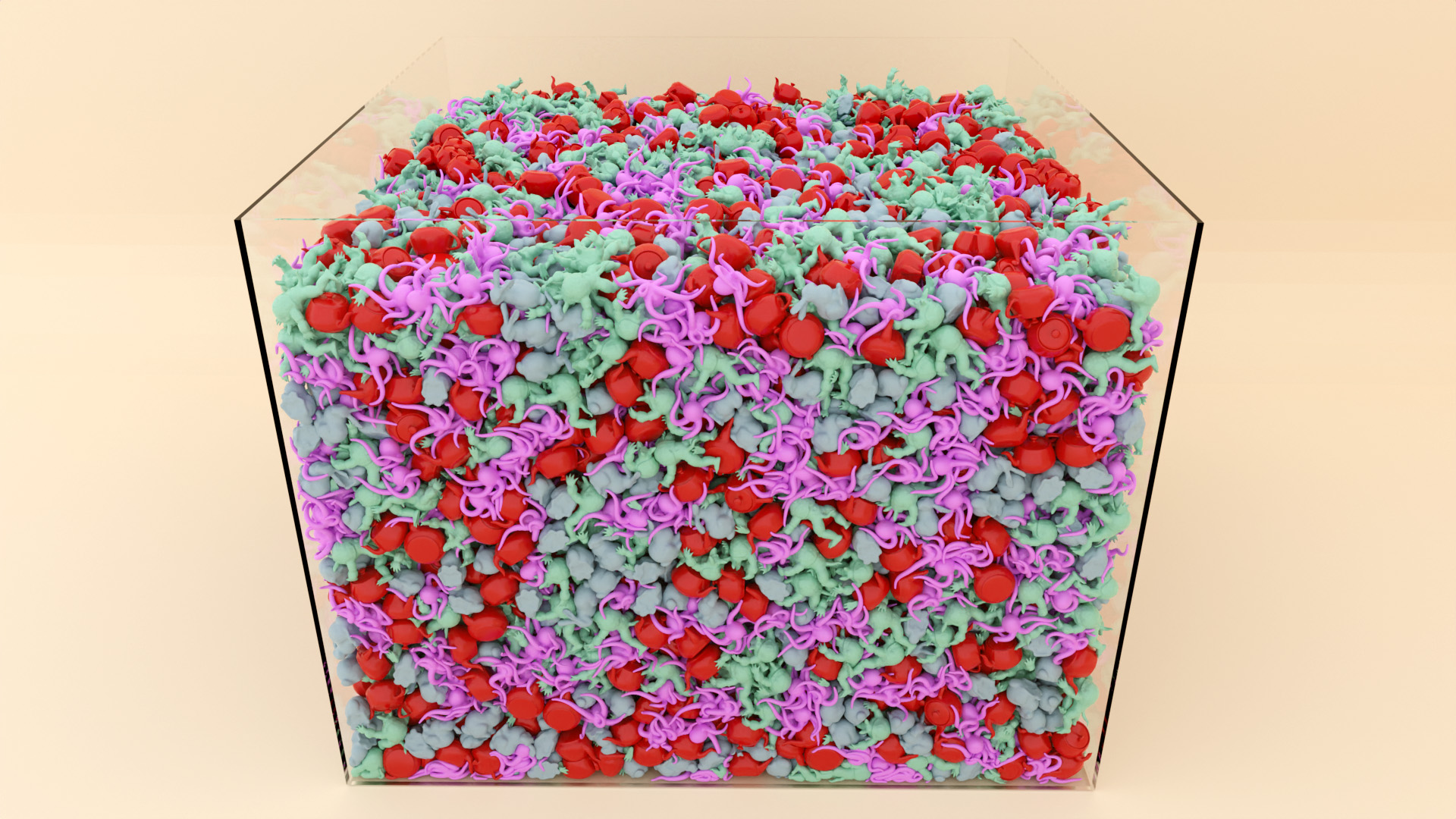}\vspace{-0.043\linewidth}\\
\hspace*{0.5\linewidth}%
\begin{minipage}{0.25\linewidth}\centering
\scriptsize\color{white} $\uparrow$ This platform\\will be removed.
\end{minipage}%
\hspace*{0.5\linewidth}\vspace{0.8em}
\caption{Simulation of 10,368 deformable objects, totaling over 36 million vertices and 124 million tetrahedra, dropped onto a platform inside a box container. Then, the platform is suddenly removed and the objects collectively fall onto the ground, forming stable piles both before and after the platform is removed.
The average and maximum computation times per time step are \del{12 and 14} \add{4.2 and 4.7} seconds, respectively, using $S=2$ substeps and $n_{\max}=60$ iterations per step. 
The final frame of this simulation is shown in \autoref{fig:teaser}.}
\label{fig:HybrridModelsDropping}
\end{figure*}

\section{Results}

We evaluate our method with elastic body dynamics qualitatively with various tests and quantitatively with direct comparisons to alternative methods.
We use Neo-Hookean\cite{smith2018stable} materials (without the logarithmic term) for our volumetric objects, StVK \cite{Volino09} for clothes, and linear springs for elastic rods.

We use a fixed frame time of $1/60$ seconds and a fixed iteration count $n_{\max}$, instead of estimating convergence after every iteration. Each frame is computed using $S$ substeps, such that ${h=1/(60S)}$ seconds. Using smaller time steps increases accuracy and reduces numerical damping with any implicit Euler method. With VBD, a smaller time step only requires fewer iterations per step for similar visual quality, but it can also achieve a smaller residual with the same number of total iterations per frame (i.e., $S n_{\max}$).
\add{The number of threads per block is set to 16 for all the experiments.}

We use no line search in our tests, as none of our tests required it for stability, and running line search can result in a noticeable drop in performance. 
In our experiments, we apply CCD only in the first iteration (i.e. $n_{\text{col}}=n_{\max}$), unless otherwise specified.

In our implementation, we handle collisions on the CPU using Intel's Embree library \cite{wald2014embree}. The two phases with parallel loops are implemented on the GPU using CUDA.
All timing results are generated on a computer with an AMD Ryzen 5950X CPU, 64GB DDR3 RAM, and an NVIDIA RTX 4090 GPU.


\subsection{Large-Scale Tests}

In \autoref{fig:teaser} we present two large-scale test, showcasing our method's performance, scalability, and stability in scenarios involving a large number of complex collisions, including stacking and rest in contact.

The first one has 216 squishy balls with tentacles, totaling 48 million vertices and 151 million tetrahedra acting as force elements, dropped into a Utah teapot. Intermediate frames of this simulation are shown in \autoref{fig:SquishyBallsToTeapot}.

The second one includes more than 10 thousand deformable objects, 
totaling over 36 million vertices and 124 million tetrahedra, dropped into a box and piled on a platform, which is then suddenly removed, making the pile collectively fall onto the ground. The intermediate frames are shown in \autoref{fig:HybrridModelsDropping}.

As can be seen in our supplemental video, both of these simulations exhibit stable motion, quickly forming static piles, while maintaining rest-in-contact behavior with over 1 million active collisions.
VBD's parallelism and fast convergence resulted in an average computation time of 40 and 25 seconds per frame in these simulations, respectively.


\begin{figure}
\centering
\fbox{\includegraphics[width=\linewidth,trim=180 0 120 20, clip]{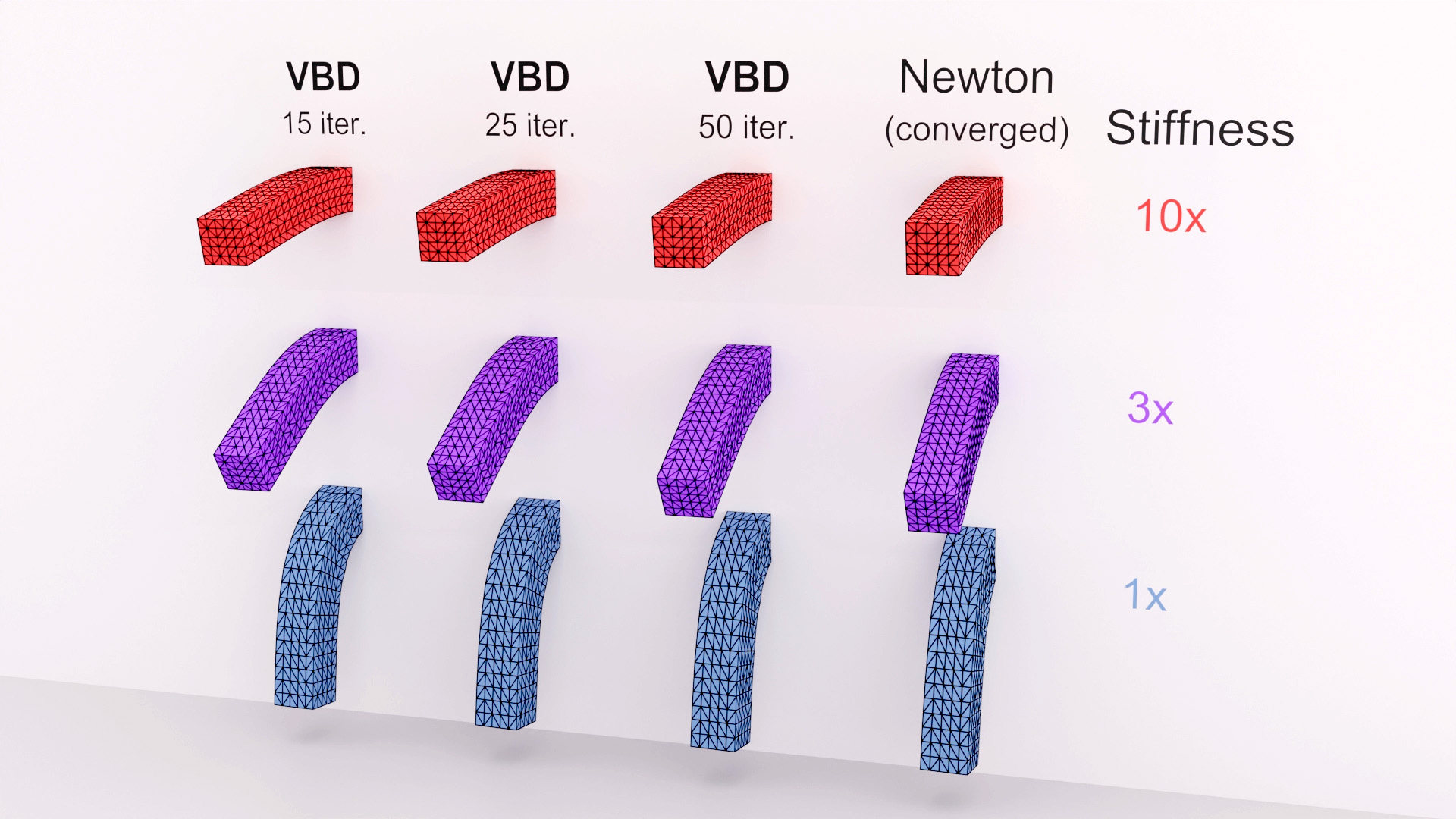}}
\caption{Visual convergence with different numbers of iterations per frame for different material stiffness (with accelerated iterations using ${\rho=0.75,0.86,0.93}$ top to bottom), simulating a beam with 463 vertices and 1.5 thousand tetrahedra.
}
\label{fig:ElasticityConvergence}
\end{figure}

\subsection{Unit Tests}

The convergence rate of VBD depends on the stiffness of the simulated system. This is demonstrated in \autoref{fig:ElasticityConvergence}, comparing VBD with different numbers of iterations per frame to the converged solution computed using Newton's method.
As expected, VBD converges slower for stiffer materials, which is common for descent-based solvers. In this example, 15 iterations are more than sufficient for the softest material, while stiffer ones require more. 
As can be seen in our supplemental video, even though VBD can qualitatively imitate the behavior of stiff materials with a relatively small number of iterations, the motion can quickly diverge from the converged solution, due to the remaining residual, unless a sufficient number of iterations are used.

\begin{figure}
\centering
\newcommand{\acap}[3]{\begin{minipage}{#1\linewidth}~\vspace{#2em}\\#3\end{minipage}}
\includegraphics[width=\linewidth,trim=0 0 150 200, clip]{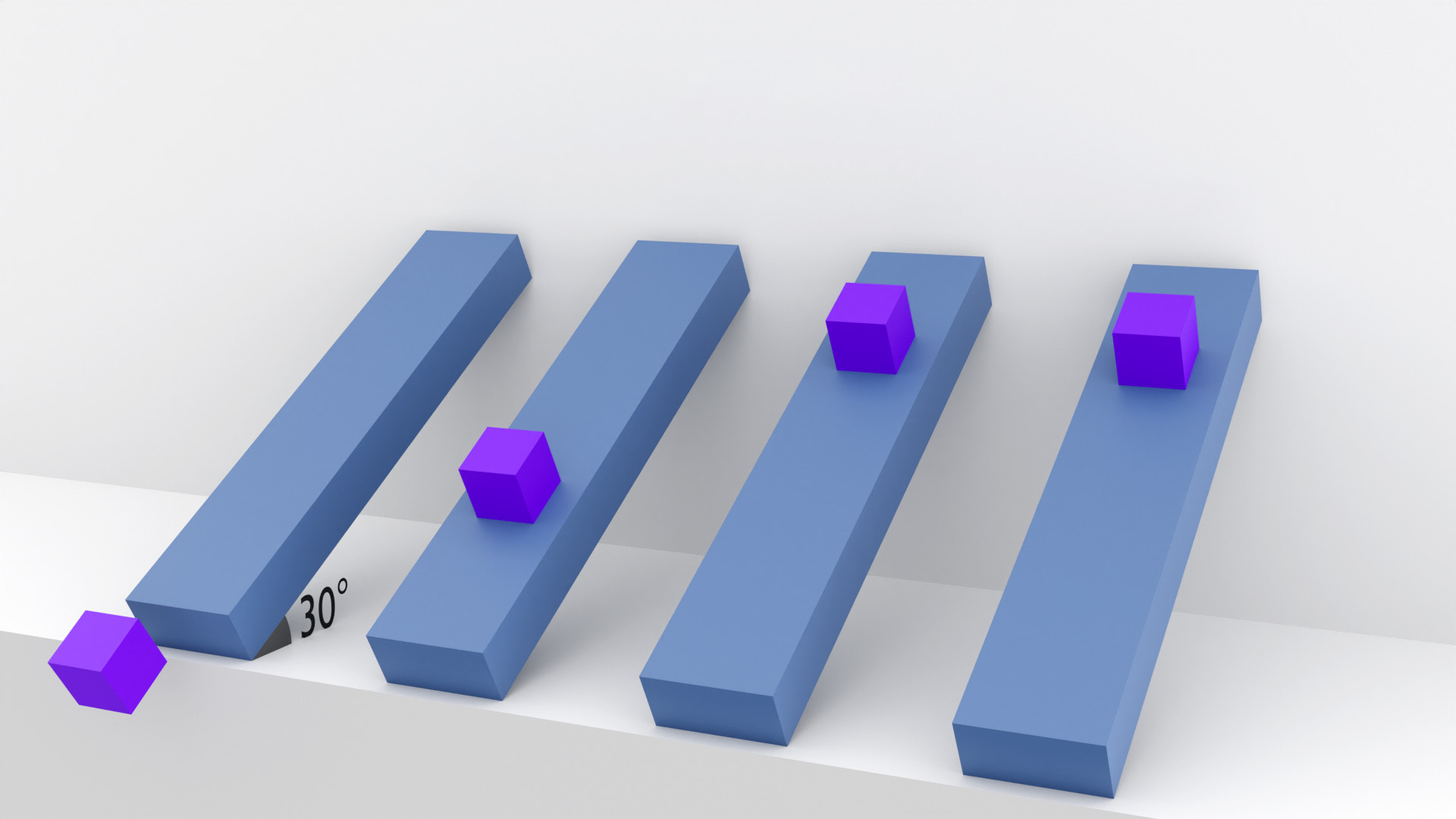}\vspace{-0.52\linewidth}\\
\hfill%
\acap{0.17}{0.0}{$\mu_c=0.0$}%
\acap{0.17}{0.2}{$\mu_c=0.3$}%
\acap{0.20}{0.4}{$\mu_c=0.6$}%
\acap{0.16}{0.8}{$\mu_c=0.9$}\vspace{0.44\linewidth}\\
\newcommand{\scap}[2]{\begin{minipage}{#2\linewidth}\centering{$\mu_c=#1$}\end{minipage}}
\newcommand{\figT}[1]{\includegraphics[width=0.25\linewidth,trim=580 0 680 260, clip]{Figures/FrictionUnitTest/#1-cc}}
\figT{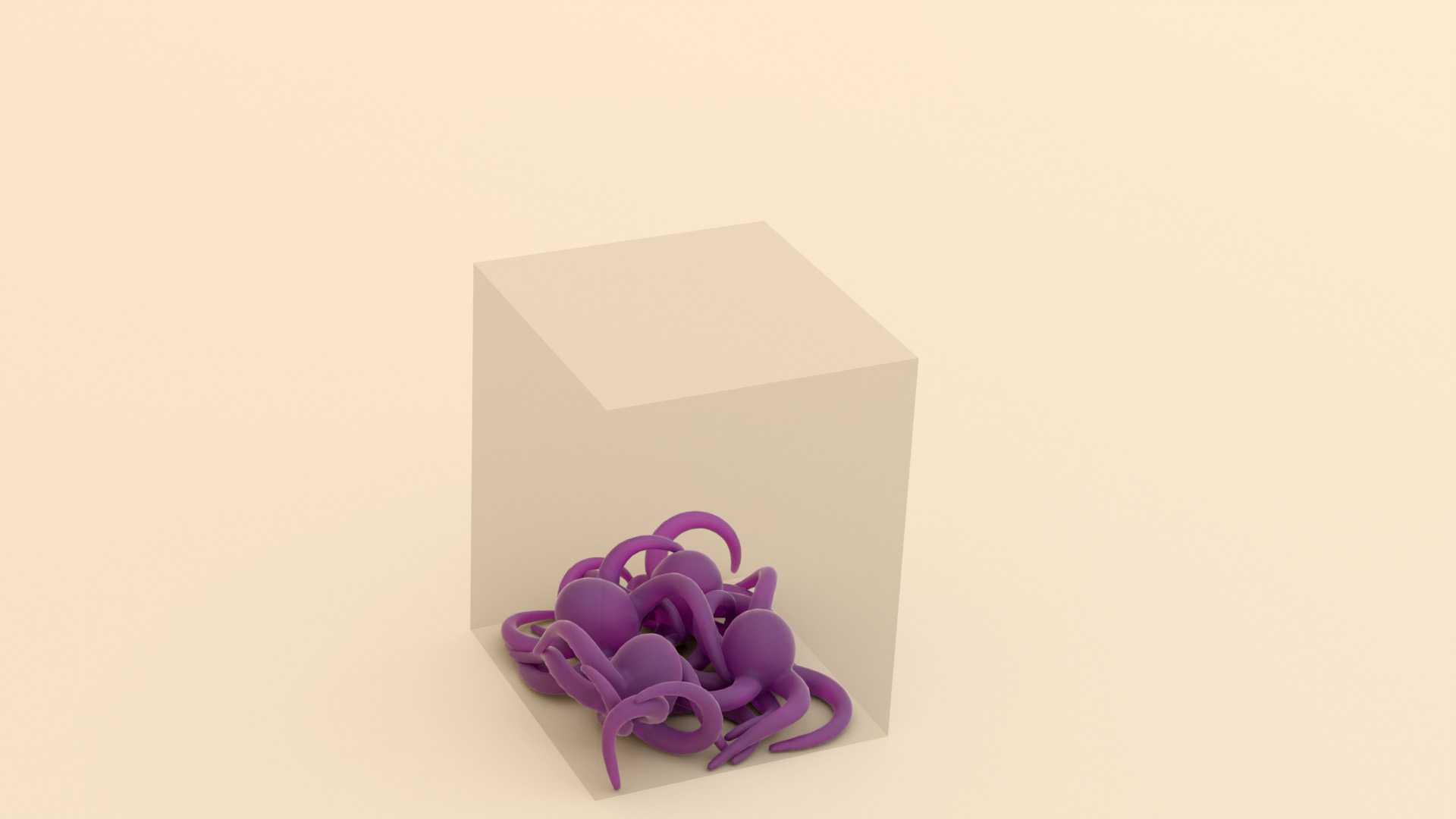}\hfill%
\figT{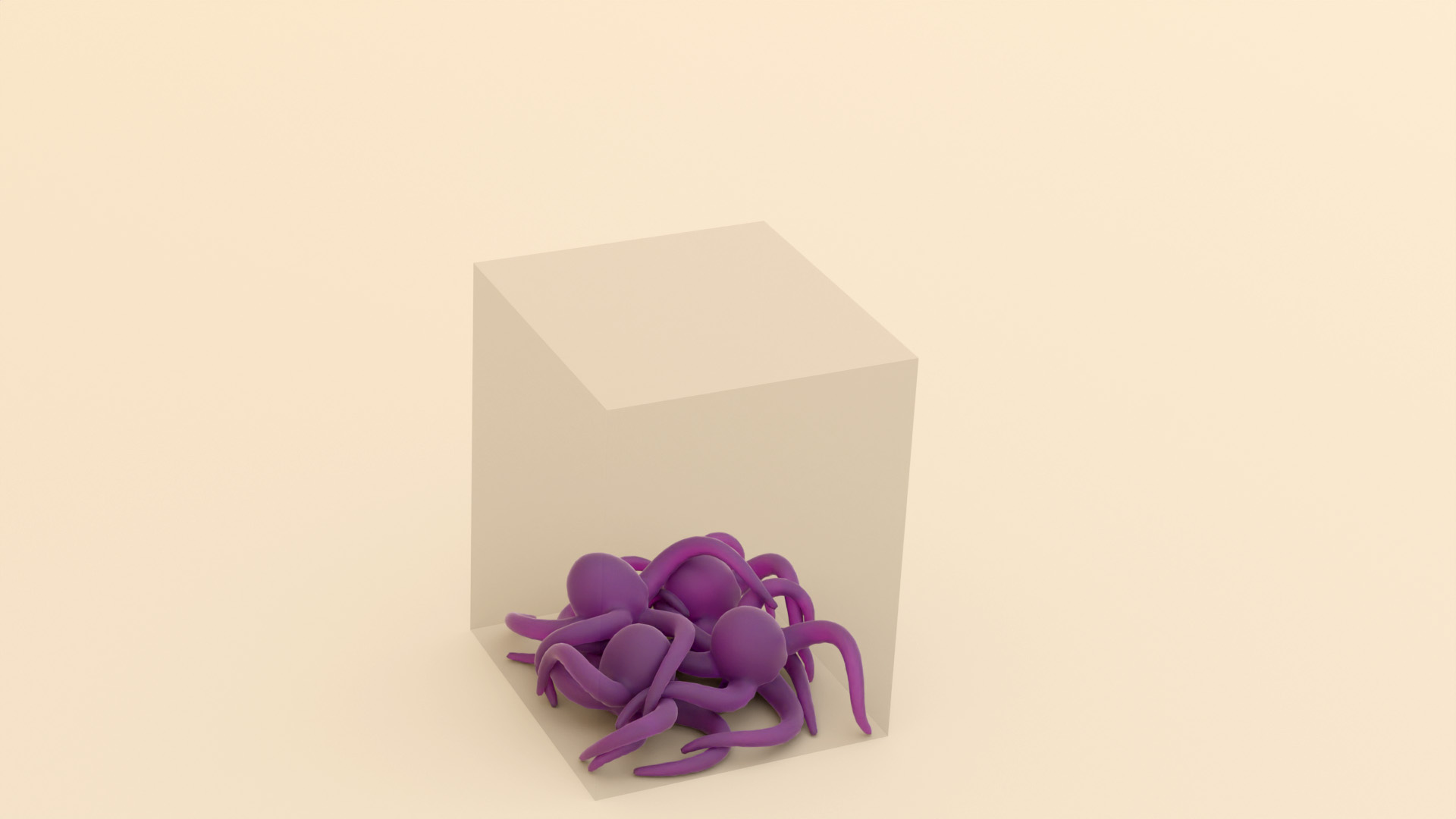}\hfill%
\figT{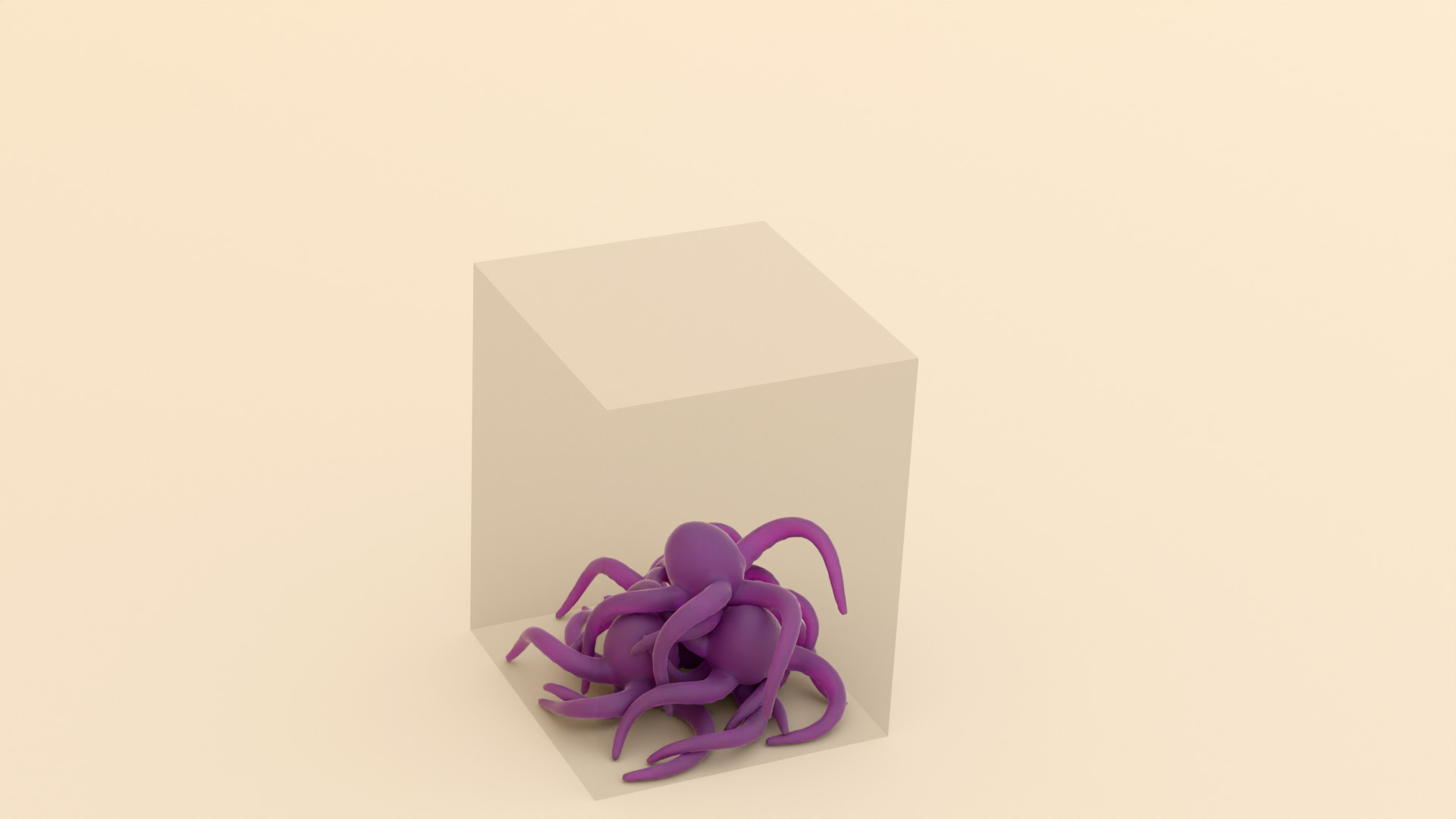}\hfill%
\figT{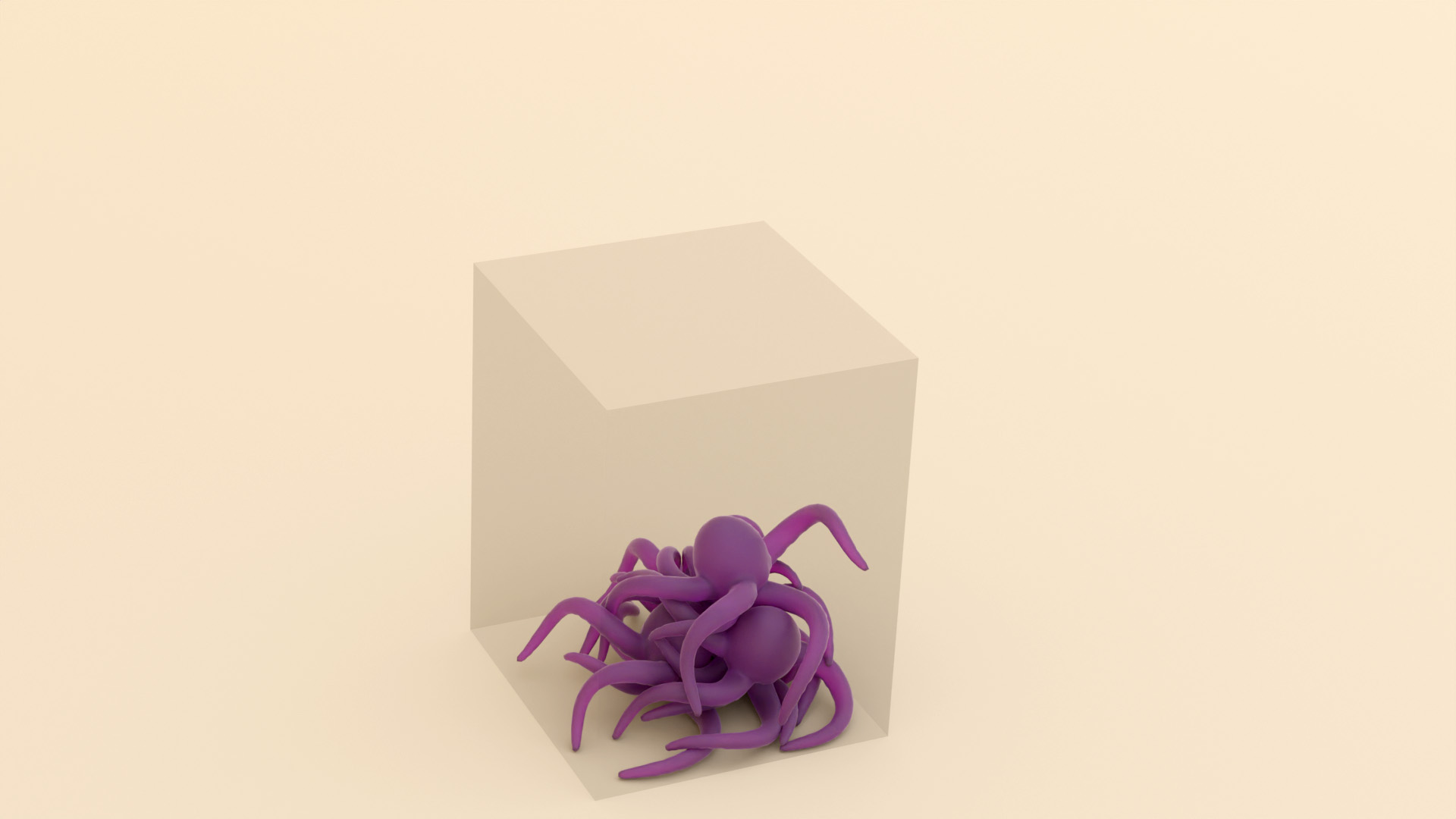}\vspace{-7.8em}
\scap{0.0}{0.25}%
\scap{0.2}{0.25}%
\scap{0.4}{0.25}%
\scap{0.6}{0.25}\vspace{6.4em}
\caption{Testing different friction coefficients $\mu_c$ for \sref{top}~an elastic cube with 400 vertices and 1.45 thousand tetrahedra, initially resting on an incline, and \sref{bottom}~4 elastic octopus models, totaling 15.6 thousand vertices and 60 thousand tetrahedra, dropped into a box.
}
\label{fig:FrictionUnitTest}
\end{figure}

We present our tests with different friction coefficients $\mu_c$ for friction forces in \autoref{fig:FrictionUnitTest}. Notice that, $\mu_c$ impacts the motion, as expected, and with sufficiently high $\mu_c$, we can properly preserve the position on an incline and form taller piles.

In our supplemental video, we also include a comparison of different damping stiffness $k_d$, showing that, despite numerical damping of implicit Euler, without damping we can preserve kinetic energy for a long time. As we increase $k_d$, the motion subsides quicker, as expected.

\subsection{Stress Tests}

We present a challenging frictional contact case in \autoref{fig:ExtremeTwist}, twisting two thin beams together. This example includes extreme deformations, generating strong material forces that compete with self-collisions and collisions between the two beams. It is simulated with collision detection occurring 
once every 5 iterations (i.e. $n_{\text{col}}=5$). Notice that VBD can properly handle such strong deformations with frictional contact.

We show two simulations in \autoref{fig:ExtremeInitialization} for stress-testing the stability of VBD under extreme deformations. The first one shows an armadillo model that is perfectly flattened and the second one is a Utah teapot model with all vertices randomly scrambled and placed on the surface of a sphere. Even though the simulations begin with these extremely unstable energy configurations, VBD quickly recovers these models without performing a line search and using 100 iterations per frame.


\begin{figure}
\centering
\newcommand{\img}[3][500 250 600 200]{\begin{subfigure}{0.327\linewidth}
\fbox{\includegraphics[width=\linewidth,trim=#1,clip]{Figures/ExtremeStretch/A0000#2-cc}}\vspace{-1.4em}\\
\small\hspace*{0.2em}\textbf{(#3)}\hspace*{0.2em}
\end{subfigure}}
\img{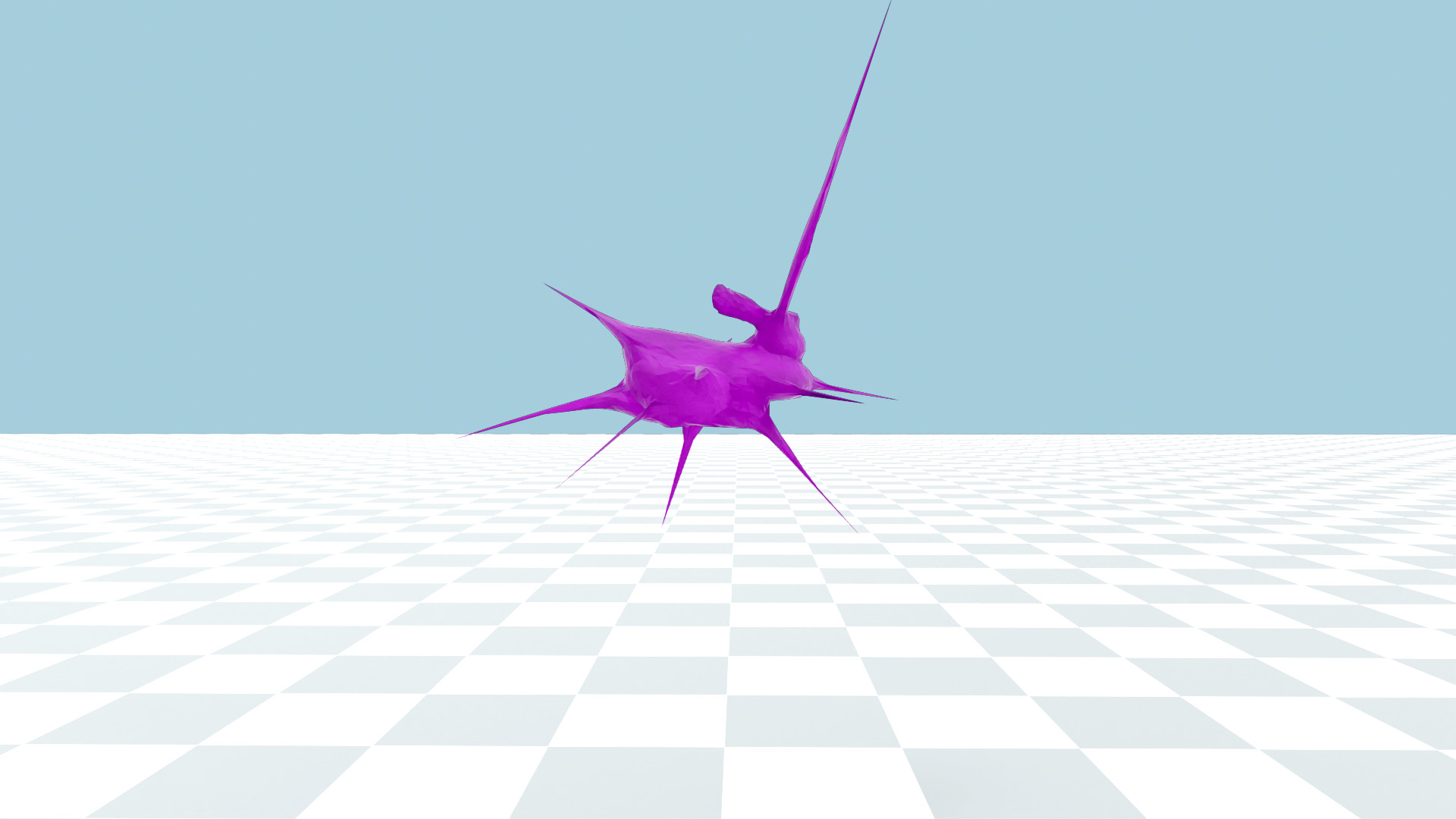}{a}\hfill%
\img{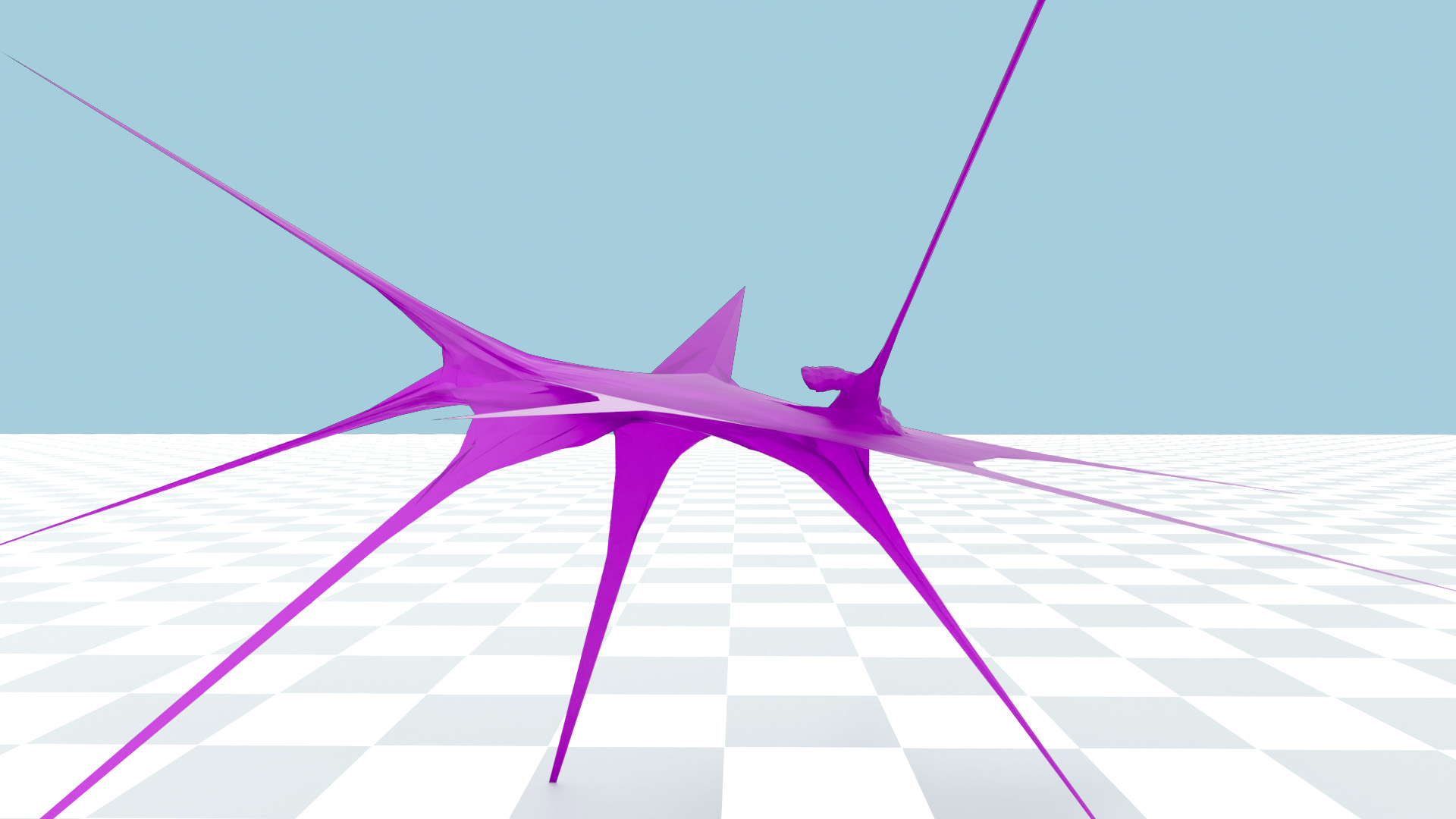}{b}\hfill%
\img{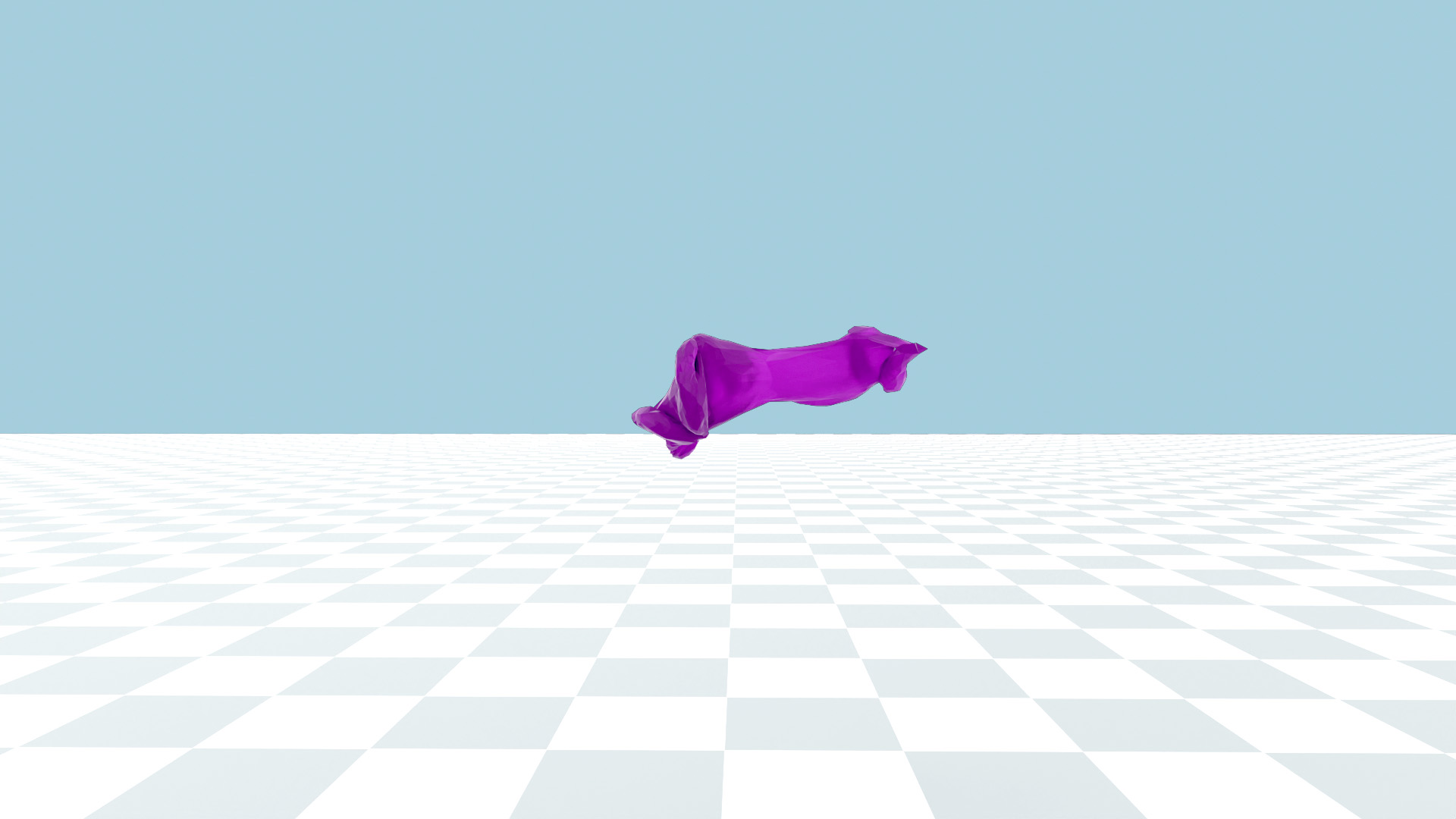}{c}\vspace{0.2em}\\
\img{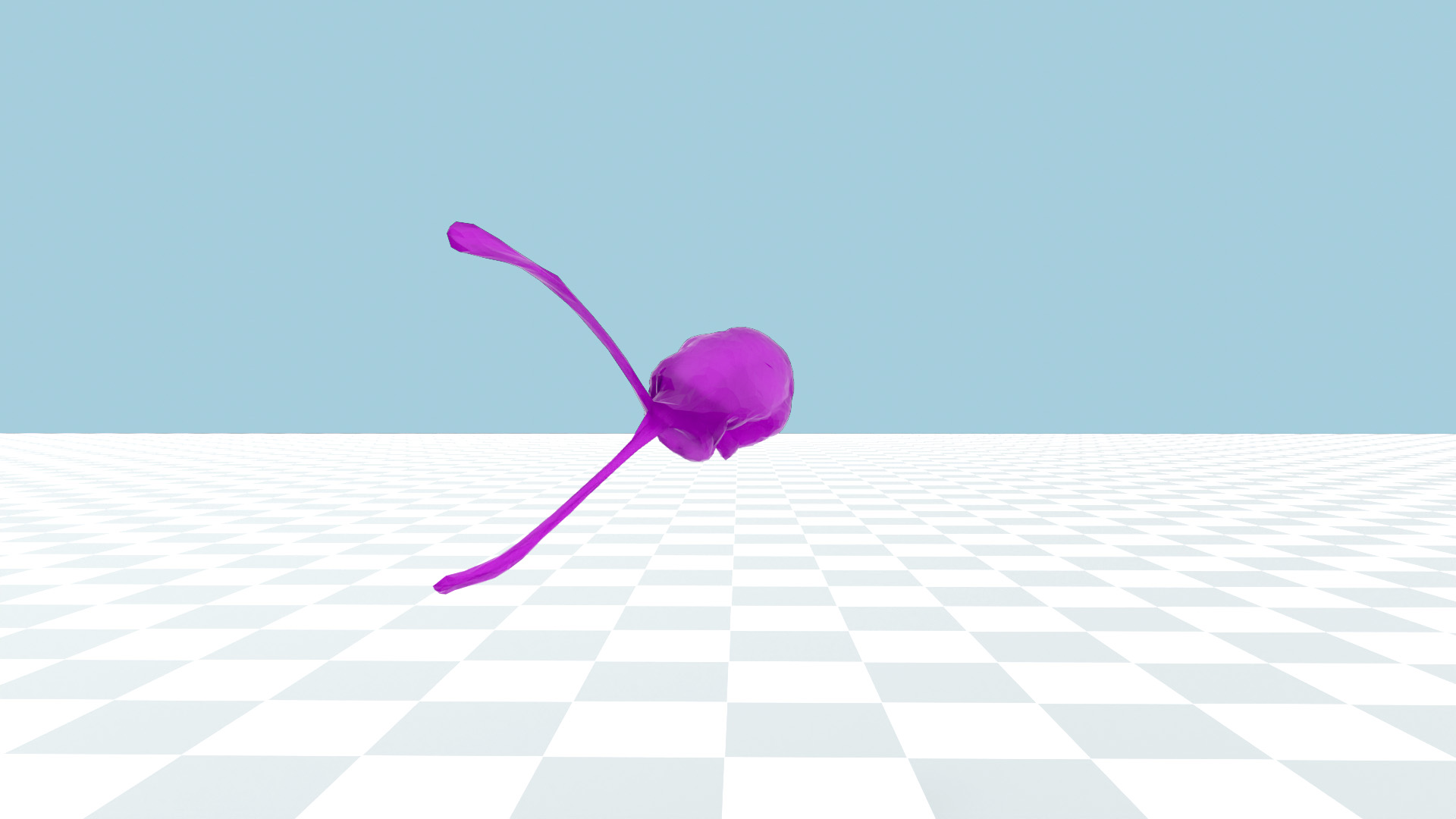}{d}\hfill%
\img[500 325 600 125]{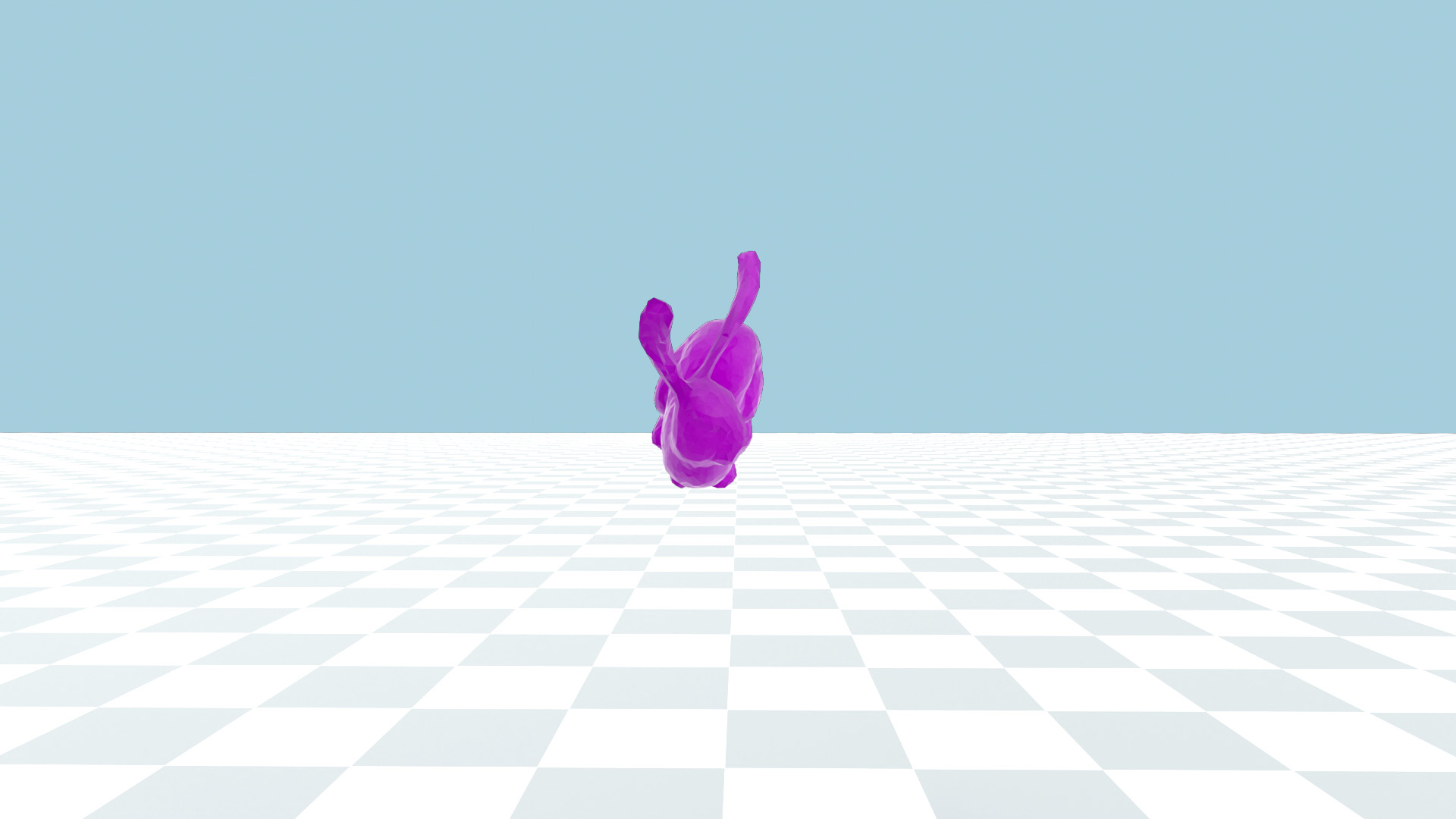}{e}\hfill%
\img[500 400 600 50]{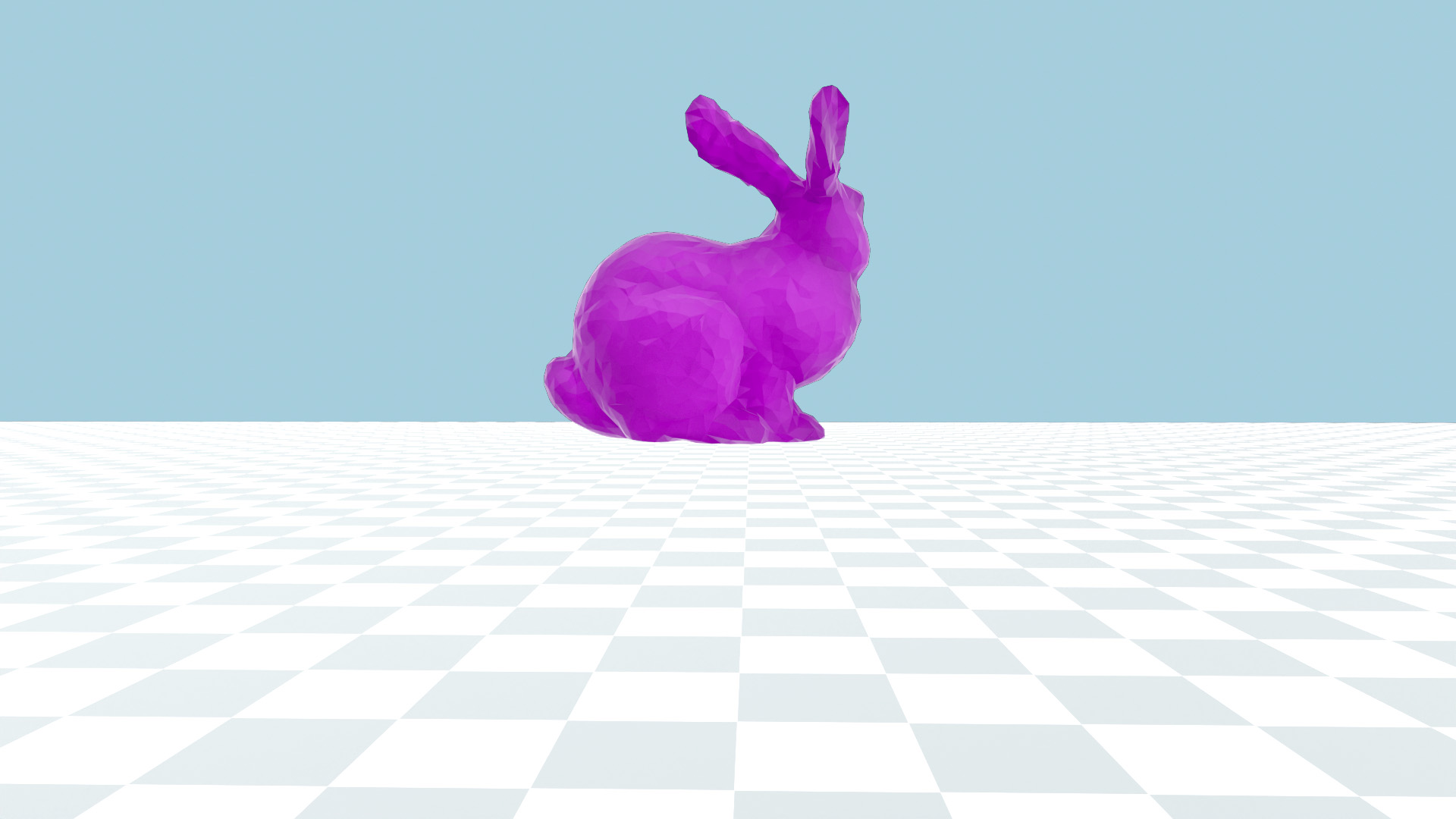}{f}\\
\caption{A stress test with extreme stretching: a Stanford bunny model with 1.8 thousand vertices and 5.9 thousand tetrahedra is stretched by slowly pulling 10 vertices away, which are then suddenly released. The model recovers its shape after going through considerable deformations and high-velocity motion, simulated with self-collisions and using $S=5$ substeps and $n_{\max}=10$ iterations per step.}
\label{fig:ExtremeStretch}
\end{figure}

Another stress test is shown in \autoref{fig:ExtremeStretch}. In this case, 10 vertices of a Stanford bunny model are first slowly pulled away, generating a state with considerable potential energy, and then suddenly released (right after \autoref{fig:ExtremeStretch}b), causing severe deformations. Once again, VBD successfully handles this challenging simulation case, involving self-collisions with high-velocity impacts, using only $n_{\max}=10$ iterations per step and $S=5$  per frame.
\begin{figure}
\centering
\newcommand{\img}[1]{\begin{subfigure}{0.33\linewidth}\centering
\fbox{\includegraphics[width=\linewidth,trim=420 220 800 100,clip]{Figures/StabilityUnderResidual/A00000#1-cc}}
\end{subfigure}}
\img{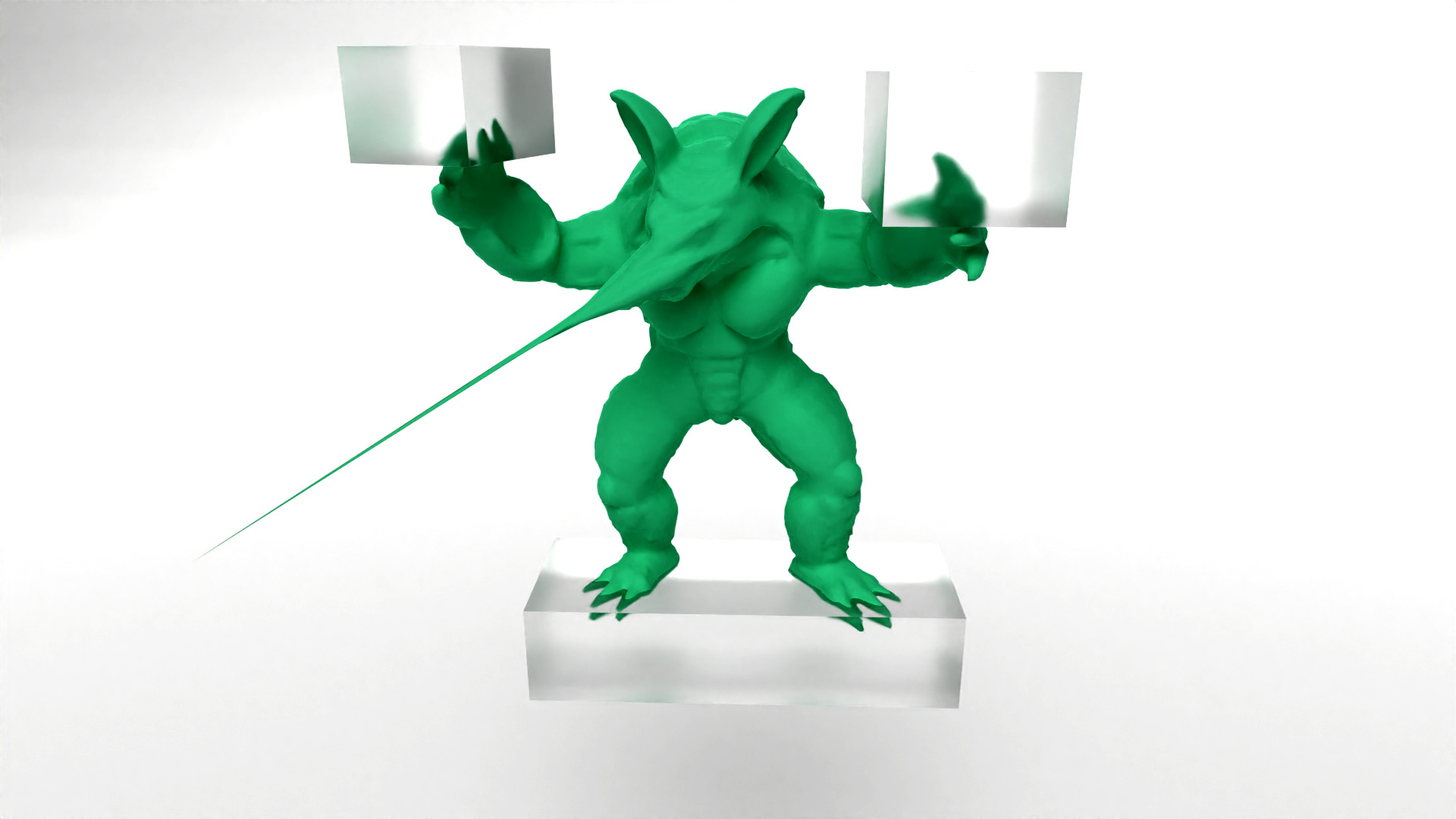}\hfill%
\img{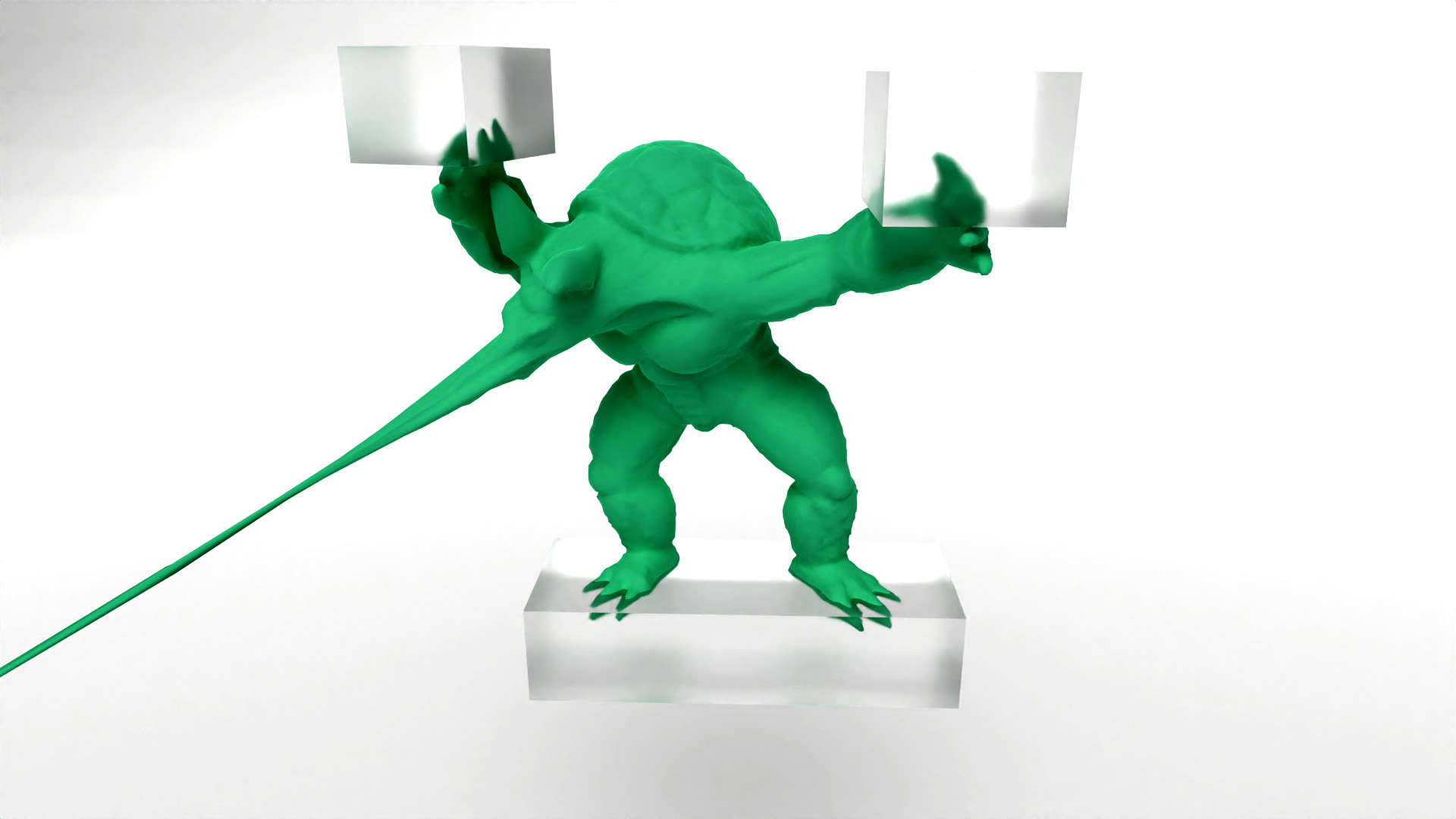}\hfill%
\img{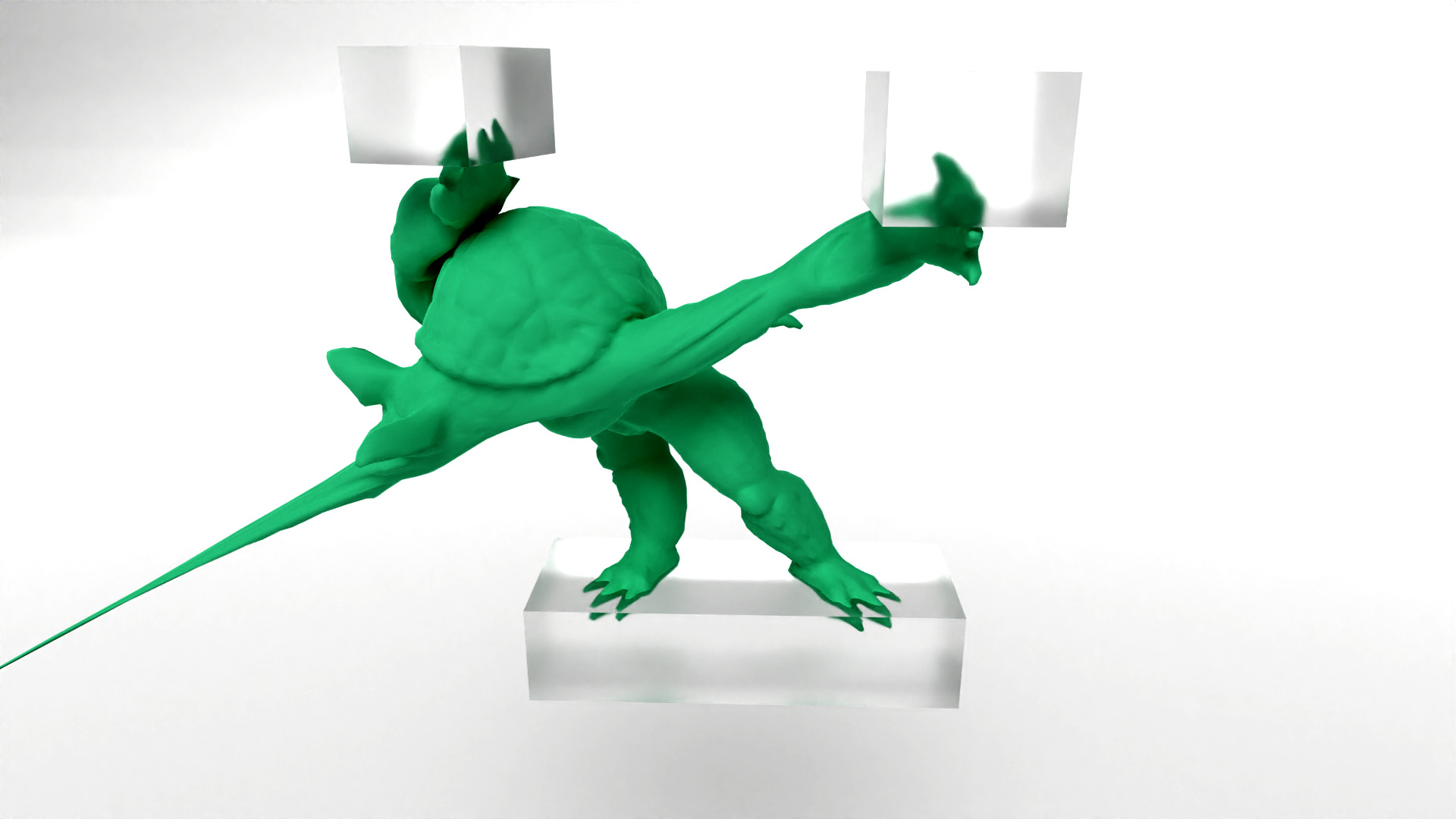}
\caption{A stress test using only a single iteration per frame (i.e. a time step of $h=1/60$ seconds and $n_{\max}=1$). One vertex on the armadillo model's nose is pulled while the finger and toe vertices are fixed. The model has 15 thousand vertices and 50 thousand tetrahedra.}
\label{fig:ExtremeStretch2}
\end{figure}
\autoref{fig:ExtremeStretch2} presents a stability test under large residuals by using only a single iteration per frame (i.e. $S=1$ and $n_{\max}=1$). Notice that our method produces stable deformations with extreme stretching, even though the simulation lacks a sufficient iteration count to properly reduce the residual at each frame.

\begin{figure*}
\centering
\newcommand{\cmpB}[1]{%
\belowbaseline[0.005\linewidth]{\fbox{\includegraphics[width=0.22\linewidth]{Figures/ConvergenceComparisonQuantative/combined_#1-cca}}}\hfill%
\belowbaseline[0pt]{\includegraphics[height=0.255\linewidth]{Figures/ConvergenceComparisonQuantative/energyByIter_armadillo2_#1}}%
\belowbaseline[0pt]{\includegraphics[height=0.255\linewidth]{Figures/ConvergenceComparisonQuantative/energyByTime_armadillo2_#1}}}
\cmpB{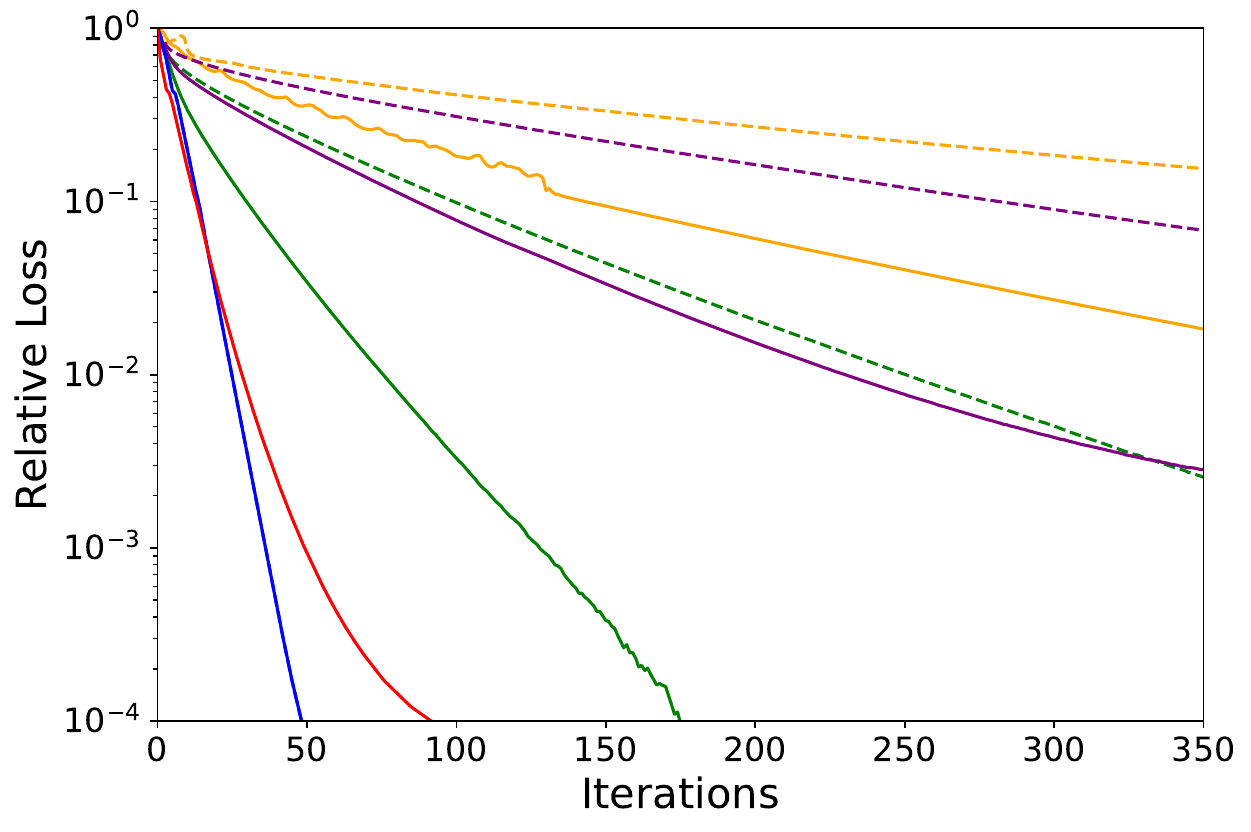}
\cmpB{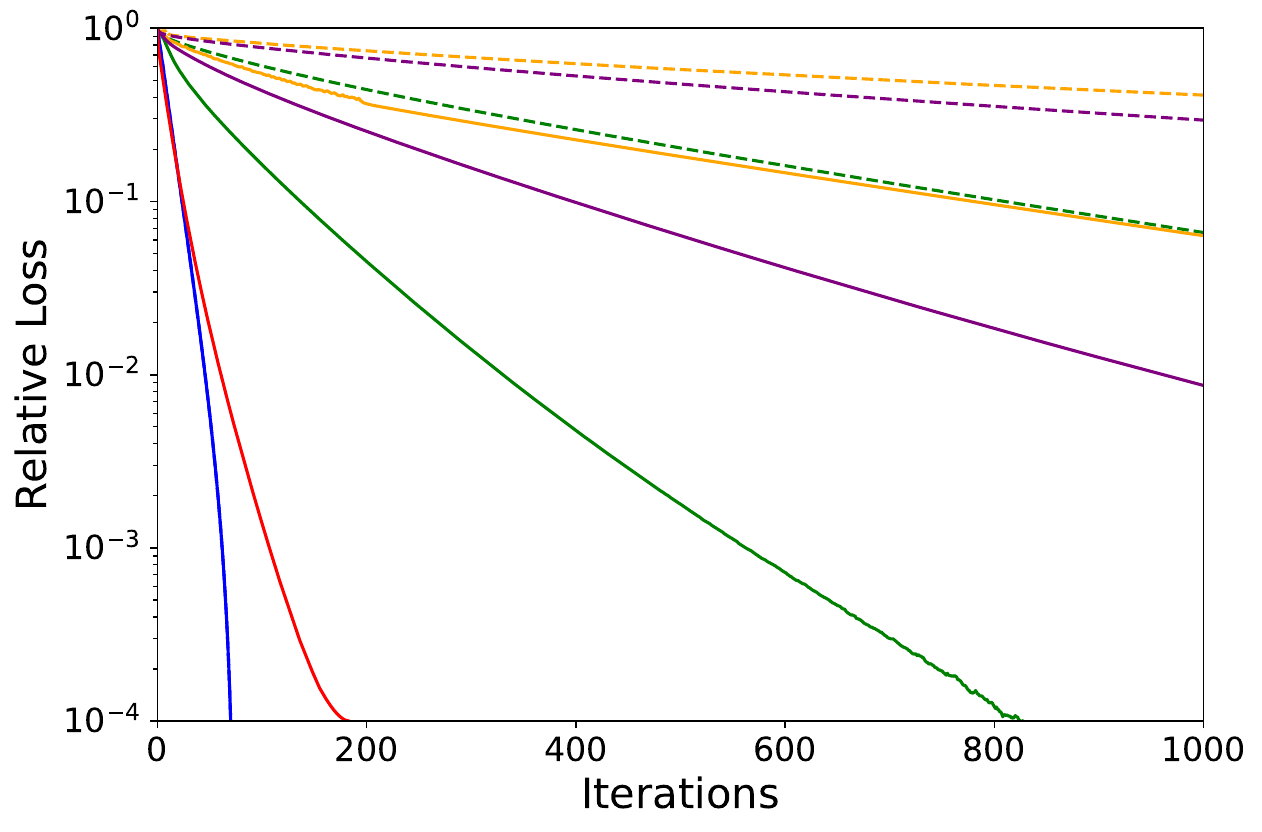}
\caption{Convergence of different descent methods for simulating an armadillo model with 15 thousand vertices and 50 thousand tetrahedra with \sref{top} a relatively soft material and \sref{bottom} a 10$\times$ stiffer material. Vertices near the top inside the glass block are fixed and the models are initially stretched, as shown on the left, by pulling down foot vertices. Then, the position constraints on foot vertices are suddenly removed, allowing the model to deform for 33 ms. The deformation is computed using a single time step of $h=33$ ms.
The graphs show relative loss over iterations and computation time. All methods are implemented on the GPU using the same framework with single precision (32-bit) floating-point numbers, except for Newton's method with Cholesky factorization, which runs on the CPU using double precision (64-bit).
Accelerated versions use $\rho=0.95$.}
\label{fig:ConvergenceComparisons}
\end{figure*}

\subsection{Convergence Rate}
To evaluate the convergence rate of VBD, we present a simple test shown in \autoref{fig:ConvergenceComparisons}, where an armadillo model that was previously stretched is suddenly released. Here, we calculate the relative loss after each iteration and compare it to alternative solvers.

The first alternative is preconditioned gradient descent (GD) \cite{wang_descent_2016} implemented on GPU within the same framework as ours. GD requires a form of line search for stability, which is implemented as testing the variational energy after every 8 iterations and, when needed, reducing the optimization step size and backtracking (following the implementation of \citet{wang_descent_2016}). We also include a version of GD that is accelerated using the Chebyshev semi-iterative approach \cite{wang_chebyshev_2015}, as our method. 
The iterations of GD are about \del{3$\times$}\add{30\%} faster than ours \add{without line search. However, it necessitates line search for stability. This makes it about 10\% slower than our VBD, which does not need line search.} \del{but}\add{Furthermore,} its 
convergence rate per iteration is considerably slower, as it corresponds to Jacobi iterations. In this example, when combined with acceleration, GD \del{outperforms}\add{performs similar to} our method without acceleration but lags significantly behind our method with acceleration.

We also compare to a version of our method that uses Jacobi iterations, called \emph{Block Jacobi}, implemented by computing the position change for all vertices in parallel and then applying the position update simultaneously to all vertices at the end of each iteration. We incorporate the same line search scheme as GD for Block Jacobi\add{, as it is necessary for stability}. Block Jacobi outperforms GD, as it uses vertex blocks for computation, which corresponds to using a diagonal Hessian block as a precondition, as opposed to just a diagonal line that GD uses. \del{It also}\add{Without line search, it} achieves \add{20\%} faster iteration times than \del{ours}\add{our VBD}, due to its perfect vertex-level parallelization (without any coloring). However, \add{combined with line search, its iterations are about 17\% slower than VBD. More importantly,} because it uses Jacobi iterations, its convergence is hindered, as compared to our Gauss-Seidel iterations.

Furthermore, we compare \del{our method's}\add{the} convergence \add{of our method} to \add{two implementations of} Newton's method\add{: first} using a direct Cholesky (LDLT) factorization solver provided by Intel's MKL library \cite{wang2014intel} \add{running on the CPU, and the second using a GPU-based conjugate gradient (CG) method.}
To ensure convergence, we do a PSD projection for each tet's Hessian of elasticity. Newton's method uses a line search for each iteration. Since its iterations are slow, the computational overhead of line search is negligible. Though the convergence of Newton's method per iteration is far superior to all others, because of its expensive computation time per iteration, in these examples it lags far behind. Nonetheless, for relatively tame experiments with less stretching and motion, and especially for highly stiff and high-resolution simulations that are much more expensive to simulate, we would expect Newton's method to eventually overtake all alternatives beyond a certain level of convergence.

\add{Finally, we compare our method to a quasi-Newton approach using Laplacian preconditioning \cite{liu_quasi-newton_2017}. We utilize a GPU-based conjugate gradient solver, similar to the one employed in our GPU-CG Newton's method's implementation. Laplacian preconditioning accelerates each linear solve, as it eliminates the need for PSD projection and involves solving a smaller system. Practically, this method demonstrates faster convergence than Newton's method, particularly in the initial stages of the optimization process. Nevertheless, it still lags behind both the accelerated and non-accelerated versions of our VBD.}
\begin{figure}
\centering
\includegraphics[width=0.85\linewidth]{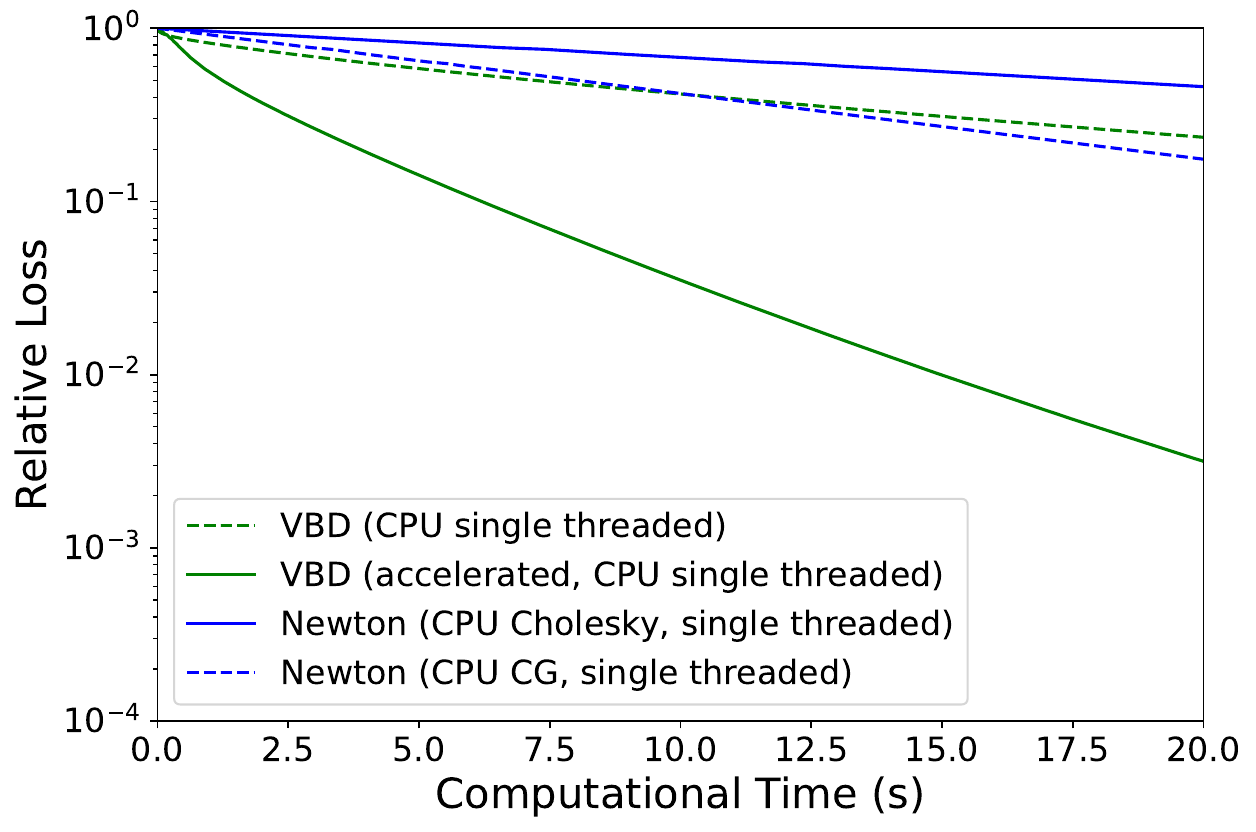}
\caption{\add{Comparing a single-threaded CPU implementation of our method with single-threaded Newton’s method (using both CG and Cholesky). The scene is identical to the bottom row of \autoref{fig:ConvergenceComparisons} .}}
\label{fig:singleThreaded}
\end{figure}
\add{A part of the performance advantages of our VBD method presented above is related to its efficiency in parallel execution on the GPU. To demonstrate its convergence in the absence of parallel computation, we include a comparison using single-threaded CPU implementations of our VBD and Newton's method with Cholesky factorization and CG. \autoref{fig:singleThreaded} shows the convergence results for the same experiment in the bottom row of \autoref{fig:ConvergenceComparisons}.
In these tests, 
our method initially demonstrates faster convergence than both versions of Newton's method. Over time, the CG-based Newton's method catches up to our VBD without Chebyshev acceleration. However, VBD with Chebyshev acceleration maintains a significant performance lead over the others. This experiment shows that the performance advantages of our method are not only due to its GPU parallelism.}

\begin{table*}[]
\caption{Performance results and simulation parameters.}
\resizebox{\linewidth}{!}{%
\begin{tabular}{|l||lll|lll|ll|ll|l|}
\hline
Experiment Name                & \multicolumn{3}{l|}{Number of}                                  & \multicolumn{3}{l|}{Material}                                                             & \multicolumn{2}{l|}{Contact \&Fiction}           & \multicolumn{2}{l|}{Simulation Parameters}  & Time per step                     \\
                               & \multicolumn{1}{l|}{Vert.} & \multicolumn{1}{l|}{Tet.}  & color & \multicolumn{1}{l|}{Type}       & \multicolumn{1}{l|}{Stiffness}                & Damping & \multicolumn{1}{l|}{$k_c$} & $\mu_c, \epsilon_v$ & \multicolumn{1}{l|}{$h$(sec.)} & Iterations & avg./max                           \\ \hline
Twisting Thin Beams (\autoref{fig:ExtremeTwist})           & \multicolumn{1}{l|}{97K}   & \multicolumn{1}{l|}{266K} & 8     & \multicolumn{1}{l|}{NeoHookean} & \multicolumn{1}{l|}{$\mu = 5e4, \lambda=1e6$} & 1e-6    & \multicolumn{1}{l|}{1e6}   & 0.1, 1e-2           & \multicolumn{1}{l|}{1/300}     & 100        & \del{120/156ms} \add{60/78ms}\\

Flattening initialization  (\autoref{fig:ExtremeInitialization})    & \multicolumn{1}{l|}{15K}   & \multicolumn{1}{l|}{50K}   & 8     & \multicolumn{1}{l|}{NeoHookean} & \multicolumn{1}{l|}{$\mu=2e6,  \lambda=1e7$}  & 1e-6    & \multicolumn{1}{l|}{NA}    & NA                  & \multicolumn{1}{l|}{1/60}      & 100        & \del{6.6/7.6ms}
\add{3.3/3.8ms}\\

Random initialization  (\autoref{fig:ExtremeInitialization}) & \multicolumn{1}{l|}{2K}    & \multicolumn{1}{l|}{8.5K}  & 8     & \multicolumn{1}{l|}{NeoHookean} & \multicolumn{1}{l|}{$\mu=2e6,  \lambda=1e7$}  & 1e-6    & \multicolumn{1}{l|}{NA}    & NA                  & \multicolumn{1}{l|}{1/60}      & 100        & \del{5.1/5.6ms}
\add{2.6/2.8ms}\\

Tetmesh Pendulum  (\autoref{fig:AdaptiveInitialization}) & \multicolumn{1}{l|}{304}   & \multicolumn{1}{l|}{755}   & 6     & \multicolumn{1}{l|}{NeoHookean} & \multicolumn{1}{l|}{$\mu=1e7,  \lambda=1e8$}  & 0       & \multicolumn{1}{l|}{NA}    & NA                  & \multicolumn{1}{l|}{1/300}     & 20         & <0.1ms \\

Squishy Ball Drops (Figure.\ref{fig:AccelerationComparison},\ref{fig:CollisionResolutionScheme},\ref{fig:CompareToXPBD_TimeStep}) & \multicolumn{1}{l|}{230K}  & \multicolumn{1}{l|}{700K}  & 8     & \multicolumn{1}{l|}{NeoHookean} & \multicolumn{1}{l|}{$\mu=2e6,  \lambda=2e7$}  & 1e-7    & \multicolumn{1}{l|}{1e7}   & 0.1, 1e-2           & \multicolumn{1}{l|}{1/120}     & 120        & \del{30/34ms}
\add{15/17ms}\\

Tearing Cloth (\autoref{fig:TopologicalChange}) & \multicolumn{1}{l|}{2500}  & \multicolumn{1}{l|}{4800}  & 3     & \multicolumn{1}{l|}{StVK}       & \multicolumn{1}{l|}{$\mu=1e4,  \lambda=1e4$}  & 1e-5    & \multicolumn{1}{l|}{NA}    & NA                  & \multicolumn{1}{l|}{1/300}     & 20         & 11.2/11.5 ms(cpu)                 \\

Dropping 216 Squshy Balls (\autoref{fig:SquishyBallsToTeapot}) & \multicolumn{1}{l|}{48M}   & \multicolumn{1}{l|}{151M}  & 9     & \multicolumn{1}{l|}{NeoHookean} & \multicolumn{1}{l|}{$\mu=2e6,  \lambda=2e7$}  & 1e-7    & \multicolumn{1}{l|}{1e7}   & 0.1, 1e-2           & \multicolumn{1}{l|}{1/240}     & 40         & \del{5.3/7.9s}
\add{3.6/3.9s}\\

Dropping 10368 Models (\autoref{fig:HybrridModelsDropping}) & \multicolumn{1}{l|}{36M}   & \multicolumn{1}{l|}{124M}  & 8     & \multicolumn{1}{l|}{NeoHookean} & \multicolumn{1}{l|}{$\mu=1e6,  \lambda=1e7$}  & 1e-7    & \multicolumn{1}{l|}{1e7}   & 0.1, 1e-2           & \multicolumn{1}{l|}{1/120}     & 60         & \del{6.3/7.7s}
\add{4.2/4.7s}\\

Beam Sagging  (\autoref{fig:ElasticityConvergence}) & \multicolumn{1}{l|}{463}   & \multicolumn{1}{l|}{1.5K}  & 6     & \multicolumn{1}{l|}{NeoHookean} & \multicolumn{1}{l|}{$\mu=1e6/3e6/1e7$}  & 1e-6    & \multicolumn{1}{l|}{NA}   & NA& \multicolumn{1}{l|}{1/300}     & 3/5/10         & avg.: \del{0.15/0.30/0.54ms}
\add{0.08/0.12/0.24ms}\\

 & \multicolumn{1}{l|}{} & \multicolumn{1}{l|}{}  &      & \multicolumn{1}{l|}{} & \multicolumn{1}{l|}{$\lambda=1e7/3e7/1e8$}  &    & \multicolumn{1}{l|}{}   & NA& \multicolumn{1}{l|}{}     &         & max: \del{0.16/0.32/0.63ms}
\add{0.08/0.16/0.31ms}\\

Cude Sliding  (\autoref{fig:FrictionUnitTest}) & \multicolumn{1}{l|}{800}   & \multicolumn{1}{l|}{2.9K}  & 6     & \multicolumn{1}{l|}{NeoHookean} & \multicolumn{1}{l|}{$\mu=1e6,  \lambda=1e7$}  & 1e-6    & \multicolumn{1}{l|}{1e7}   & 0/0.3/0.6/0.9, 1e-2 & \multicolumn{1}{l|}{1/300}     & 10         & \del{0.20/0.31ms}
\add{0.10/0.17ms}\\

Octopi Stacking (\autoref{fig:FrictionUnitTest}) & \multicolumn{1}{l|}{15.6K} & \multicolumn{1}{l|}{60K}   & 8     & \multicolumn{1}{l|}{NeoHookean} & \multicolumn{1}{l|}{$\mu=1e6,  \lambda=1e7$}  & 1e-6    & \multicolumn{1}{l|}{1e7}   & 0/0.1/0.4/0.6, 1e-2 & \multicolumn{1}{l|}{1/300}     & 10         & \del{1.6/2.0ms}
\add{1.1/1.3ms}\\


Extreme Stretch  (\autoref{fig:ExtremeStretch2}) & \multicolumn{1}{l|}{1.8K}  & \multicolumn{1}{l|}{5.9K}  & 8     & \multicolumn{1}{l|}{NeoHookean} & \multicolumn{1}{l|}{$\mu=2e6,  \lambda=1e7$}  & 1e-6    & \multicolumn{1}{l|}{1e7}   & 0.2, 1e-2           & \multicolumn{1}{l|}{1/300}     & 10         & \del{0.86/1.02ms}
\add{0.36/1.02ms}\\

2 Cube colliding (\autoref{fig:CompareToXPBD_MassRatio}) & \multicolumn{1}{l|}{800}   & \multicolumn{1}{l|}{2.9K}  & 6     & \multicolumn{1}{l|}{NeoHookean} & \multicolumn{1}{l|}{$\mu=1e6,  \lambda=1e7$}  & 1e-6    & \multicolumn{1}{l|}{1e7}   & 0.3, 1e-2           & \multicolumn{1}{l|}{1/300}     & 10         & \del{0.21/0.27ms}
\add{0.16/0/20ms}\\

\hline
\end{tabular}
}
\end{table*}

\subsection{Comparisons to XPBD}

Our method has an entirely different formulation than XPBD, but there are some strong similarities, as both methods operate with position updates using Gauss-Seidel iterations. Here we provide two direct comparisons to highlight some important differences.

\begin{figure}
\centering
\newcommand{\img}[7]{\begin{subfigure}{0.245\linewidth}\centering
\includegraphics[width=\linewidth,trim=500 200 500 0,clip]{Figures/CompareToXPBD_TimeStep/#11-cc}\\
\includegraphics[width=\linewidth,trim=400 100 400 100,clip]{Figures/CompareToXPBD_TimeStep/#12-cc}
\caption{#3\\$h=1/#4$ sec.\\$n_{\max}=#5$\\$n_{\text{col}}=#6$\\\emph{($#7$ sec./frame)}}\label{fig:CompareToXPBD_TimeStep:#2}
\end{subfigure}}
\img{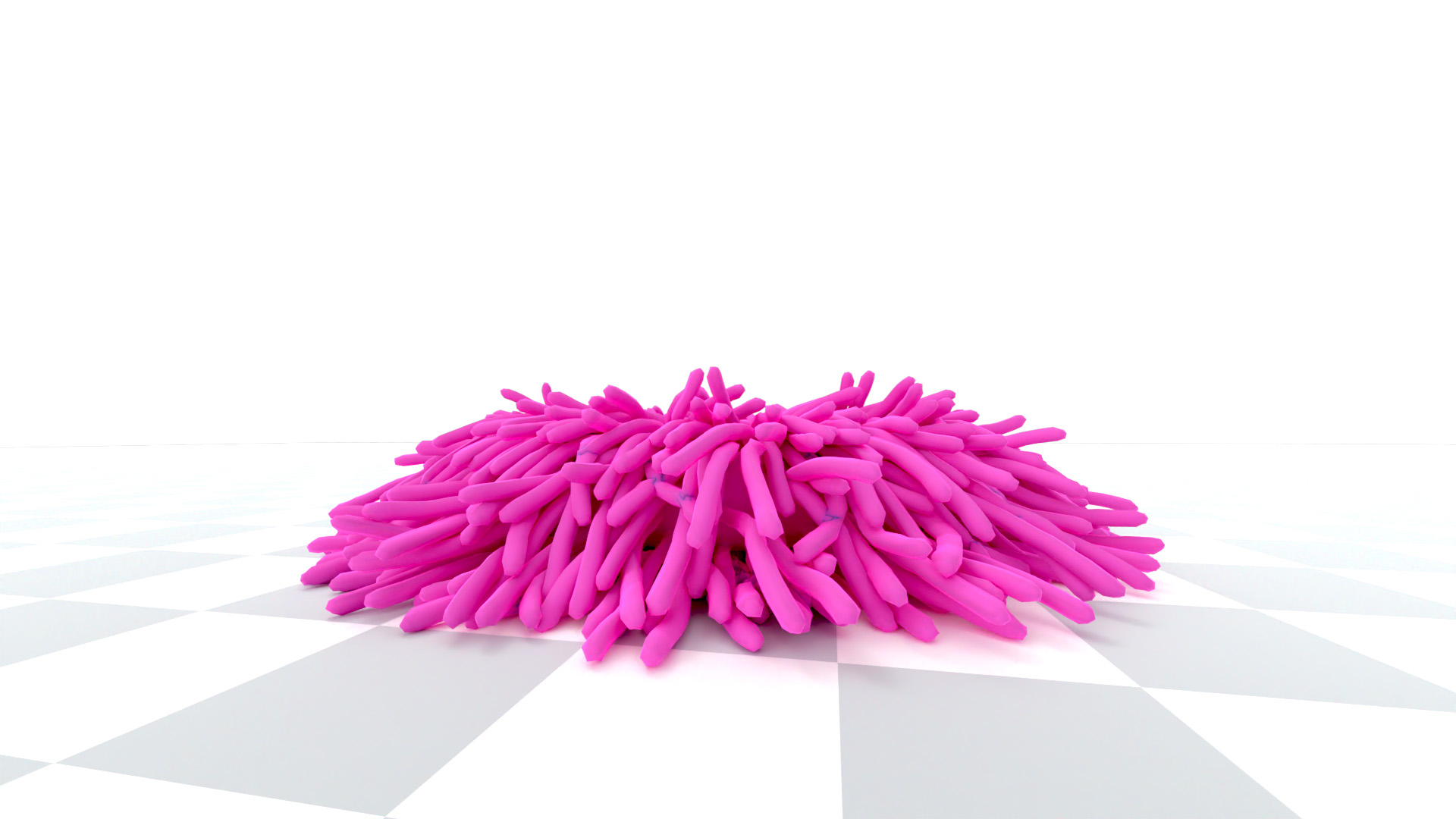}{a}{XPBD}{120}{120}{120}{0.32}\hfill%
\img{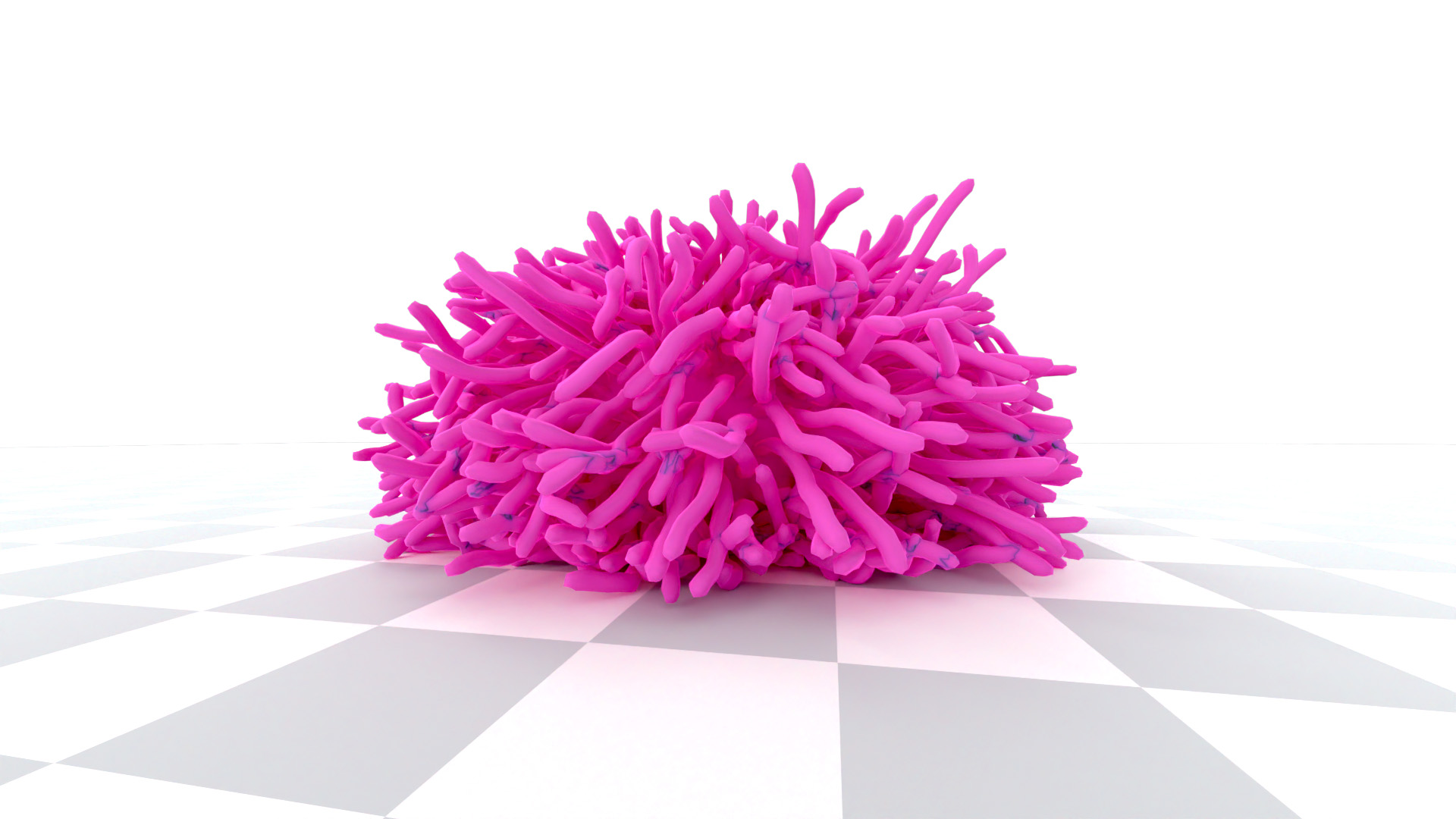}{b}{XPBD}{3000}{5}{125}{0.35}\hfill
\img{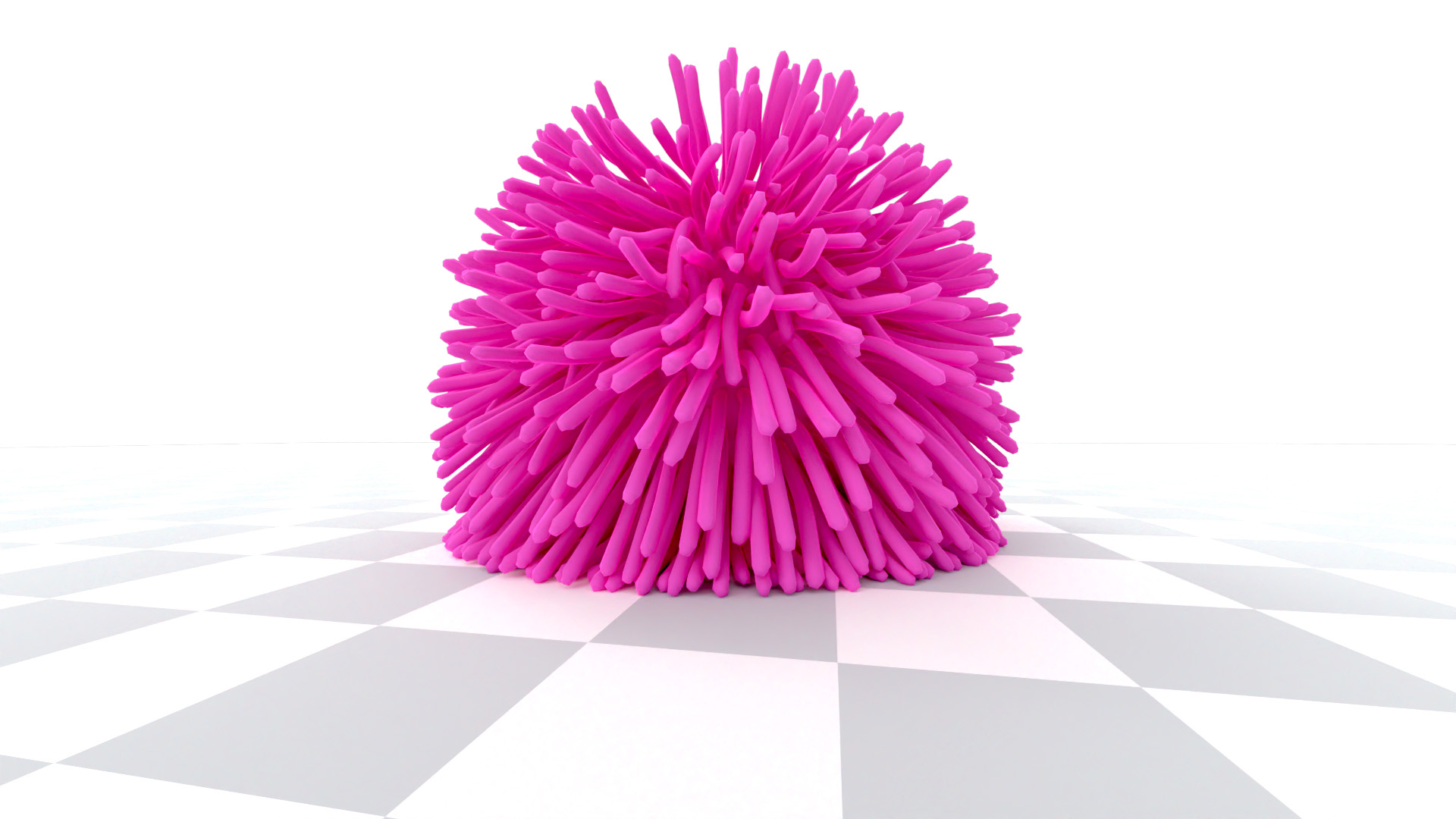}{c}{XPBD}{3000}{5}{5}{3.1}\hfill%
\img{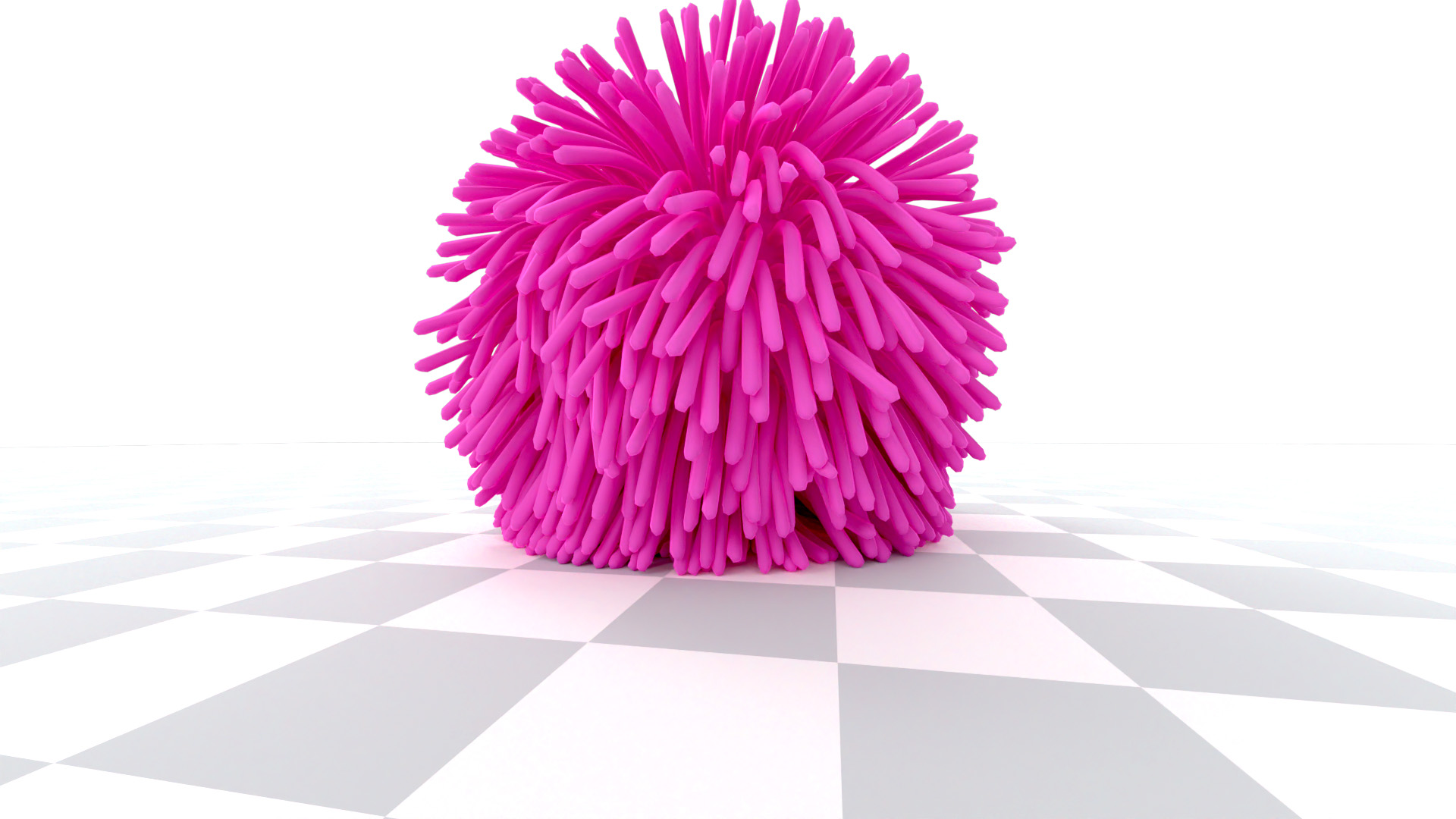}{d}{\textbf{VBD (ours)}}{120}{120}{120}{0.031}
\caption{A squishy ball with tentacles, comprising 230 thousand vertices and 700 thousand tetrahedra, dropped on the ground, simulated using \sref{a}~XPBD with a large time step and 240 iterations per frame, \sref{b}~XPBD with a $25\times$ smaller time step and 250 total iterations per frame, \sref{c}~XPBD with the same small time step and iteration count but with $25\times$ more frequent collision detection, and \sref{d}~VBD with a large time step and 240 iterations per frame. Comparing \sref{a} and \sref{d}, VBD is faster than XPBD with the same settings. XPBD's solution approaches VBD as the time step decreases, but it also requires more frequent collision detection to achieve a visually similar result to VBD.
}
\label{fig:CompareToXPBD_TimeStep}
\end{figure}

XPBD replaces the Hessian matrix and uses only the Hessian of inertia potential. This omission is justified by using a small time step, because the significance of the inertia potential increases quadratically as the time step decreases. 
Nonetheless, with complex examples, the impact of this approximation can be severe, even with small time step.
This is demonstrated in \autoref{fig:CompareToXPBD_TimeStep} with a challenging collision-rich scenario involving a squishy ball with tentacles dropped to the ground. Comparing XPBD with 120 iterations per step (\autoref{fig:CompareToXPBD_TimeStep:a}) to our method with the same number of iterations (\autoref{fig:CompareToXPBD_TimeStep:d}), we can see that our method not only achieves a more stable animation, it also performs faster because of its improved parallelism, as compared to XPBD. Reducing the time step helps XPBD even when using a similar total number of iterations per frame (\autoref{fig:CompareToXPBD_TimeStep:b}). However, simply reducing the time step is not sufficient in this case, as XPBD also needs more frequent collision detection (\autoref{fig:CompareToXPBD_TimeStep:c}). Using collision detection with the same frequency as ours while taking small time step (\autoref{fig:CompareToXPBD_TimeStep:b}) leads to collisions that are detected too late and cause stability issues in this case. This is not only because the collisions that are detected too late are deeper, but also because smaller time step lead to higher vertex velocities when resolving stiff collisions.


\begin{figure}
\centering
\newcommand{\figA}[3]{\begin{subfigure}{0.32\linewidth}\centering
\fbox{\includegraphics[width=\linewidth,trim=450 300 550 100,clip]{Figures/CompareToXPBD_MassRatio/#1}}
\caption{#2}\label{fig:CompareToXPBD_MassRatio:#3}
\end{subfigure}}
\figA{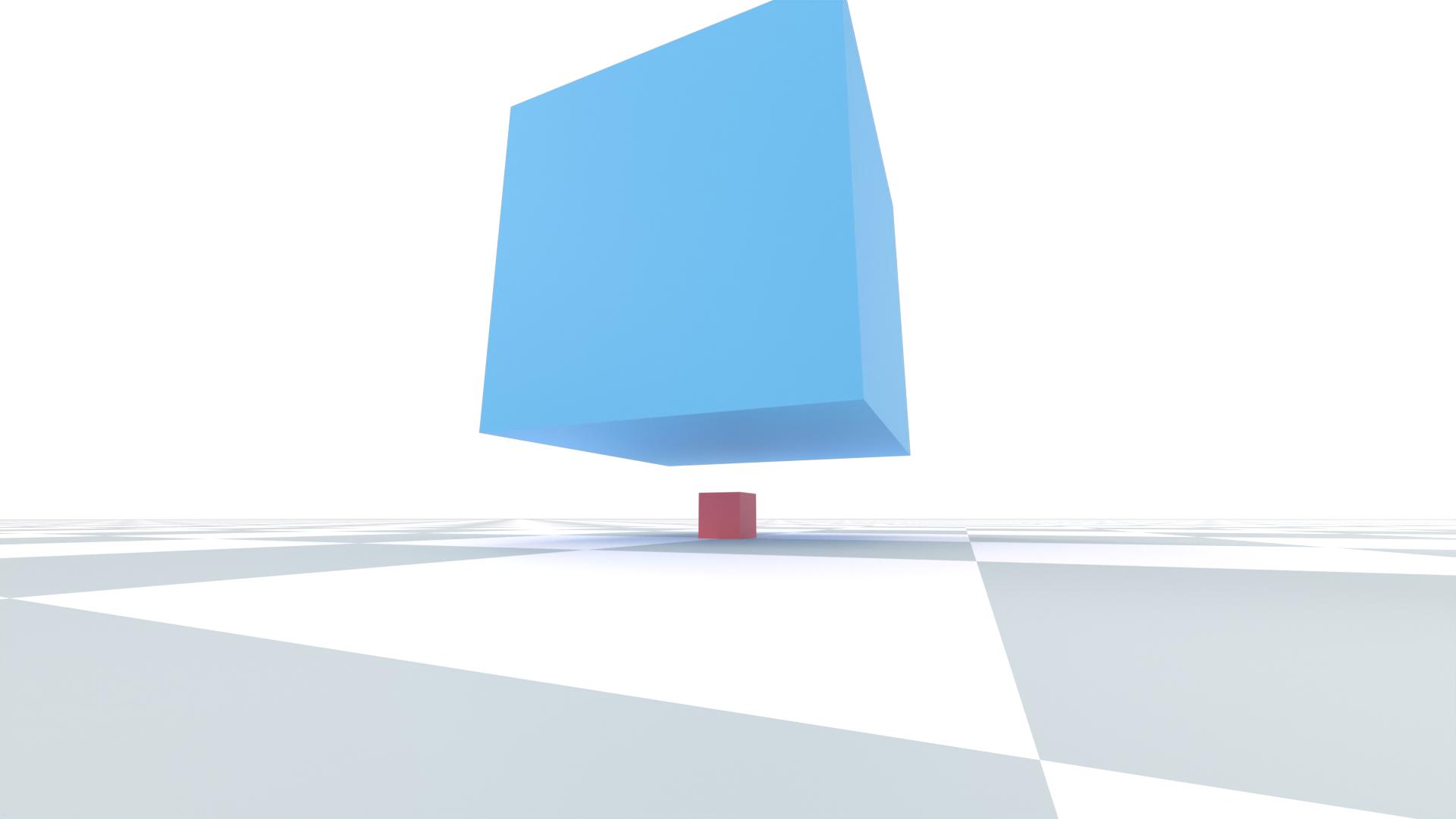}{Initial State}{a} \hfill%
\figA{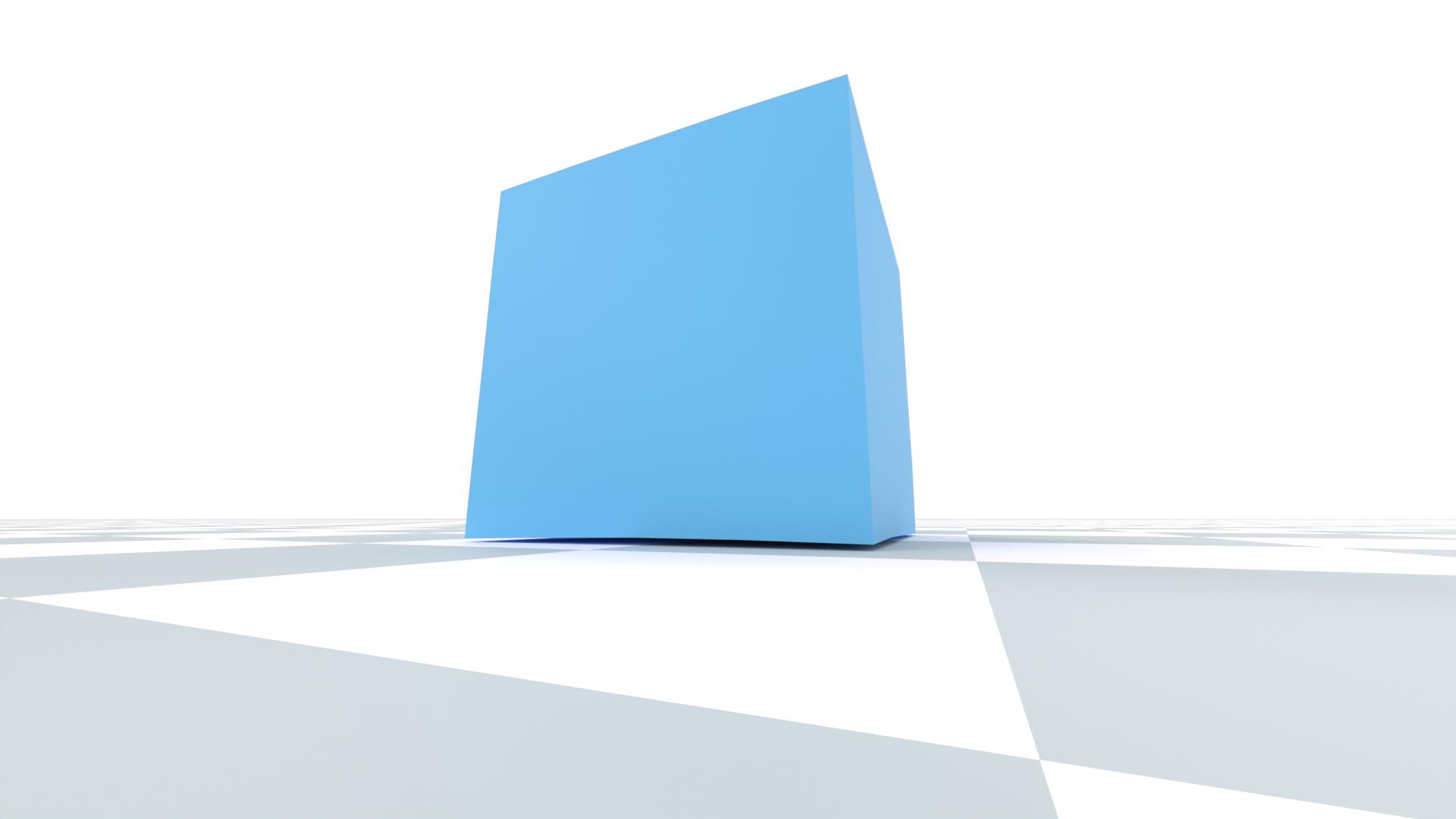}{XPBD}{b} \hfill%
\figA{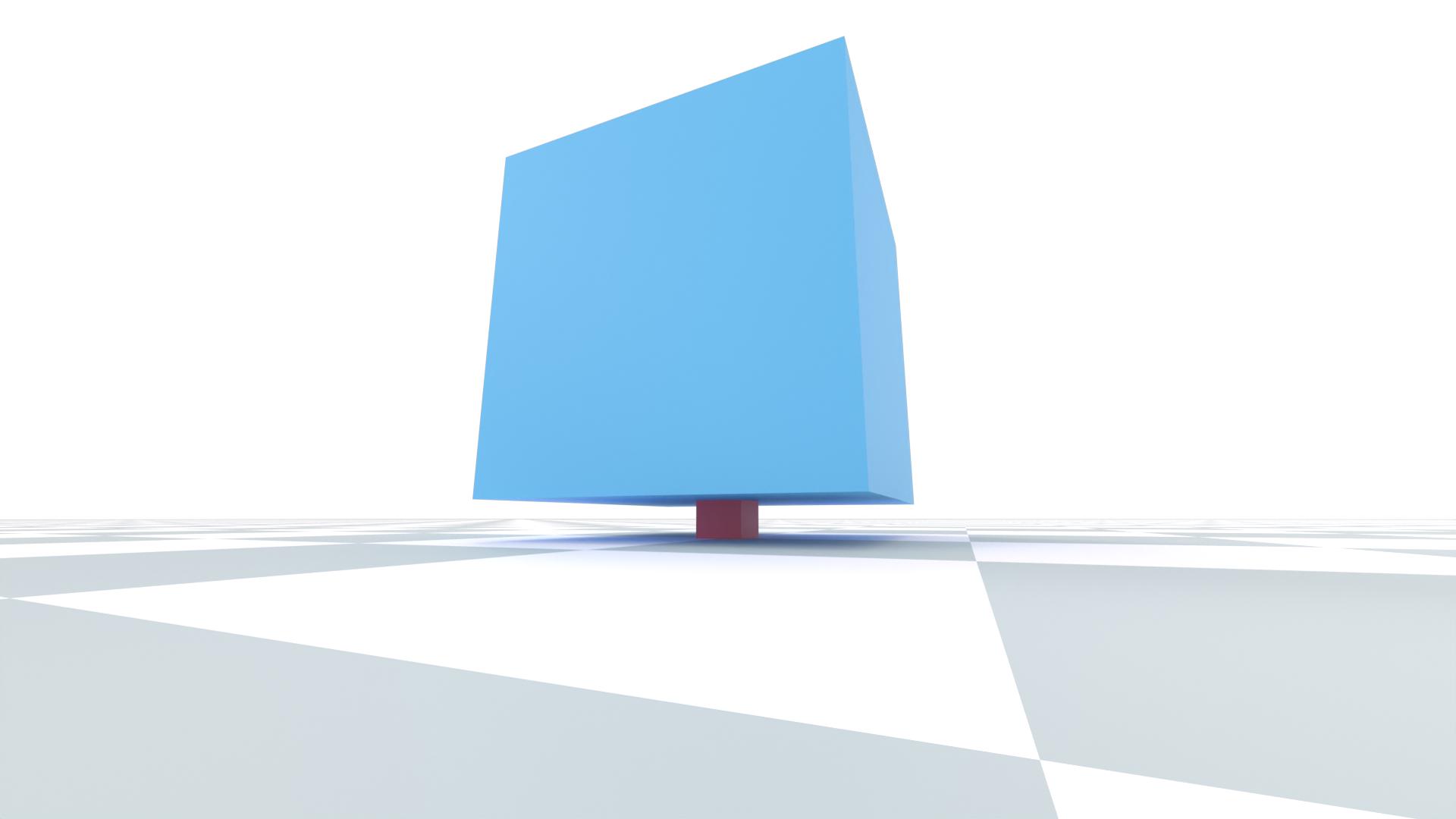}{\textbf{VBD (ours)}}{c} 
\caption{Dropping a large and heavy elastic cube onto a smaller and much lighter box with a mass ratio of 1:2000. Each cube has 400 vertices and 1.5 thousand tetrahedra.
}
\label{fig:CompareToXPBD_MassRatio}
\end{figure}

One of the fundamental challenges of XPBD is handling high mass ratios. This is demonstrated with a simple example in \autoref{fig:CompareToXPBD_MassRatio}, where a large and heavy elastic cube is dropped onto a smaller and much lighter cube, with a mass ratio of 1:2000. In this example, XPBD's collision constraints, even with infinite stiffness, cannot overcome the mass ratio and the smaller cube is entirely crushed upon contact. This is because of the dual formulation of XPBD \cite{macklin_primaldual_2020}. Our method, on the other hand, has no such difficulties with handling high mass ratios.


%% file: 05_Discussion.tex
\section{VBD for Other Simulation Systems}
\label{sec:othersims}
We have described our method in \autoref{sec:vbd} in the context of elastic body dynamics. Yet, VBD is not limited to such simulations and can be used to solve various optimization problems. Here, we consider some other example simulation systems and briefly discuss how our method can be applied. This is not intended as an exhaustive list but merely as examples that could guide the reader to discern how their specific simulation problem could utilize VBD.

\subsection{Particle-Based Simulations}
Particle-based simulations can easily use VBD by simply replacing the vertices in our description above with particles. Since VBD needs the Hessian of the force element energies, implementations would require computing the derivatives of all forces acting on a particle wrt. its position.

Parallelizing particle-based simulations also involves additional considerations.
Mass-spring type simulations, such as peridynamics \cite{Levine14}, can use our parallelization approach with vertex coloring.
However, simulations involving disjoint or loosely-joined particles, such as particle-based fluid simulation
\cite{Muller03,Takahashi15,Peer15}, would not only require recoloring at each time step but also using a conservative neighborhood definition (including potential position change within a time step) for coloring, since position updates can alter the set of neighboring particles that interact with each particle.
\begin{figure}
\centering
\newcommand{\img}[1]{\includegraphics[width=0.495\linewidth,trim=500 0 360 0, clip]{Figures/MassSpring/#1_combined-cc}}
\newcommand{\spc}{\hspace*{1em}}
\newcommand{\scap}[3]{\spc#1\hfill#2\spc\vspace{#3em}}
\img{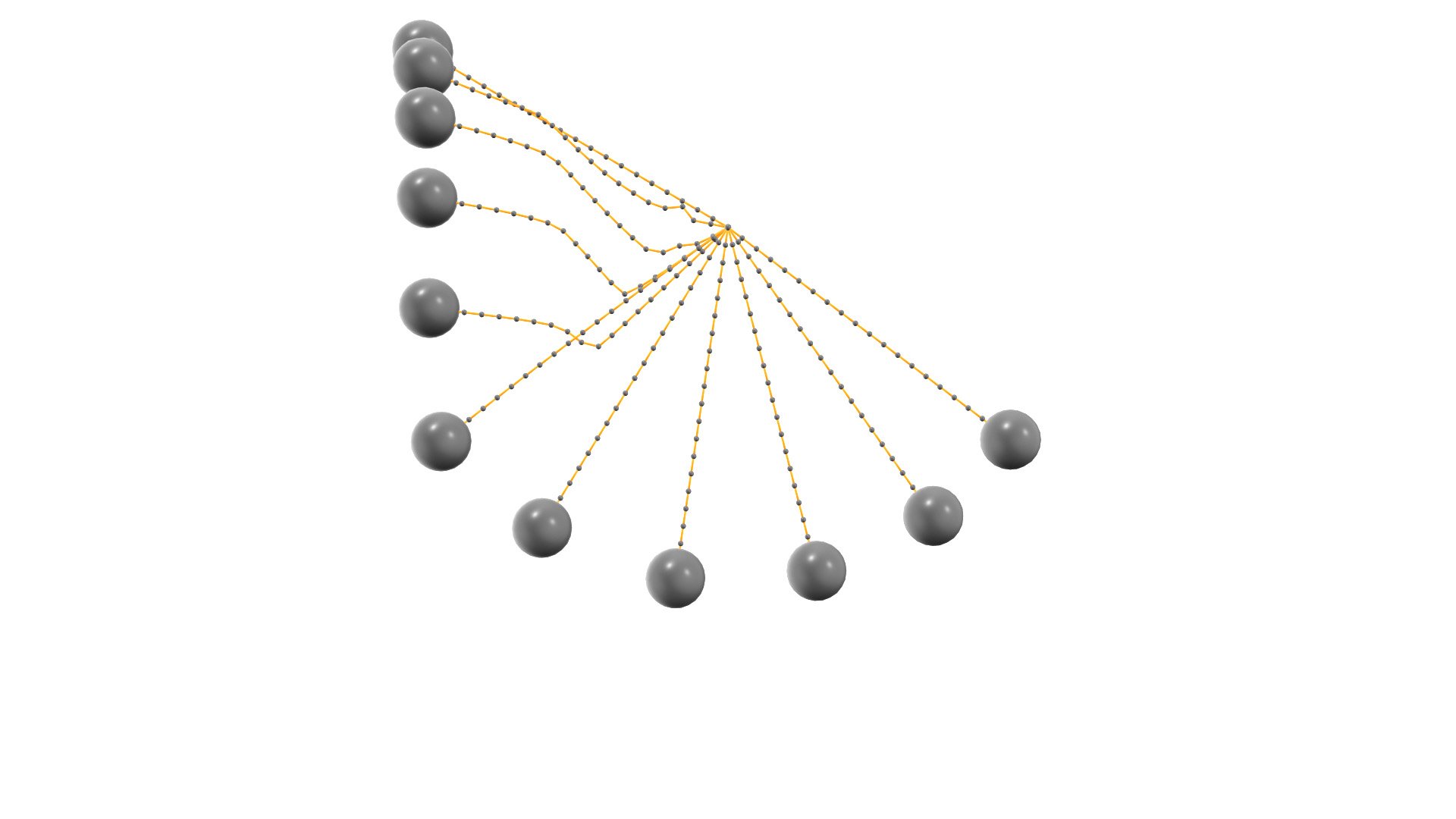}\hfill%
\img{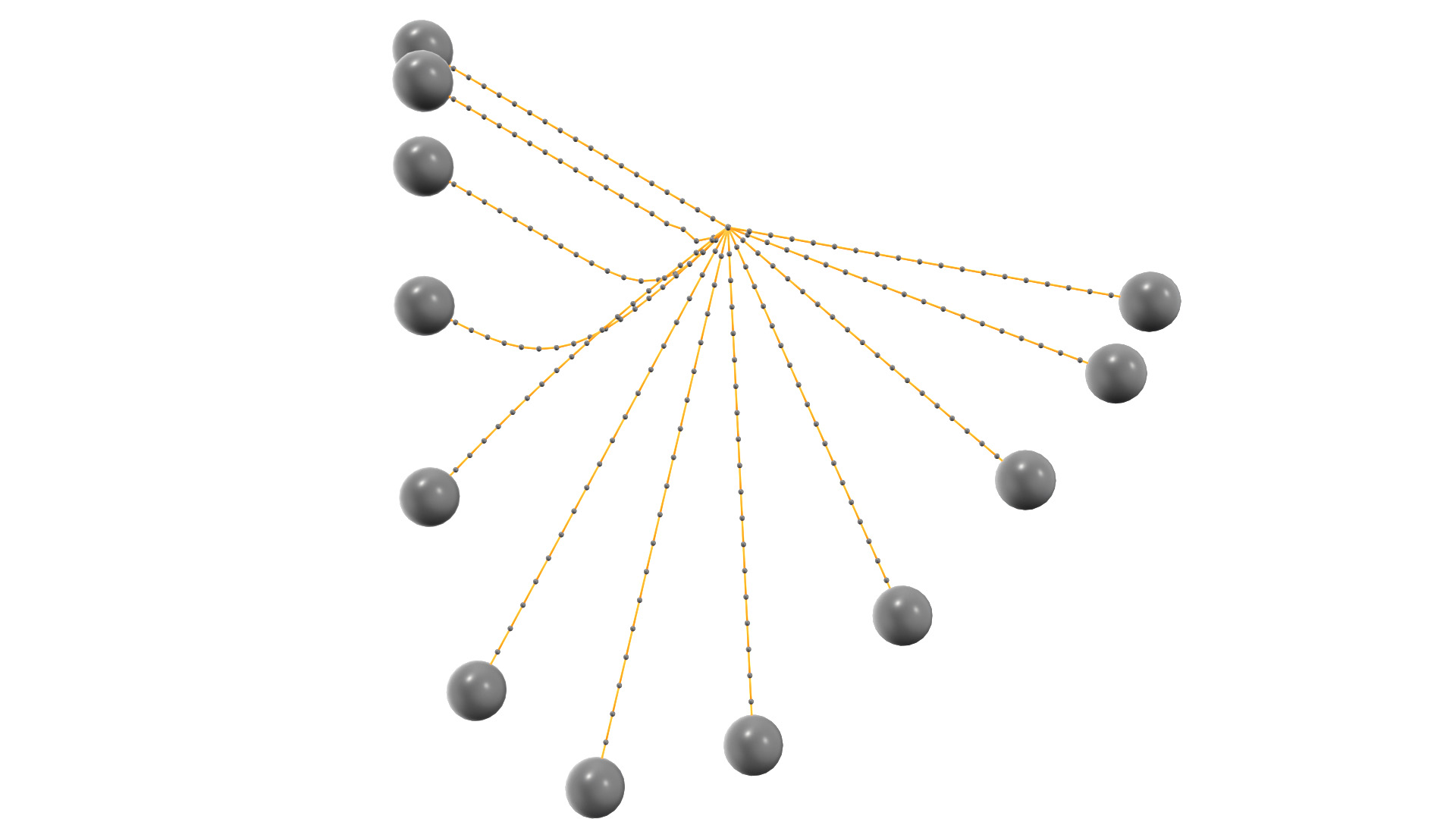}\vspace{-3em}
\small\scap{stiff}{soft}{0}\\
\scap{springs}{springs}{1}\\
\caption{20 particles attached with springs, forming a swinging chain, simulated using VBD with a $S=1$ substep and $100$ iterations per step. The particle on one end of the chain is fixed and the particle on the other end has 1000$\times$ more mass than the others. \sref{Left}~using sufficiently stiff springs, they expand no more than 0.7\% of their rest lengths, despite the substantial mass difference. \sref{Right}~using 100$\times$ less stiff springs, the chain undergoes a visible expansion as it swings.}
\label{fig:massspring}
\end{figure}
\autoref{fig:massspring} shows a simple example where 20 particles, including one that is 1000$\times$ heavier, are connected with springs of two different stiffness, simulated using VBD.

\subsection{Rigid Body Simulation}

For handling rigid body simulations with VBD, we can replace each vertex in our formulation with an entire rigid body, using the variational formulation of rigid body dynamics \cite{Ferguson:2021:RigidIPC}. Unlike a vertex that has only 3 DoF, a rigid body also has rotational DoF, resulting in 6 DoF. Therefore, in our local system, we must solve a larger problem, where $\vec{x}_i \in  \mathbb{R}^{6}$ and $\Vec{H}_i \in \mathbb{R}^{6 \times 6}$, including Hessians of all force elements wrt. all 6 DoF of $\vec{x}_i$. Note that, in this case, these force elements are not internal material forces, but external forces acting on the rigid body, due to collisions or other constraints.

Other than this additional complexity, we can follow the same procedure with VBD. Parallelization with coloring depends on the nature of the rigid body simulation. For example, pre-coloring, as we used in our examples for elastic bodies might work for problems like a rigid body chain. For disjoint rigid bodies interacting through collisions only, dynamic recoloring might be needed.

Articulated rigid bodies can be handled by defining joint constraints with an elastic potential. 
Infinitely stiff constraints are also possible, but VBD cannot guarantee that they will be satisfied using a fixed number of iterations. Another alternative is 
hard constraints can be introduced by
reducing the total DoF in the system and replacing the vertices in our formulation with an articulated rigid body, having more than 6 DoF. Obviously, this would lead to an even larger local system, requiring modifications to the variational formulation. 

\begin{figure}
\centering
\includegraphics[width=\linewidth,trim=100 200 100 120, clip]{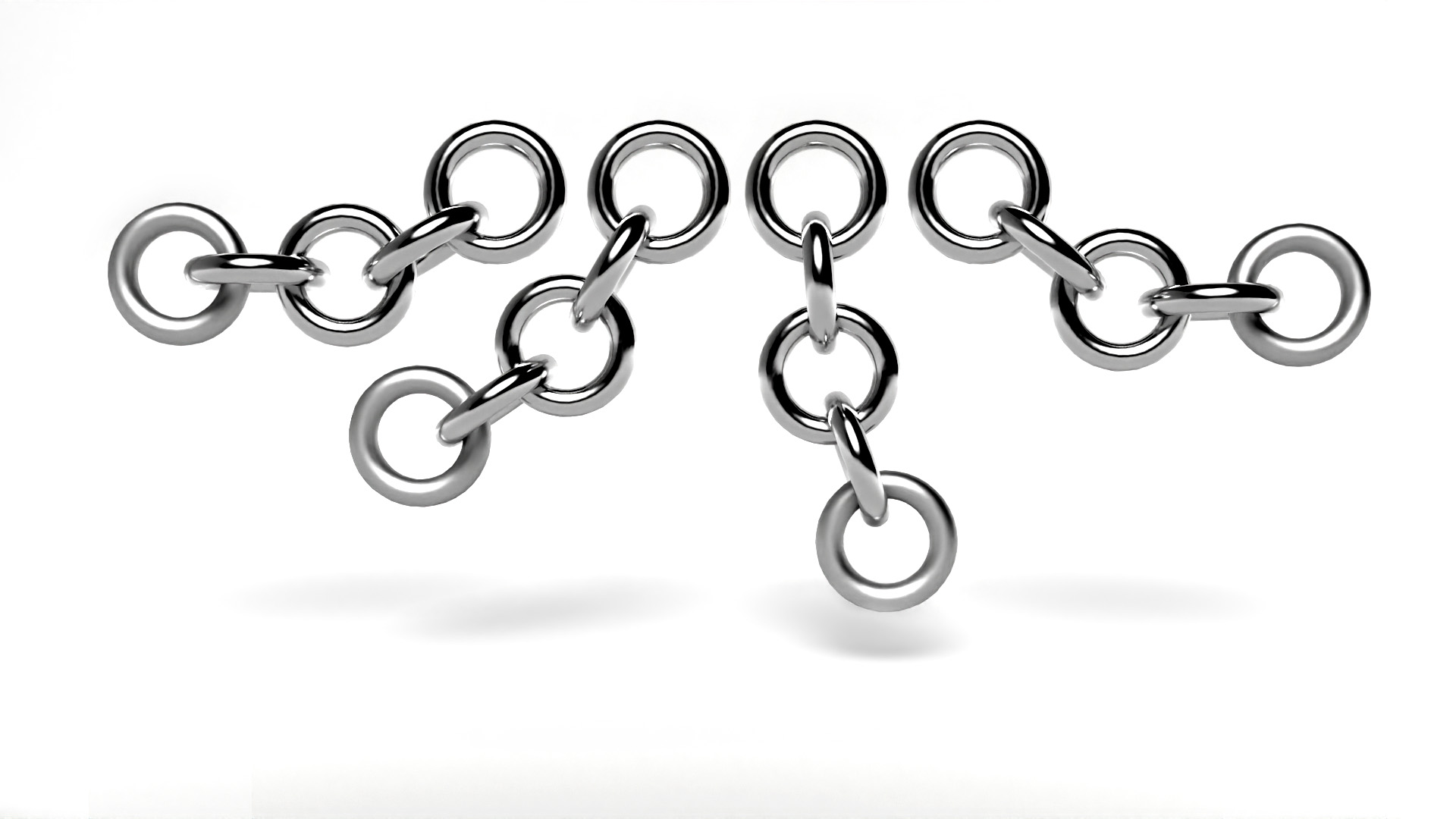}
\caption{5 rigid bodies, each with 6 DoF, forming a chain through collisions, simulated using our VBD formulation for rigid body dynamics.}
\label{fig:rigidbody}
\end{figure}

\begin{figure}
\centering
\newcommand{\img}[3]{\begin{subfigure}{0.498\linewidth}\centering
\includegraphics[width=\linewidth]
{Figures/TeapotOfRigid/#1}
\end{subfigure}}
\img{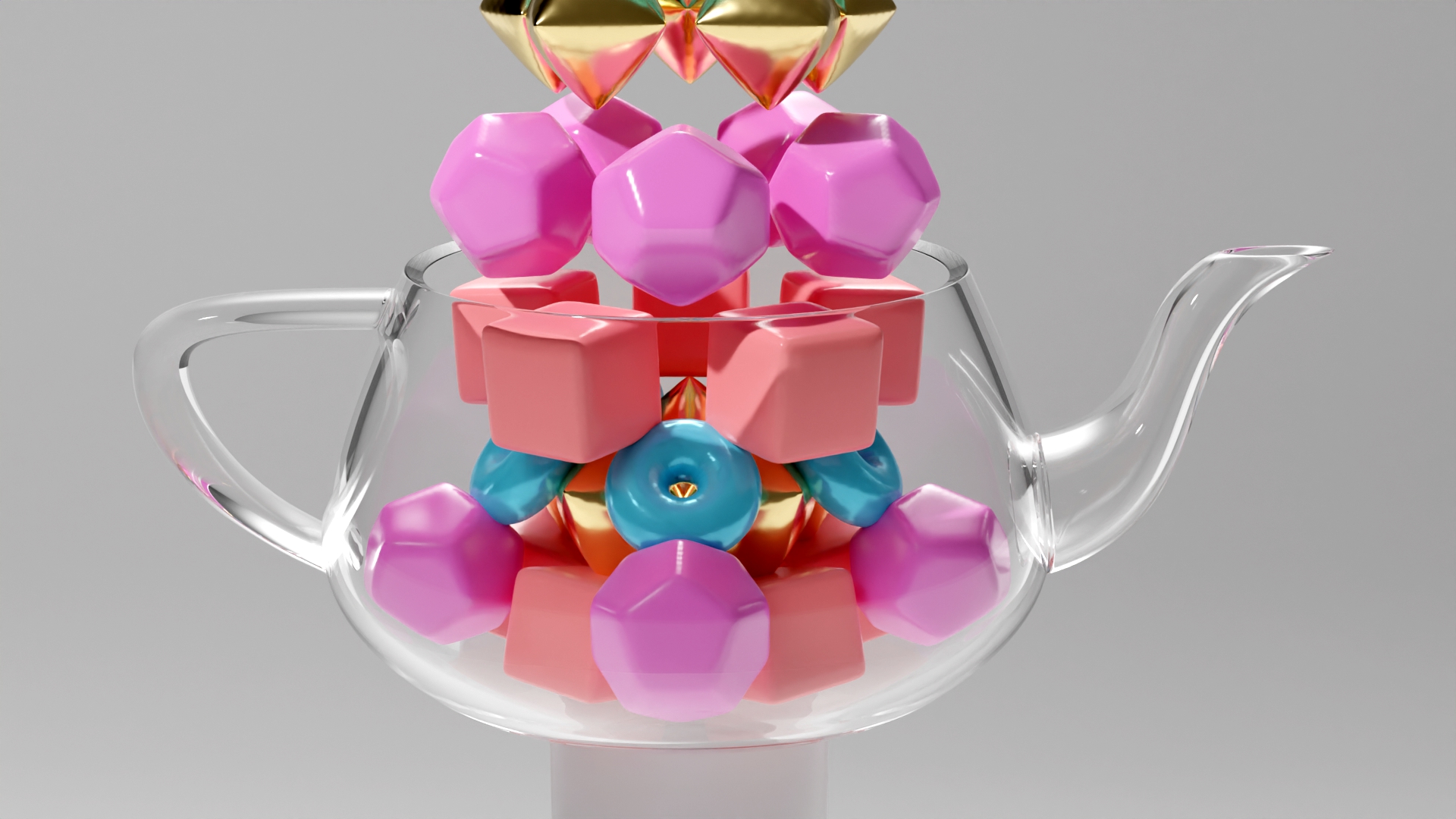}{180 0 150}{}\hfill%
\img{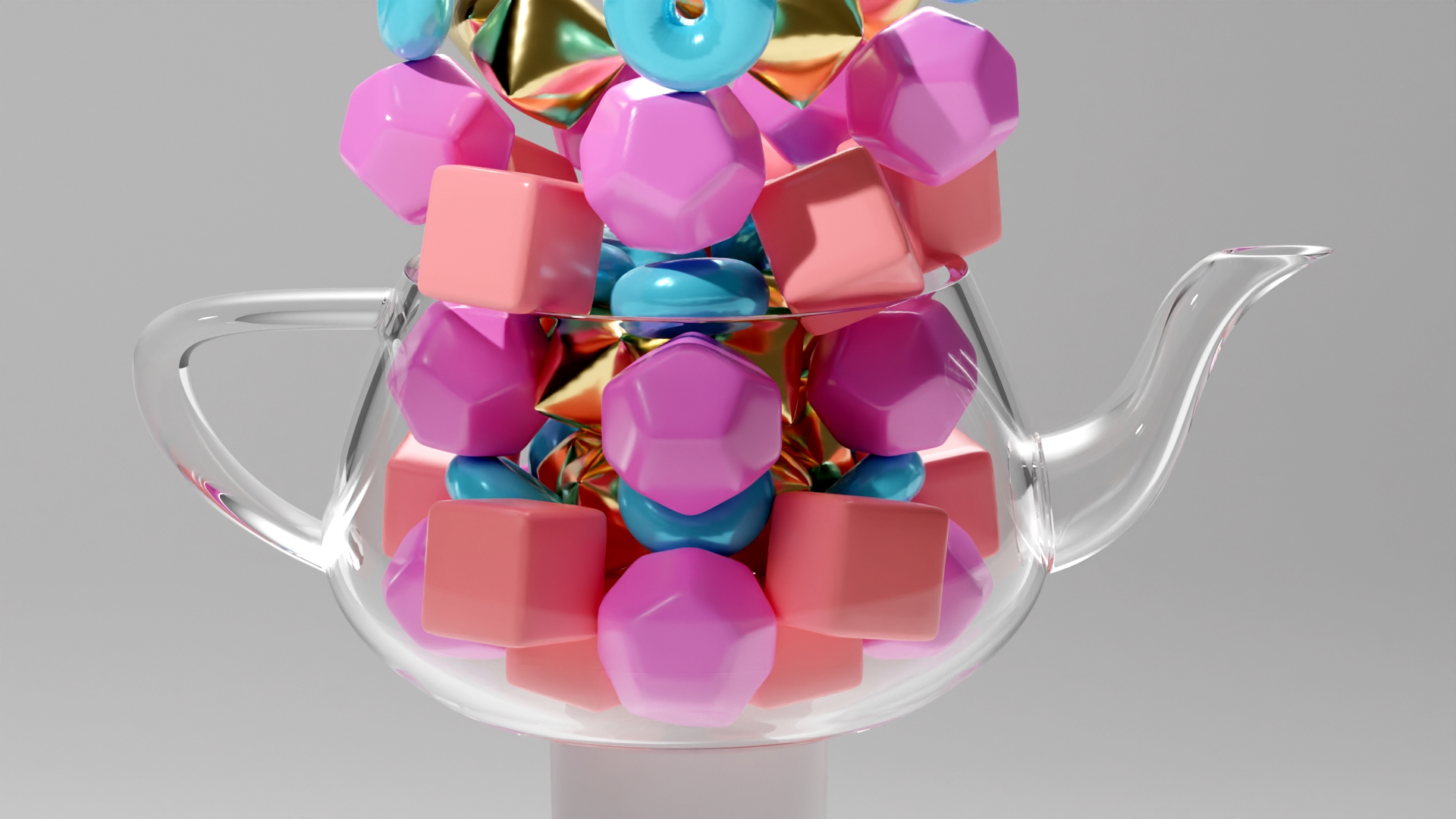}{180 0 150}{}
\img{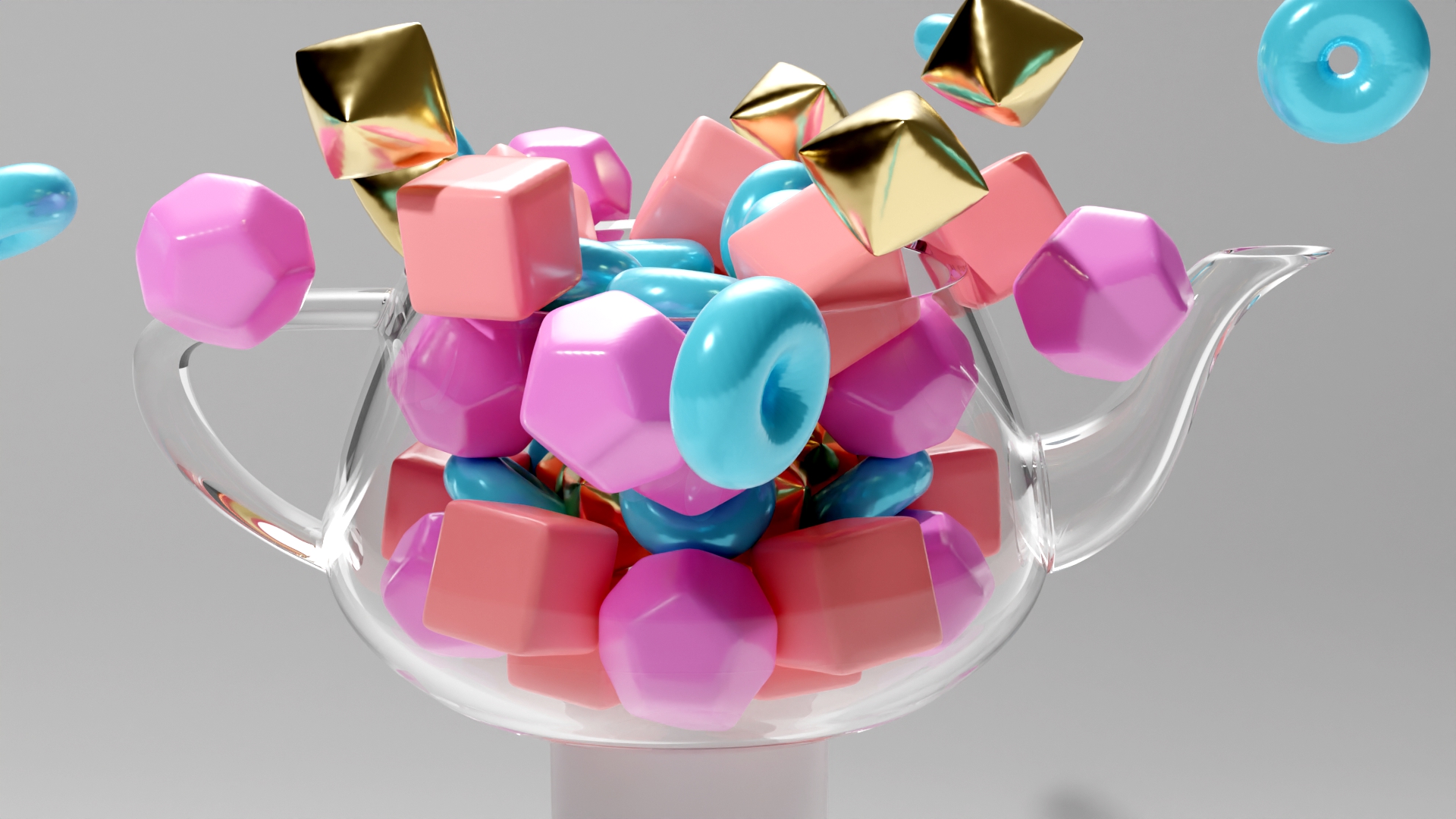}{180 0 150}{}\hfill%
\img{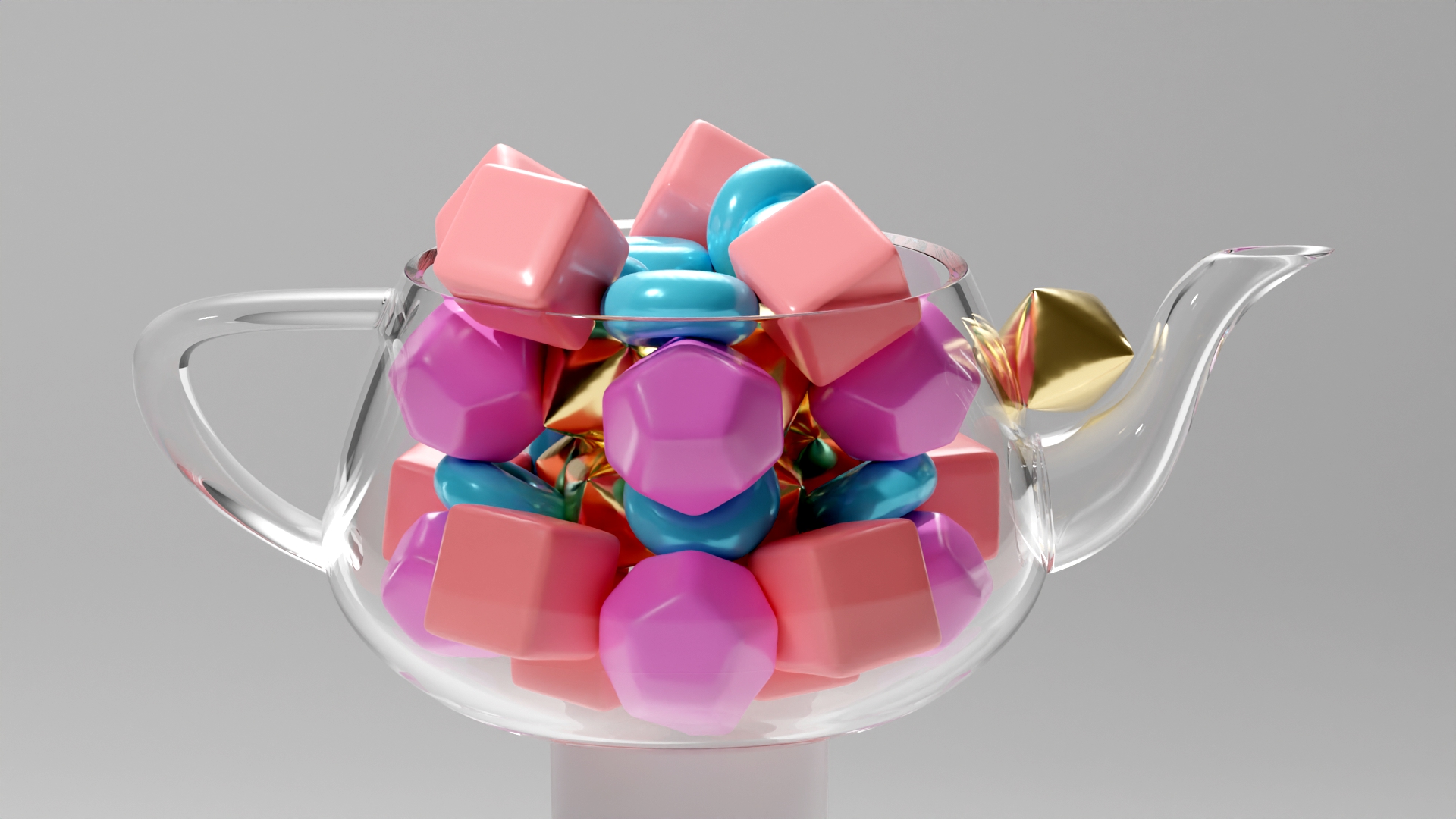}{180 0 150}{}
\caption{Dropping 60 rigid bodies into a Utah teapot, showcasing collisions and frictional contact. Remarkably, one rigid body stays on the spout due to friction. }
\label{fig:RigidToTeapot}
\end{figure}

Example rigid body simulations are shown in \autoref{fig:rigidbody} and \autoref{fig:RigidToTeapot}, simulated using VBD, as described above.

Note that since our collision formulation is based on penetration potential, it corresponds to penalty forces. We leave the exploration of handling impulse-based collisions \cite{Mirtich96} with VBD to future work.

\subsection{Unified Simulations}

Unified simulation systems are useful for handling scenarios that involve different material types. Typical unified simulation systems use a fundamental building block, such as a particle, to represent all supported materials \cite{Muller04,Solenthaler07,Becker09,Martin10,Macklin14}.
We can form a unified simulation system using VBD without representing all materials using the same building block.
For any simulation system described above, we can combine it with another, provided that we can define the information exchange as an energy potential.
For example, when the interactions take place as collisions, we can easily join rigid body simulations with elastic bodies or particles via the collision potential.
Joint constraints with elastic potential would be another easy way to combine different simulation systems. The advantage here is that a large rigid body, for example, can be represented as a single object with just 6 DoF, as opposed to using multiple building blocks that are constrained to move as a rigid construction.
This way, a heterogeneous collection of representations can be joined within the same integrator using VBD.

On the other hand, this form of defining a unified simulation system may be challenging for other types of information exchange, such as evaluating buoyancy. Exploring such problems would be another interesting direction for future research.

%% file: 06_Conclusion.tex
\section{Discussion}

\add{%
We derive our method as a block coordinate descent method for variational time integrators, which offers optimization techniques like PSD projection and line search. 
The fact that we do not require those techniques to guarantee stability actually makes our method a more general solver of nonlinear equations. 
When line search is not used, our method can effectively manage \textit{non-conservative forces}, such as friction, the same as how it handles conservative forces. 
In other words, our method allows for a seamless transition between block coordinate descent and block Gauss-Sediel \cite{hageman1975aspects, Grippo2000BlockGS}.  
While we do not practically utilize these optimization techniques derived from the descent view, they remain available options for users.}

VBD is a descent-based method that operates through local iterations. Therefore, it may not be a good solution for problems that would benefit from a global treatment.

The speed of information travel with VBD depends on the connections of vertices and the number of iterations used. A perturbation applied on a vertex can impact other vertices of a connected chain through force elements at most as far as the number of colors within a single iteration. Therefore, VBD is not ideal for high-resolution stiff systems, as it may require too many iterations for a perturbation of a vertex to travel across the system. In such cases, a global solution using Newton's method may prove to be more effective.



Our collision formulation for VBD is based on penetration potential. Therefore, it cannot guarantee penetration-free results. In fact, penetrations are almost never completely resolved, as some amount of penetration is needed to maintain some collision force. Exploring penetration-free collisions with VBD would be an interesting direction for future research.

In addition, defining a similar collision energy for codimensional objects, particularly for self-collisions, can be a challenge.

\begin{figure}
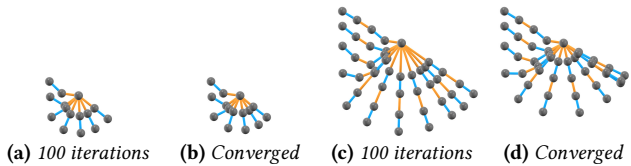

\centering
\newcommand{\img}[5]{\begin{subfigure}{0.245\linewidth}
\includegraphics[width=\linewidth,trim=540 #4 540 #5, clip]{Figures/MassSpring/Rendered_#1_combined#2-cc}\vspace{-0.5em}
\caption{#3}
\end{subfigure}}
\newcommand{\imgA}[2]{\img{#1}{2}{#2}{380}{320}}
\newcommand{\imgB}[2]{\img{#1}{4}{#2}{100}{250}}
\imgA{100iters}{100 iterations}\hfill%
\imgA{converged}{Converged}\hfill%
\imgB{100iters}{100 iterations}\hfill%
\imgB{converged}{Converged}
\caption{A chain of particles connected with soft springs (orange) and 10,000$\times$ stiffer (blue) springs. Simulations with VBD using \sref{a,c}~100 iterations per frame fail to converge and result in excessive extensions, as compared to \sref{b,d}~convreged results.}
\label{fig:stiffnessratios}
\end{figure}

VBD is a primal solver \cite{macklin_primaldual_2020}, so it can easily handle high mass ratios (see \autoref{fig:AdaptiveInitialization}, \ref{fig:CompareToXPBD_MassRatio}, and \ref{fig:massspring}), but it struggles with high stiffness ratios. This is shown in \autoref{fig:stiffnessratios} using a stiffness ratio of 1:10000, where VBD has poor convergence behavior.

\section{Conclusion}

We have presented vertex block descent, an efficient iterative descent-based solution for optimization problems, and described how it can be used for physics-based simulations with implicit Euler integration defined through a variational formulation.
We have explained all essential details of elastic body dynamics using VBD, including handling damping, constraints, collisions, and friction. We have defined an adaptive initialization technique, enabled by VBD's formulation, and discussed how to use momentum-based acceleration to improve convergence. We have also presented effective methods for parallelization using VBD, considering dynamically introduced/removed force elements, and explained how its vertex-level computation improves the parallelization of its Gauss-Seidel iterations.

Our results show that VBD can handle highly complex simulation cases (\autoref{fig:teaser}), it remains stable under extreme stress tests (\autoref{fig:ExtremeTwist}, \ref{fig:ExtremeInitialization}, \ref{fig:ExtremeStretch}, and \ref{fig:ExtremeStretch2}), and offers fast convergence (\autoref{fig:ConvergenceComparisons}). 


In addition, we have summarized how VBD can be used for other types of simulation problems, such as particle systems and rigid bodies, including unified simulations. We have also mentioned some related future research directions and discussed VBD's limitations.